%% file: P7DMLine.tex
\documentclass[aps,prd,superscriptaddress,floatfix,nofootinbib,notitlepage]{revtex4-1}

\usepackage{graphicx}
\usepackage{amsmath}
\usepackage{longtable}
\usepackage{aas_macros}
\usepackage{xcolor}

\graphicspath{{./figures/}}

\usepackage[sort&compress]{natbib}
\usepackage{hyperref}
\input{sty_sty.tex}

\begin{document}

\title{Search for Gamma-ray Spectral Lines with the \fermi\ Large Area Telescope and Dark Matter Implications}

\date{\today}

\input{authors.tex}

\begin{abstract}
\pagebreak
Weakly Interacting Massive Particles (WIMPs) are a theoretical class
of particles that are excellent dark matter candidates.  WIMP
annihilation or decay may produce essentially monochromatic
\gammaRays\ detectable by the \Fermi~Large Area Telescope (LAT) against the astrophysical \gammaRayHyph\ emission of the Galaxy.  
We have searched for spectral lines in the energy range 5--300~GeV using 3.7~years 
of data, reprocessed with updated instrument calibrations and an improved 
energy dispersion model compared to the previous \Fermi-LAT Collaboration line 
searches.  We searched in five regions selected to optimize sensitivity to different 
theoretically-motivated dark matter density distributions.  We did not find any globally significant lines in our {\em a priori} 
search regions and present 95\% confidence limits for 
\newText{annihilation cross sections of self-conjugate WIMPs}
and decay lifetimes. Our most significant
fit occurred at 133~GeV in our smallest search region and had a local
significance of 3.3 \newText{standard deviations}, which translates
to a global significance of \newText{1.5 standard deviations}.  We discuss potential
systematic effects in this search, \newText{ and examine the feature at 133~GeV
in detail.   We find that both the use of reprocessed data and of additional information in the energy
dispersion model contribute to the reduction in significance of the
line-like feature near 130~GeV relative to significances reported in other
works.   We also find that the feature is narrower than the LAT energy resolution at the 
level of 2 to 3 standard deviations, which somewhat disfavors the interpretation of the 133~GeV feature 
as a real WIMP signal.}
  
\end{abstract}
\maketitle

\input{intro.tex}

\input{Method.tex}

\input{Fitting.tex}

\input{Systematic_Errors.tex}

\input{Results.tex}

\clearpage
\input{135Feature.tex}

\input{Discussion.tex}

\begin{acknowledgments}
The \textit{Fermi} LAT Collaboration acknowledges generous ongoing support
from a number of agencies and institutes that have supported both the
development and the operation of the LAT as well as scientific data analysis.
These include the National Aeronautics and Space Administration and the
Department of Energy in the United States, the Commissariat \`a l'Energie Atomique
and the Centre National de la Recherche Scientifique / Institut National de Physique
Nucl\'eaire et de Physique des Particules in France, the Agenzia Spaziale Italiana
and the Istituto Nazionale di Fisica Nucleare in Italy, the Ministry of Education,
Culture, Sports, Science and Technology (MEXT), High Energy Accelerator Research
Organization (KEK) and Japan Aerospace Exploration Agency (JAXA) in Japan, and
the K.~A.~Wallenberg Foundation, the Swedish Research Council and the
Swedish National Space Board in Sweden.

Additional support for science analysis during the operations phase is gratefully
acknowledged from the Istituto Nazionale di Astrofisica in Italy and the Centre National d'\'Etudes Spatiales in France.

\newText{We would also like to thank Christoph Weniger for providing the limit values used in \Figref{SigmaVUL}.}
\end{acknowledgments}

\clearpage

\appendix

\input{Data_Reprocessing.tex}

\input{ROI_Optimization.tex}

\input{Formalism.tex}

\input{App_Systematic_Errors.tex}

\input{Results_Tables.tex}
\clearpage

\bibliography{P7DMLine}

\end{document}

%% file: sty_sty.tex
\newcommand{\gammaRay}{$\gamma$ ray}
\newcommand{\gammaRays}{$\gamma$ rays}
\newcommand{\gammaRayHyph}{$\gamma$-ray}

\newcommand{\stools}{\emph{ScienceTools}}
\newcommand{\stool}[1]{\emph{#1}}

\newcommand{\gtmktime}{\stool{gtmktime}}

\newcommand{\selection}[1]{\texttt{#1}}

\newcommand{\irf}[1]{\texttt{#1}}
\newcommand{\psix}{\irf{Pass~6}}
\newcommand{\pseven}{\irf{Pass~7}}
\newcommand{\psevenrep}{\irf{Pass~7REP}}
\newcommand{\peight}{\irf{Pass~8}}
\newcommand{\aeff}{\ensuremath{A_{\rm eff}}}

\newcommand{\fov}{{\rm Fo\kern-1ptV}}

\newcommand{\tobs}{\ensuremath{t_{\rm obs}}}
\newcommand{\probE}{\ensuremath{P_{\rm E}}}

\newcommand{\Secref}[1]{Sec.~\ref{#1}}

\newcommand{\Appref}[1]{App.~\ref{#1}}
\newcommand{\Tabref}[1]{Tab.~\ref{tab:#1}}

\newcommand{\Figref}[1]{Fig.~\ref{fig:#1}}
\newcommand{\Figrefs}[2]{Figs.~\ref{fig:#1} and \ref{fig:#2}}

\newcommand{\Eqref}[1]{Eq.~\eqref{#1}}

\newcommand{\Sectionref}[1]{Section~\ref{#1}}
\newcommand{\Tableref}[1]{Table~\ref{tab:#1}}
\newcommand{\Figureref}[1]{Figure~\ref{fig:#1}}

\newcommand{\Fermi}{{\textit{Fermi}}}
\newcommand{\fermi}{\Fermi}

\input{sty_thisPaperStyle.tex}

%% file: sty_thisPaperStyle.tex
\newcommand{\Etrue}{\ensuremath{E}}
\newcommand{\Egamma}{\ensuremath{E_{\gamma}}}
\newcommand{\Ereco}{\ensuremath{E'}}
\newcommand{\Epivot}{\ensuremath{E_{0}}}

\newcommand{\newText}{}

\newcommand{\pTrue}{\ensuremath{\hat{p}}}
\newcommand{\pReco}{\ensuremath{\hat{p}^{\,\prime}}}
\newcommand{\vTrue}{\ensuremath{\hat{v}}}

\newcommand{\intROI}{\ensuremath{\int^{ROI}}}
\newcommand{\intPTrue}{\ensuremath{\int}}
\newcommand{\intFOV}{\ensuremath{\int^{FOV}}}
\newcommand{\intTime}{\ensuremath{\int}}
\newcommand{\intEnergy}{\ensuremath{\int}}
\newcommand{\intPE}{\ensuremath{\int}}

\newcommand{\OmegaVhat}{\ensuremath{\Omega_{\hat{v}}}}

\newcommand{\sVec}{\ensuremath{\vec{s}}}

\newcommand{\GamBkg}{\ensuremath{\Gamma_{\rm bkg}}}

\newcommand{\nSig}{\ensuremath{n_{\rm sig}}}
\newcommand{\nBkg}{\ensuremath{n_{\rm bkg}}}

\newcommand{\nPred}{\ensuremath{n^{\rm pred}}}
\newcommand{\nPredSig}{\ensuremath{n^{\rm pred}_{\rm sig}}}
\newcommand{\nPredBkg}{\ensuremath{n^{\rm pred}_{\rm bkg}}}
\newcommand{\nObs}{\ensuremath{n^{\rm obs}}}

\newcommand{\SSig}{\ensuremath{S_{\rm sig}}}
\newcommand{\SBkg}{\ensuremath{S_{\rm bkg}}}

\newcommand{\ISig}{\ensuremath{I_{\rm sig}}}
\newcommand{\IBkg}{\ensuremath{I_{\rm bkg}}}

\newcommand{\CSig}{\ensuremath{C_{\rm sig}}}
\newcommand{\CBkg}{\ensuremath{C_{\rm bkg}}}

\newcommand{\FSig}{\ensuremath{F_{\rm sig}}}
\newcommand{\FBkg}{\ensuremath{F_{\rm bkg}}}

\newcommand{\wSig}{\ensuremath{w_{\rm sig}}}
\newcommand{\wBkg}{\ensuremath{w_{\rm bkg}}}
\newcommand{\wROI}{\ensuremath{w^{\rm ROI}}}

\newcommand{\slocal}{\ensuremath{s_{\rm local}}}
\newcommand{\sglobal}{\ensuremath{s_{\rm global}}}

\newcommand{\plocal}{\ensuremath{p_{\rm local}}}
\newcommand{\pglobal}{\ensuremath{p_{\rm global}}}

\newcommand{\smax}{\ensuremath{s_{\rm max}}}
\newcommand{\ntrial}{\ensuremath{n_{\rm t}}}

\newcommand{\Deff}{\ensuremath{D_{\rm eff}}}

\newcommand{\Exposure}{\ensuremath{\mathcal{E}}}

\newcommand{\Rgc}{\ensuremath{R_{\rm GC}}}

\newcommand{\sigE}{\ensuremath{\sigma_{E}}}

\def\SPSB#1#2{\rlap{\textsuperscript{{#1}}}_{#2}}

\ifdefined\bwfigures
\newcommand{\colSw}[1]{#1_bw} 
\else
\newcommand{\colSw}[1]{#1} 
\fi

\newlength{\enumindent}
\newcounter{fermicnt}
% ...
\newenvironment{fermienumerate}{
  \begin{list}{\arabic{fermicnt}.}{\usecounter{fermicnt}%
  \setlength{\leftmargin}{0pt}%
  \setlength{\labelwidth}{0pt}}%
  \setlength{\itemindent}{\enumindent}%
  \setlength{\labelsep}{\enumindent}%
  \setlength{\topsep}{0pt}%
  \setlength{\itemsep}{0pt}%
}{
  \end{list}%
}

\newenvironment{fermiitemize}[1][$\bullet$]{%
  \begin{list}{#1}{%
  \setlength{\leftmargin}{0pt}%
  \setlength{\labelwidth}{0pt}}%
  \setlength{\itemindent}{\enumindent}%
  \setlength{\labelsep}{\enumindent}%
  \setlength{\topsep}{0pt}%
  \setlength{\itemsep}{0pt}%
}{%
  \end{list}%
}

\newlength{\twocolfigwidth}
\newlength{\onecolfigwidth}
\newlength{\twothirdscolfigwidth}
\newcommand{\onewidepanel}[3]{%
  \begin{figure}[#1]
    \centering
    \includegraphics[width=\twocolfigwidth,clip=true,trim=0 5 0 5]{#2}
    
    #3
  \end{figure}
}

\newcommand{\onepanel}[3]{%
  \begin{figure}[#1]
    \centering
    \includegraphics[width=\onecolfigwidth]{#2}
    
    #3
  \end{figure}
}

\newcommand{\twopanel}[4]{%
  \begin{figure}[#1]
    \centering
    \includegraphics[width=\onecolfigwidth]{#2}\hfill%
    \includegraphics[width=\onecolfigwidth]{#3} 

    #4
  \end{figure}
}

\newcommand{\threepanelvert}[5]{%
  \begin{figure}[#1]
    \centering
    \includegraphics[width=\onecolfigwidth]{#2} \\
    \includegraphics[width=\onecolfigwidth]{#3} \\
    \includegraphics[width=\onecolfigwidth]{#4} 

    #5
  \end{figure}
}

\newcommand{\fourpanel}[6]{%
  \begin{figure}[#1]
    \centering
    \includegraphics[width=\onecolfigwidth]{#2}\hfill%
    \includegraphics[width=\onecolfigwidth]{#3}\\
    \includegraphics[width=\onecolfigwidth]{#4}\hfill%
    \includegraphics[width=\onecolfigwidth]{#5}

    #6
  \end{figure}
}

%% file: authors.tex
\author{M.~Ackermann}
\affiliation{Deutsches Elektronen Synchrotron DESY, D-15738 Zeuthen, Germany}
\author{M.~Ajello}
\affiliation{Space Sciences Laboratory, 7 Gauss Way, University of California, Berkeley, CA 94720-7450, USA}
\author{A.~Albert}
\email{albert.143@osu.edu}
\affiliation{Department of Physics, Center for Cosmology and Astro-Particle Physics, The Ohio State University, Columbus, OH 43210, USA}
\author{A.~Allafort}
\affiliation{W. W. Hansen Experimental Physics Laboratory, Kavli Institute for Particle Astrophysics and Cosmology, Department of Physics and SLAC National Accelerator Laboratory, Stanford University, Stanford, CA 94305, USA}
\author{L.~Baldini}
\affiliation{Universit\`a  di Pisa and Istituto Nazionale di Fisica Nucleare, Sezione di Pisa I-56127 Pisa, Italy}
\author{G.~Barbiellini}
\affiliation{Istituto Nazionale di Fisica Nucleare, Sezione di Trieste, I-34127 Trieste, Italy}
\affiliation{Dipartimento di Fisica, Universit\`a di Trieste, I-34127 Trieste, Italy}
\author{D.~Bastieri}
\affiliation{Istituto Nazionale di Fisica Nucleare, Sezione di Padova, I-35131 Padova, Italy}
\affiliation{Dipartimento di Fisica e Astronomia ``G. Galilei'', Universit\`a di Padova, I-35131 Padova, Italy}
\author{K.~Bechtol}
\affiliation{W. W. Hansen Experimental Physics Laboratory, Kavli Institute for Particle Astrophysics and Cosmology, Department of Physics and SLAC National Accelerator Laboratory, Stanford University, Stanford, CA 94305, USA}
\author{R.~Bellazzini}
\affiliation{Istituto Nazionale di Fisica Nucleare, Sezione di Pisa, I-56127 Pisa, Italy}
\author{E.~Bissaldi}
\affiliation{Institut f\"ur Astro- und Teilchenphysik and Institut f\"ur Theoretische Physik, Leopold-Franzens-Universit\"at Innsbruck, A-6020 Innsbruck, Austria}
% ...

%% file: intro.tex
\section{INTRODUCTION}\label{sec:intro}

Cosmological studies indicate that $\sim 27\%$ of the energy density of the Universe is non-baryonic dark matter (DM)~\cite{Ade:2013lta}.  
While substantial astrophysical evidence exists for DM through its gravitational interaction, little has been determined about the composition of the DM or its properties.  In a popular class of models~\cite{Martin:1997ns,Chung:2003fi,Pape:2006ar}, the DM is a weakly interacting massive particle (WIMP), denoted by $\chi$. 
In many models \newText{WIMP pairs} can annihilate into a photon ($\gamma$) and a second particle ($X$), for example, $\gamma \gamma$, $\gamma Z$, or $\gamma H$.  \newText{(See~\cite{Jungman:1995df,Bertone:2005xv} for reviews on WIMPs and indirect DM detection.)} 
Since DM is strongly constrained to be electrically neutral, it has no direct coupling to photons.  Thus the process $\chi \chi \rightarrow \gamma X$ occurs only through higher order loops resulting in a branching fraction that is only $\sim{10^{-4}}-{10^{-1}}$~\cite{REF:Bergstrom:1997fh,REF:Matsumoto:2005ui,REF:Ferrer:2006hy,REF:Gustafsson:2007pc,REF:Profumo:2008yg}. If a WIMP annihilates to $\gamma X$ the photons are monochromatic with rest-frame energy

\begin{equation}\label{eq:LineEn}
\Egamma = m_{\chi} \left( 1-\frac{m^2_{X}}{4m^2_{\chi}} \right).
\end{equation}

\noindent An intrinsic broadening occurs if $X$ is an unstable particle like $Z$. In the case of $X$ being a second photon, the \gammaRayHyph\ line appears at the mass of the WIMP particle.  WIMP decay could also produce a monochromatic signal~\cite{Bertone:2007aw,Gustafsson:2013gca} (e.g., $\chi\rightarrow\gamma\nu$~\cite{REF:Ibarra:2007wg}).  \newText{Additionally, \gammaRays\ created in WIMP annihilations via internal bremsstrahlung could produce a sharp spectral feature~\cite{Bringmann:2007nk}, but this channel is not considered in this search.} We assume WIMPs in the Milky Way are non-relativistic ($v\sim10^{-3}c$), therefore these signals should be approximately monochromatic in the lab frame as well. In this paper we present a search for monochromatic \gammaRays\ from WIMP annihilation or decay. 

The {\em Fermi Gamma-ray Space Telescope} (\Fermi\/) with its main instrument, the Large Area Telescope (LAT)~\cite{REF:2009.LATPaper}, is exploring the \gammaRayHyph\ sky in the energy range 20~MeV to above 300~GeV.  Previous searches by the LAT Collaboration for \gammaRayHyph\ lines were published using 11 months and 2 years of LAT data~\cite{REF:2010.LineSearch, REF:2012.LineSearch}.   For the search presented here, we use 3.7 years of LAT data that have been reprocessed with updated calibrations.  Additionally, two analysis improvements enhance the sensitivity of this search relative to our previous papers: (i) we included an event-by-event estimate of the energy reconstruction quality in our parametrization of the energy dispersion and (ii) we selected regions of interest (ROIs) \textit{a priori} to maximize the sensitivity based on different DM density profiles.

\newText{Detections of a line-like feature at 130~GeV have been reported in the literature.  This feature is reported to be strongly correlated with the Galactic center region~\cite{Bringmann:2012vr,REF:Weniger:2012tx,Tempel:2012ey,Su:2012ft},
and also with nearby galaxy clusters~\cite{Hektor:2012kc}, and unassociated LAT sources~\cite{Su:2012zg,Hektor:2012jc}. The feature has not been seen in the vicinity of nearby dwarf galaxies~\cite{GeringerSameth:2012sr}.  However such a signal is expected to be much fainter than in the Galactic center.  Potential instrumental effects and a similar feature detected in the bright \gammaRayHyph\ emission from cosmic-ray (CR) interactions in Earth's upper atmosphere (the Limb) have also been discussed~\cite{Whiteson:2012hr,Hektor:2012ev,Finkbeiner:2012ez}.  A systematic investigation of the spatial morphology of the 130~GeV feature and other line-like features in the Galactic plane is presented in~\cite{Boyarsky:2012ca}.  In addition to the results from our search for \gammaRayHyph\ lines, we also include a detailed investigation of this feature in the Galactic center region and the Limb.}

\Sectionref{sec:method} describes \newText{the LAT instrument and} the event selections used for this analysis. \Sectionref{sec:method_roi} describes the choice of ROIs, and \Secref{sec:method_2DPDF} describes the development of the energy dispersion model. \Sectionref{sec:fitting} presents the fitting procedure. \Sectionref{sec:systematics} summarizes the  instrumental and methodological uncertainties associated with this search.  \Sectionref{sec:results} presents the fitting results and derives upper limits for DM annihilation and decay assuming several potential distributions of DM.  \Sectionref{sec:133GeV_Feature} describes studies performed specifically to explore the line-like feature at 133~GeV detected with moderate local significance in our smallest search region. Finally, \Secref{sec:discussion} discusses our results and conclusions.

%% file: Method.tex
\section{LAT INSTRUMENT AND EVENT SELECTION}\label{sec:method}

\newText{The LAT is a pair-conversion telescope, which converts \gammaRays\ to $e^{+}e^{-}$ pairs that are tracked in the instrument.   The data analysis is
event-based; individual events are reconstructed and their energies and
directions are estimated from the reconstructed data.    Rates of CR backgrounds can exceed the \gammaRayHyph\ rates by
factors of up to $10^4$,
requiring powerful event selection criteria to obtain relatively pure
\gammaRayHyph\ samples.} 

\newText{The LAT consists of three detector subsystems:  a
tracker/converter to promote pair conversion and measure the
directions of the resulting particles, a calorimeter composed of 8.6
radiation lengths of CsI(Tl) scintillation crystals that provides an
energy resolution of $\Delta E / E \sim 10\%$ at 100~GeV, and an
anticoincidence detector of plastic scintillator tiles that
surrounds the tracker and is key in CR background rejection.   
The tracker comprises 18 $x$-$y$ layers of silicon-strip detectors; 
the front 12 layers are interleaved with thin (3\% of a radiation
length) tungsten converter foils, then the next 4 layers are interleaved with
thick (18\% of a radiation length) foils, and the final 2 layers have no converter foils.  
Detailed descriptions of the LAT and of its performance can
be found elsewhere~\cite{REF:2009.LATPaper,REF:2012.P7Perf}.}

\newText{Iterations of the LAT event reconstruction and classification algorithms have been grouped into so called ``Passes''.  The 
first five ``Passes'' occurred before launch.  For the first three years of the mission, data were processed with the \psix\ 
version of the algorithms.   Since then, the data have been processed with \pseven, which consists of the same event reconstruction
algorithms, but the event classification criteria were updated to
account for knowledge gained since launch.  (Before 
switching to \pseven\ the LAT Collaboration also reprocessed all of the original \psix\ data with the \pseven\ algorithms, 
so as to provide a single, coherent data set.)  Finally, in 2012 and 2013, we reprocessed the data using almost exactly 
the same \pseven\ algorithms, but with updated calibration constants in the reconstruction algorithms to make the \psevenrep\ data sets. More details about the data reprocessing are provided in \Appref{app:pass_7_rep} and \cite{REF:Bregoen:2013p7rep}.   
All of these data, as well as more information about recommended usage,
are publicly available from the Fermi Science Support Center~\footnote{The LAT photon data are available at \url{http://fermi.gsfc.nasa.gov/ssc/data/access/}}.}

\newText{Each Pass of the algorithms implements several different event selection criteria that are optimized for different types of analyses.  
In \pseven, the LAT Collaboration implemented four nested event selections that provide varying levels of CR background rejection.  
The names of the event selections, as well as the types of analyses
they are optimized for, are listed in \Tabref{event_class}.   The nomenclature convention for the various event selections is to provide the
Pass version and the name of the event selection criteria (e.g.,
\irf{P7REP\_CLEAN}).   Associated with each event selection are instrument response functions (IRFs)
that parametrize the LAT performance.   As our understanding of the
instrument improves, from time to time the LAT Collaboration updates
the IRFs for the various event selections.   The IRF names indicate
which data set they are associated with, as well as a version number
(e.g., \irf{P7REP\_CLEAN\_V10}).  More details about the event reconstruction, event selection criteria, and IRFs can be found in 
\cite{REF:2012.P7Perf}\footnote{Performance details for
  all the iterations of the event reconstruction and classification
  algorithms used since launch are available at
  \url{http://www.slac.stanford.edu/exp/glast/groups/canda/lat_Performance.htm}}.}

\begin{table}[ht]
  \caption{\label{tab:event_class}\newText{Event selections in \pseven\ and
    \psevenrep\ iterations of the LAT event reconstruction and
    classification algorithms.}}
 \begin{center}
  \begin{tabular}{llc}
    \hline\hline
  \pseven\ Selection & \psevenrep\ Selection & Recommended Use /  Notes \footnote{\newText{The selections are nested; each is a strict
    subset of the previous one.}}\footnote{\newText{Although the selection criteria are 
    identical between the \pseven\ and \psevenrep\ versions, the
    events selected differ due to changes in the calibration
    constants used during event reconstruction.}} \\
  \hline
  \irf{P7\_TRANSIENT} & \irf{P7REP\_TRANSIENT} & Analysis of short-duration ($< 200$ s) transient sources \\
  \irf{P7\_SOURCE} & \irf{P7REP\_SOURCE} & Analysis of point sources and regions of bright diffuse emission \\
  \irf{P7\_CLEAN} & \irf{P7REP\_CLEAN} & Analysis of regions of faint diffuse emission \\
  \irf{P7\_ULTRACLEAN} & \irf{P7REP\_ULTRACLEAN} & Nearly identical
  selection as \irf{CLEAN} for energies above a few GeV \\
  \hline\hline
  \end{tabular}
  \end{center}
\end{table}

\newText{As discussed in \Secref{subsec:method_event}, we use only the
\irf{P7REP\_CLEAN} event selection for the line search.   For
certain studies of potential systematic biases, we compare the 
\irf{P7REP\_CLEAN} sample against either the \irf{P7REP\_TRANSIENT} 
or \irf{P7REP\_SOURCE} sample.   Finally, as part of our examination 
of the feature near 130~GeV we compare the \irf{P7REP\_CLEAN} sample
with the \irf{P7\_CLEAN} sample used in previous papers~\citep{Bringmann:2012vr,REF:Weniger:2012tx,Tempel:2012ey,Su:2012ft}.}

\subsection{Event selection}\label{subsec:method_event}

We searched for the presence of \gammaRayHyph\ lines between 5 and 300~GeV; to include spectral sideband 
regions in the energy ranges for all the fits (see \Secref{subsec:fitting_method}), we extracted data in the range 2.6--541~GeV.  

We used the \irf{P7REP\_CLEAN}\ event selection for data acquired
between 2008 August 4 and 2012 April 18. 
We used this more selective event class for this analysis because the CR background 
contamination in the \irf{P7REP\_SOURCE} class can dominate over the diffuse \gammaRayHyph\ 
contribution at high Galactic latitudes.   We sought to minimize CR background contamination 
because Monte Carlo (MC) studies have shown that reconstructing CRs (and especially protons and 
other hadrons) under the assumption that they are \gammaRays\ can produce a variety of spectral 
features (see \Secref{sec:systematic_errors_backgrounds}).   
Further discussion about the CR background contamination in \irf{P7SOURCE}\ and \irf{P7CLEAN} can be found 
in~\cite{REF:2012.P7Perf}; the results change little for the
reprocessed \irf{P7REP\_SOURCE} and \irf{P7REP\_CLEAN} event
selections.    \newText{The \gammaRayHyph\ effective collecting area (or simply ``effective area'') on-axis for the \irf{P7REP\_CLEAN} event selection ranges from
6500 to 7200 cm$^2$ over the energy range of interest.}

We selected both a Celestial dataset (for the line search) and a dataset corresponding to the Limb (as a control region), see \Tabref{event_samples}.   The Limb is 
a very bright \gammaRayHyph\ source of secondary \gammaRays\
produced by CR interactions in the upper atmosphere.   \Figureref{LimbSchematic} shows a schematic of the geometry
for \gammaRayHyph\ production in the Limb, as well as the definitions of the zenith angle ($\theta_{\rm z}$), spacecraft rocking angle
($\theta_{\rm r}$), and \gammaRayHyph\ incidence angle ($\theta$). 

\onepanel{ht}{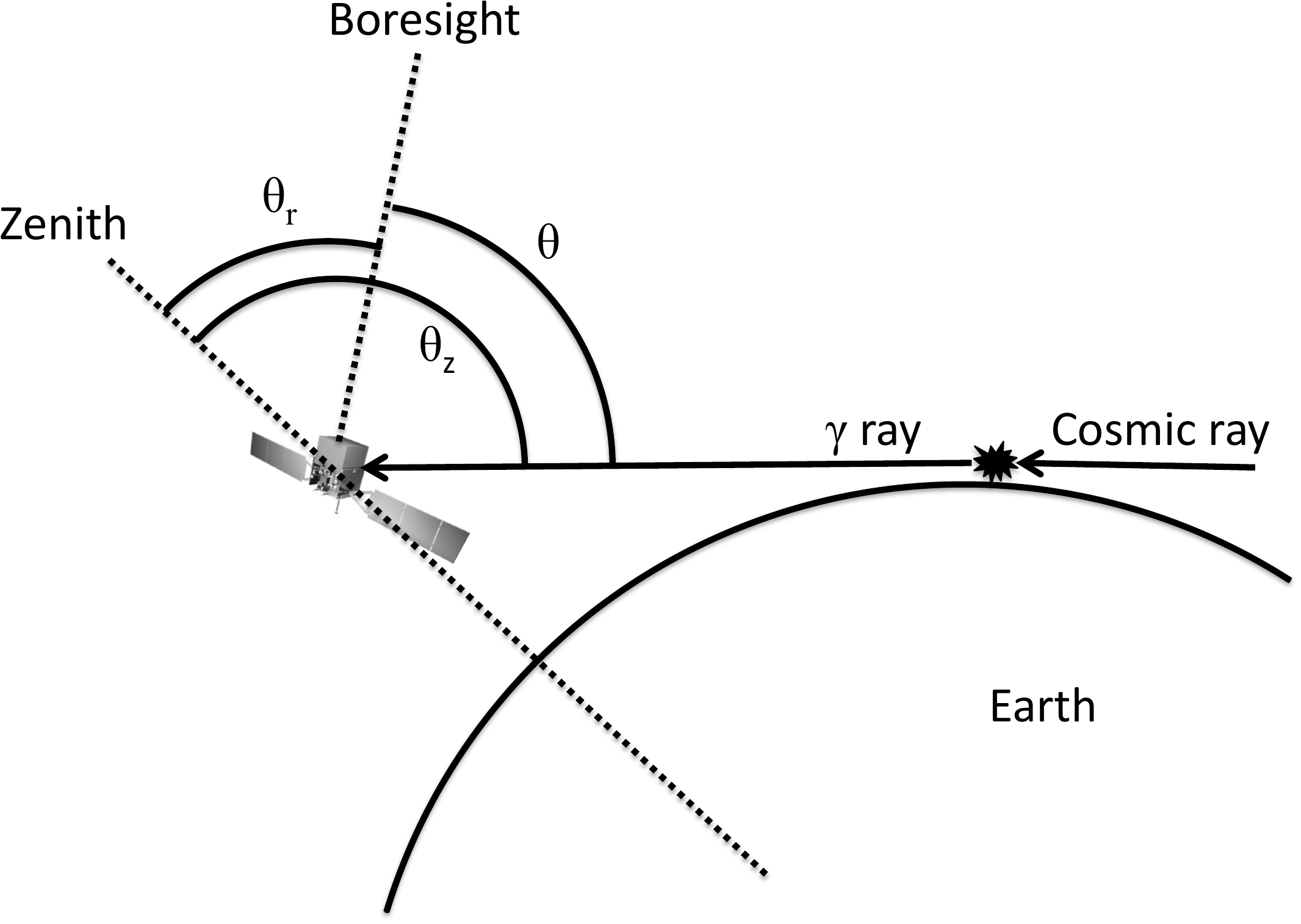}
{\caption{\label{fig:LimbSchematic}Schematic of Limb \gammaRayHyph\
    production by CR interactions in the Earth's atmosphere, showing
    the definitions of the zenith angle ($\theta_{\rm z}$), the spacecraft
    rocking angle ($\theta_{\rm r}$) and the incidence angle ($\theta$).  Dashed line starts at the center of the Earth.  Not drawn to scale.}}

For the Celestial dataset, we removed the Limb \gammaRays\ by
selecting only events with $\theta_z < 100^{\circ}$.  We also only used data collected when a \newText{small} fraction of the LAT field-of-view (FOV) subtended the Limb by removing times when $|\theta_{\rm r}| > 52^\circ$. 

For the Limb dataset we selected a narrow range of zenith angles ($111^\circ < \theta_z < 113^\circ$).  One should note that while
\Fermi\ is in normal survey mode~\cite{REF:2012.P7Perf}, the Limb is fairly far off-axis ($\theta > 60^\circ$), near the edge of the LAT FOV. This means that the events in the Celestial dataset have a quite different $\theta$ distribution than events from the Limb collected during survey mode observations. 
Since the effective area and energy resolution of the LAT depend strongly on $\theta$, it is important to use a Limb control dataset collected when \Fermi\ was not in normal survey mode, but rather was slewed toward the Limb. 
Therefore, for the Limb dataset we reversed the rocking angle
criterion and selected times when $|\theta_{\rm r}| > 52^\circ$.
This represents $\sim 0.3\%$ of the livetime of the 3.7 year Celestial
dataset.   Because of the extreme brightness of the limb 
the contamination from Celestial \gammaRays\ is small;  it is $< 6\%$ at
3~GeV, decreases with energy, and is $< 0.5\%$ for all energies $> 6$~GeV.

The initial steps of the data reduction and all of the exposure
calculations were performed with the LAT \stools\footnote{The \stools\
and documentation are available at \url{http://fermi.gsfc.nasa.gov/ssc/data/analysis/scitools/overview.html}} version 09-29-00
using the \irf{P7REP\_CLEAN\_V10}\ IRFs.   The
\irf{P7REP\_CLEAN\_V10}\ IRFs will not be the set of IRFs recommended
for use with \irf{P7REP\_CLEAN}\ data.  \newText{The recommended IRFs
  for use with the reprocessed data will be publically released in the
  fall of 2013.} The differences
between \irf{P7REP\_CLEAN\_V10}\ and subsequent versions of the 
\irf{P7REP\_CLEAN}\ IRFs are very small above 5~GeV, and we have
verified that their use does not significantly change the results presented in this paper.

In order to limit the contribution to the Celestial dataset from discrete \gammaRayHyph\ sources, we applied an energy-dependent mask 
around the 527 point sources in the \newText{second \Fermi-LAT source catalog (2FGL catalog)}~\cite{REF:2011.2FGL} detected with greater than 10$\sigma$ significance above
1~GeV. The energy scaling of the 68\%
containment angle ($\theta_{68}$) of the LAT point-spread function
(PSF) can be modeled as $\theta_{68}(E) = \sqrt{c_{0}^2 (E/1~{\rm
    GeV})^{-2 \beta}+c_{1}^2}$~\cite{REF:2012.P7Perf}.  We performed
an \newText{effective-area}-weighted average of the flight-derived \irf{P7CLEAN\_V6} PSF over incidence angle to obtain the parameters $c_{0}=0.881$,
$c_{1}=0.2016$ and $\beta=0.817$, which give $\theta_{68} = 0.31^\circ$
and $\theta_{68} = 0.20^\circ$ at 5~GeV and 300~GeV, respectively. We used a source mask radius of $2\times \theta_{68}(E)$.  In each of the ROIs (see \Secref{sec:method_roi}), this masking removed $\sim1.5\%$ of the solid angle 
and $\sim10\%$ of the events.  We estimate that the residual contamination from point sources in our energy range constitutes $\lesssim 10\%$ of the events in our Celestial dataset.

Our event selection criteria for the Celestial and Limb datasets are
summarized in \Tabref{event_samples}.  Note that we included events
through 2012 September in our Limb dataset to take advantage of events
collected \newText{during a week-long targeted 
pointing that included appreciable amounts of time with $|\theta_{\rm r}| > 52^\circ$}.

\begin{table}[ht]
  \caption{\label{tab:event_samples}Summary table of data selections.}
  \begin{center}
  \begin{tabular}{lcc}
    \hline\hline
  Selection & Celestial data & Limb data\\
  \hline
  Observation Period & 2008 August 4--2012 April 4\ & \ 2008 August 4--2012 October 6 \\
  Mission Elapsed Time\footnote{\Fermi\ Mission Elapsed Time is defined as seconds since 2001 January 1, 00:00:00 UTC.} (s) & $[239557447,356434906]$ & $[239557447,371176784]$ \\
  Energy range (GeV) & $[2.6,541]$ & $[2.6,541]$ \\
  Zenith range (deg) & $\theta_{\rm z} < 100 $ & $111< \theta_{\rm z} < 113$ \\
  Rocking angle range (deg)\footnote{Applied by selecting on 
    \texorpdfstring{\selection{ROCK\_ANGLE}}{ROCK_ANGLE} 
    \newText{with the \gtmktime\ {\emph{ScienceTool}}}.} & $|\theta_{\rm r}| < 52$ & $|\theta_{\rm r}| > 52$  \\
  Data quality cut\footnote{Standard data quality selection:
    \texorpdfstring{\selection{DATA\_QUAL == 1 \&\& LAT\_CONFIG ==
        1}}{DATA_QUAL == 1 \&\& LAT_CONFIG == 1}  
    \newText{with the \gtmktime\ {\emph{ScienceTool}}}.} & Yes & Yes \\
  Source masking (see text) & Yes & No \\
  \hline\hline
  \end{tabular}
  \end{center}
\end{table}

\subsection{Simulated datasets}\label{sec:MC}

\newText{To model the response of the LAT} we used several simulated datasets created with a GEANT4-based~\cite{REF:Agostinelli:2002hh}
MC simulation of \gammaRayHyph\ interactions with the LAT and analyzed using the same event reconstruction algorithms as are applied to the data.
We relied on a few particular simulated datasets: (i) the ``all-gamma'' dataset~\cite{REF:2012.P7Perf}, 
an isotropic distribution of \gammaRays\ with an $E^{-1}$ spectrum
\newText{used to generate the standard IRFs};  (ii) ``isotropic monochromatic'' datasets, i.e., isotropic distributions of 
\gammaRays\ at specific energies \newText{used to generate our model
  for the energy dispersion};  (iii) an ``all-sky background'' dataset, where the simulation used the \Fermi\ pointing history and the source 
model included all 2FGL catalog sources, diffuse emission from the
Galaxy and isotropic emission\footnote{Specifically, {\em
    \texorpdfstring{gal\_2yearp7v6\_v0.fits}{gal_2yearp7v6_v0.fits}}
  and {\em \texorpdfstring{iso\_p7v6clean.txt}{iso_p7v6clean.txt}},
  available at
  \url{http://fermi.gsfc.nasa.gov/ssc/data/access/lat/BackgroundModels.html}}.

\section{REGIONS OF INTEREST}\label{sec:method_roi}  

We have developed a set of five ROIs optimized for sensitivity to WIMP
annihilation or decay and four reference models for the distribution of DM
in the Galaxy. The details of the optimization procedure are described in \Appref{app:method_roi}. 

For the distribution of DM in the Galaxy, we consider four
smooth parametrizations.  The Navarro-Frenk-White (NFW) profile~\cite{REF:1996ApJ...462..563N},

\begin{equation}
  \rho(r) = \frac{\rho_s}{(r/r_s)(1+r/r_s)^2}
\end{equation}

\noindent with $r_s=20$~kpc has been found to characterize the smooth
distribution of DM in simulated halos.  The Einasto profile,

\begin{equation}
  \rho(r) = \rho_s \exp\{ -(2/\alpha)[(r/r_s)^\alpha - 1]\}
\end{equation}

\noindent with $r_s=20$ kpc and $\alpha=0.17$ is favored by more recent cold
dark matter (CDM) simulations~\cite{REF:2010MNRAS.402...21N}.  We
additionally consider an isothermal profile with a central core,

\begin{equation}
  \rho(r) = \frac{\rho_s}{1+(r/r_s)^2}
\end{equation}

\noindent with $r_s=5$~kpc~\cite{REF:1980ApJS...44...73B}.  Finally, adiabatic
contraction of the DM halo due to infall of baryonic matter in the Galactic center region could result in DM density profiles with a
much steeper central slope than either the NFW or Einasto
profile~\cite{REF:2006PhRvD..74l3522G}.  We take as a representative
of this class of models a contracted NFW profile defined by

\begin{equation}
  \rho(r) = \frac{\rho_s}{(r/r_s)^{\gamma}(1+r/r_s)^{3-\gamma}}
\end{equation}

\noindent with $\gamma=1.3$.  For all profiles we determine the normalization of the
profile density ($\rho_s$) by fixing the DM density at the solar radius
\newText{$\rho(r_\odot=8.5$~kpc$)$} = 0.4~GeV~cm$^{-3}$~\cite{REF:2010JCAP...08..004C,Salucci:2010qr}.

We defined a set of five ROIs, circular regions of radius \Rgc\
centered on the Galactic center with $|b| < 5^{\circ}$ and $|l| > 6^{\circ}$ masked, which were optimized for each of the DM density profiles considered. For annihilating DM models
we use $\Rgc = 3^{\circ}$ (R3, optimized for the contracted NFW profile), $16^{\circ}$ (R16, optimized for the Einasto profile), $41^{\circ}$ (R41, NFW) and
$90^{\circ}$ (R90, optimized for the Isothermal profile), 
while for decaying DM models we use $\Rgc =180^{\circ}$ (R180).  
We did not apply a source mask for the R3 dataset, so we limited the
search in R3 to energies greater than 30~GeV (see \Appref{app:method_roi}).  Above this energy, the composite \gammaRayHyph\ flux from point sources is much less than the integral flux of the Galactic diffuse emission rate in R3.

\Tableref{roi_jval} summarizes the optimized ROI that was used for
each DM halo profile and its associated astrophysical J-factor (\newText{i.e., the integral along the line of sight of $\rho(r)^{2}$ for DM annihilation or $\rho(r)$ for DM decay;} see \Appref{app:method_roi}).  We note that the point-source masking reduced the annihilation J-factor by $<10\%$ in 
each ROI, except for R3 where no point source masking was applied. 
The counts map of the 3.7 year Celestial dataset in the R180 ROI with outlines of the other four ROIs is shown in \Figref{ROIs}.

\begin{table}[ht]
  \caption{\label{tab:roi_jval}Summary of optimized ROIs and J-factor values for each of the four
    DM density profiles considered for both annihilating or decaying WIMPs.}
\begin{center}
\begin{tabular}{ccccc}
\hline\hline
&\multicolumn{2}{c}{Annihilation}&\multicolumn{2}{c}{Decay}\\
Profile& ROI& J-factor &ROI&J-factor \\
 & & (10$^{22}$ GeV$^{2}$ cm$^{-5}$) & &  (10$^{23}$ GeV cm$^{-2}$) \\
\hline
NFW Contracted  & R3   &   13.9  & R180 & 2.42\\
Einasto         & R16  &   8.48  & R180 & 2.49\\
NFW             & R41  &   8.53  & R180 & 2.46\\
Isothermal      & R90  &   6.94  & R180 & 2.80\\
\hline\hline
\end{tabular}
\end{center}
\end{table}

\onewidepanel{ht}{\colSw{Figure_2}}{
\caption{\label{fig:ROIs} Counts map for the line search dataset
  binned in $1^{\circ} \times 1^{\circ}$ spatial bins in the R180
  ROI, and plotted in Galactic coordinates using the Hammer-Aitoff
  projection.  The energy range is 2.6--541~GeV and the most-significant
  2FGL sources have been removed using an energy-dependent
  mask (see text).  Also shown are the outlines of
  the other ROIs (R3, R16, R41, and R90) used in this search.}}

\section{MODELING OF THE ENERGY DISPERSION}\label{sec:method_2DPDF}

The algorithms for reconstructing LAT events provide three
estimates of the event energy: one based on a parametric correction of the raw energy measured by the calorimeter, a second based on a maximum likelihood fit using the correlations between the raw energy in the calorimeter and other event properties and a third based on a fit to the shower profile in the calorimeter~\cite{REF:2009.LATPaper}.  The likelihood-based method
 was found to create narrow features in the LAT energy
response that could mimic line-like spectral features, which is the
main reason why previous spectral line searches performed by the LAT 
Collaboration with the \psix\ datasets used the shower profile energy estimate exclusively~\cite{REF:2010.LineSearch,REF:2012.LineSearch}.
In the \pseven\ version of the event-level analysis the result of the likelihood method
is ignored and we use a classification tree analysis to select which of the other two methods is more likely to provide the 
best energy estimate on an event-by-event basis. The corresponding
estimate is the energy assigned.  We note that above a few GeV the
shower profile method is typically more accurate than the parametric correction
method (the former being selected by the classification tree
analysis for $\sim 80\%$ of the events above 10 GeV).

The energy assignment algorithm also performs a classification tree analysis to estimate the probability that the energy estimate is within the nominal $68\%$ containment band for events of that energy and incidence angle (\probE)~\footnote{Available as
\texttt{CTBBestEnergyProb} in the extended event files available at the
\Fermi\ Science Support Center at \url{http://fermi.gsfc.nasa.gov/ssc/data/access/}, and
described at \url{http://fermi.gsfc.nasa.gov/ssc/data/analysis/documentation/Cicerone/Cicerone_Data/LAT_Data_Columns.html\#ExtendedFile}}.

To model the signal from a \gammaRayHyph\ line, we used a parametrization of the effective energy dispersion of the instrument, i.e., the
probability density $\Deff(\Ereco; \Etrue, \sVec)$ to measure an energy \Ereco\ for a \gammaRay\ of (true) energy \Etrue\ and other event
parameters, \sVec. The fraction of the electromagnetic shower
contained in the calorimeter can vary significantly event to event.  In
general, the energy dispersion depends on $\theta$ and the
\gammaRayHyph\ conversion point in the instrument, among other
quantities.  
Furthermore, the $\theta$-distribution of the observing time varies across the sky, causing corresponding changes in the effective energy dispersion.  These considerations are discussed in more detail in \Appref{app:Likelihood_Formalism}, in particular in \Secref{app:Formalism_Correction_Terms}. 

When fitting essentially monochromatic lines (i.e., the intrinsic spectrum is much narrower that the instrumental resolution),  for a given line energy, \Egamma, we expect the distribution of observed energies for a line signal, $\CSig(\Ereco)$, to follow the effective energy dispersion, \Deff; so that 

\begin{equation}
\CSig(\Ereco|\Egamma,\sVec) = \nSig \int \Deff(\Ereco;\Etrue,\sVec)
\delta(\Egamma -\Etrue) dE = \nSig \Deff(\Ereco;\Egamma,\sVec),
\end{equation}

\noindent where \nSig\ is the number of observed signal events, which we treat
as a free parameter in the fitting (see
\Secref{sec:fitting})\footnote{This assumption breaks down when the
  intrinsic width of the \gammaRayHyph\ emission becomes a sizable fraction of the LAT energy resolution.  
  In practical terms, this applies for final states with unstable particles such as $Z \gamma$, in particular for \gammaRayHyph\ energies 
  at the low end of our search range.  We discuss the implications of this in \Secref{sec:systematic_errors_decay_width}.}.

Following the approach used in previous line searches published by the LAT Collaboration, we use a sum of Gaussians to parametrize the energy
dispersion at any given energy, averaging over the LAT FOV and
combining events \newText{that convert in the front or back sections of the
tracker~\cite{REF:2012.LineSearch}}.  One
notable improvement relative to our previous studies is that the parametrization
$\Deff(\Ereco;\Etrue,\probE)$ used in this work includes the energy
reconstruction quality estimator, \probE.   Specifically, we modeled the energy dispersion in 10 \probE\ bins of 0.2 from 0.1 to 0.5, bins of 0.1 from 0.5 to 0.7, and bins of 0.05 from 0.7 to 1.  The \irf{P7REP\_CLEAN}\ event class only
includes events with $\probE>0.1$.  

The energy dispersion in each \probE\ bin was modeled with a triple Gaussian function 

\begin{equation}\label{eq:TripGaus}
\Deff(\Ereco;\Etrue,\probE) = \sum_{k=1}^3
\frac{a_k}{\sigma_k\sqrt{2\pi}}e^{-((\Ereco/\Etrue) - (1+\mu_k))^2/2\sigma_k^2}\;,
\end{equation}

\noindent where $a_3 = 1-a_2-a_1$.   To avoid degeneracy between the
Gaussians, we constrain the ranges of the $\sigma_i$ to ensure that
$\sigma_1 > \sigma_2 > \sigma_3$. 

We explicitly determined energy dispersion model parameters for \Etrue\ values of 5, 7, 10, 20, 50, 100, 200 and 300~GeV using
``isotropic monochromatic'' \gammaRayHyph\ MC simulations (see
\Secref{sec:MC}) at each of those energies.  The systematic uncertainties associated with using these simulations to derive our model are discussed in \Secref{sec:systematic_errors_edispmodel}.  When fitting for a spectral line at \Egamma, we interpolated the appropriate energy dispersion parameters.  The resulting energy dispersion models at
$\Etrue = 100$~GeV in all 10 \probE\ bins are shown in \Figref{100GeV_Line}.  The bias and 68\% containment of our energy dispersion model as a
function of \Etrue\ are shown in \Figref{Cont68}.  The bias is the fractional deviation of the energy dispersion peak from the true energy.

\onepanel{ht}{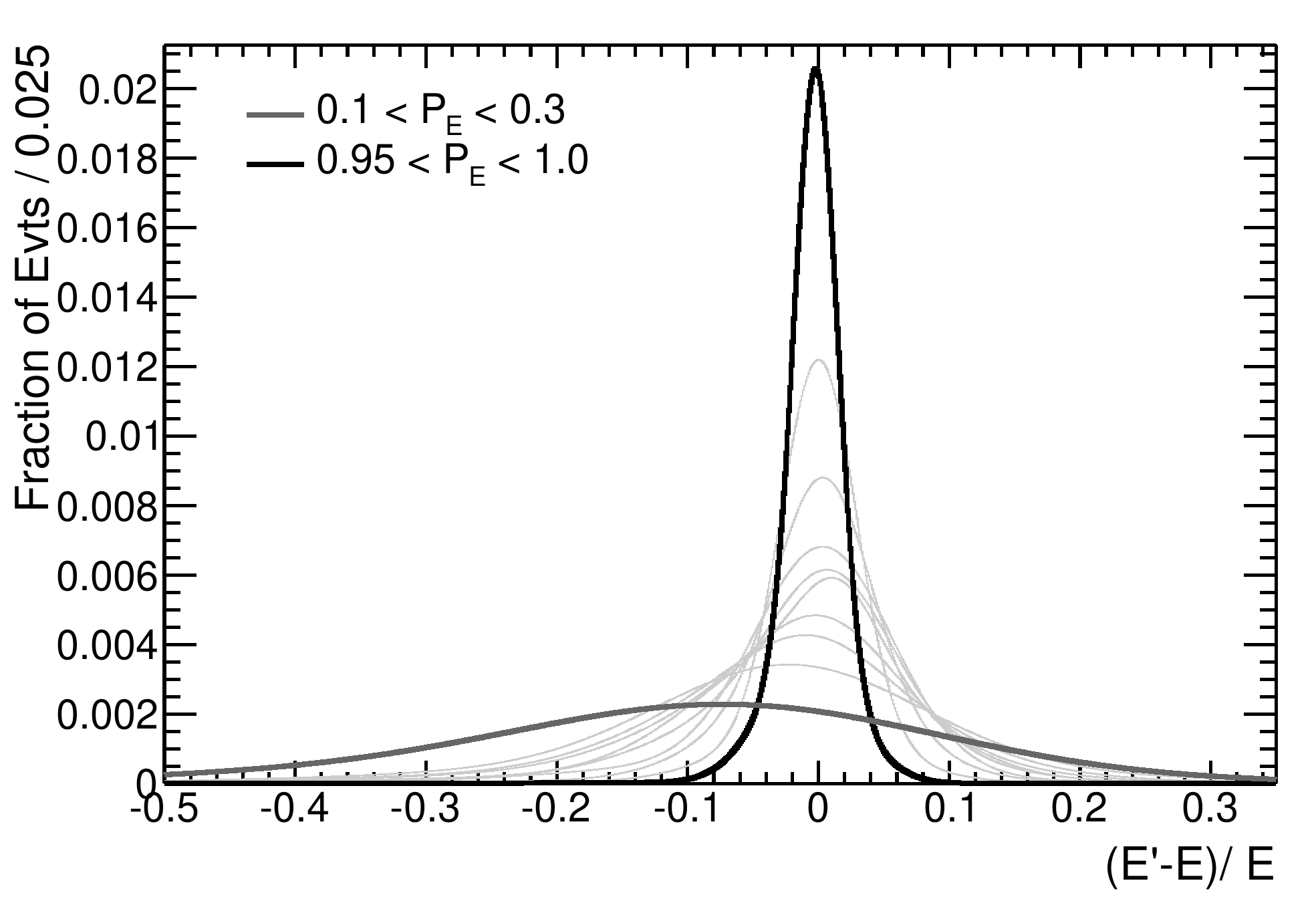}{\caption{\label{fig:100GeV_Line}Energy dispersion model for 100~GeV in all 10 \probE\ bins.  The darker lines show the distributions for the smallest and largest values of \probE,
  while the thinner gray lines show the models for the intermediate
  \probE\ bins.}}

\onepanel{ht}{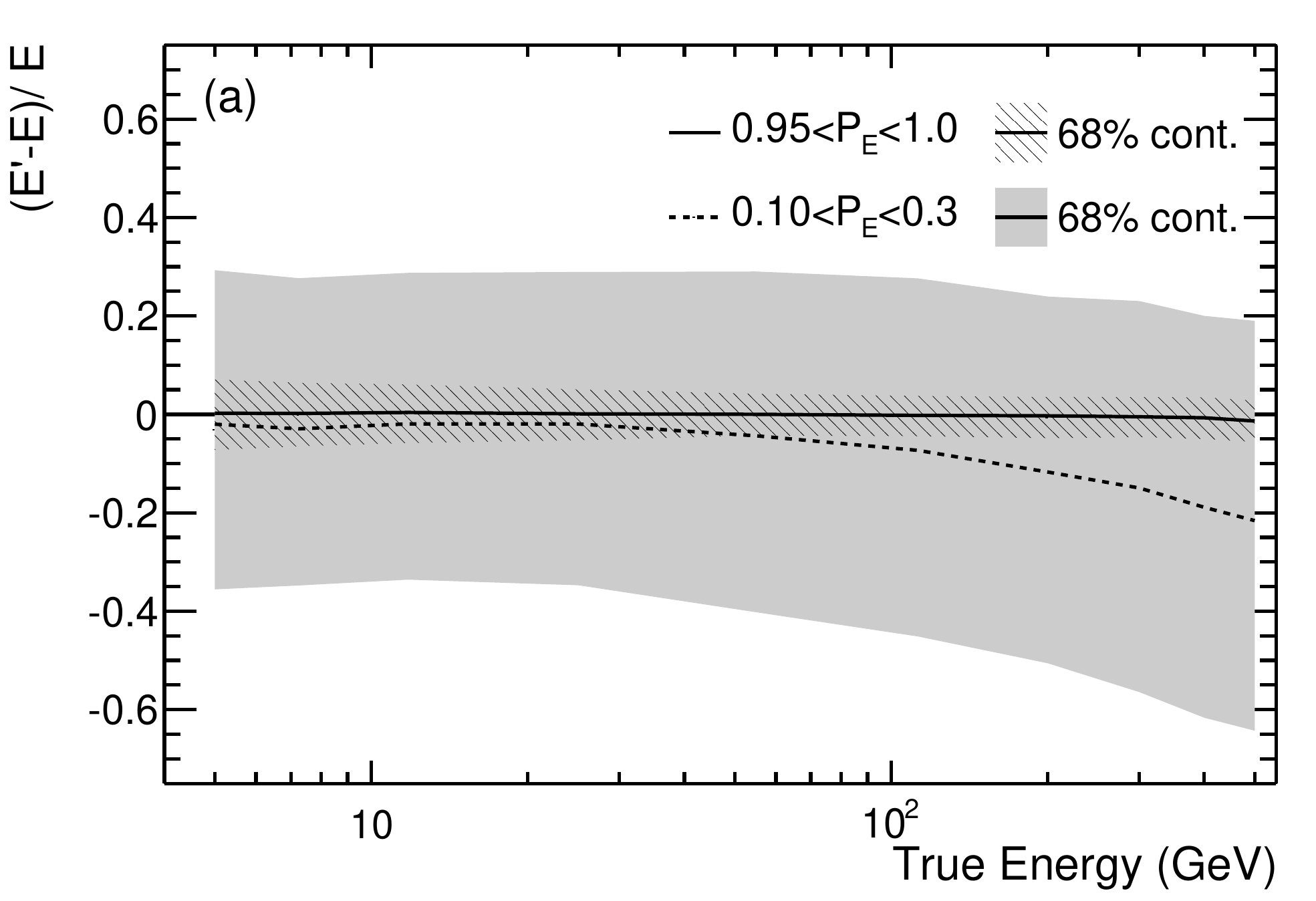}
{\caption{\label{fig:Cont68}The bias of the energy dispersion for the lowest (highest) \probE\ bins is indicated by the dashed (solid) line.  The shaded regions shows the 68\% containment for the highest and lowest \probE\ bins.}}

The distribution of \probE\ depends on energy
and ROI.  \Figureref{ProbDist_DataVsMC} shows the distribution of \probE\ in 
the \irf{P7CLEAN}\ and \irf{P7REP\_CLEAN}\ datasets and the ``all-sky background''
\gammaRayHyph\ MC for $|b|>10^{\circ}$.   While the agreement between
MC and data is good, overall, there is clearly some discrepancy in the
upper half, in $\log(E)$, of our energy range that has been reduced by the reprocessing.

\twopanel{ht}{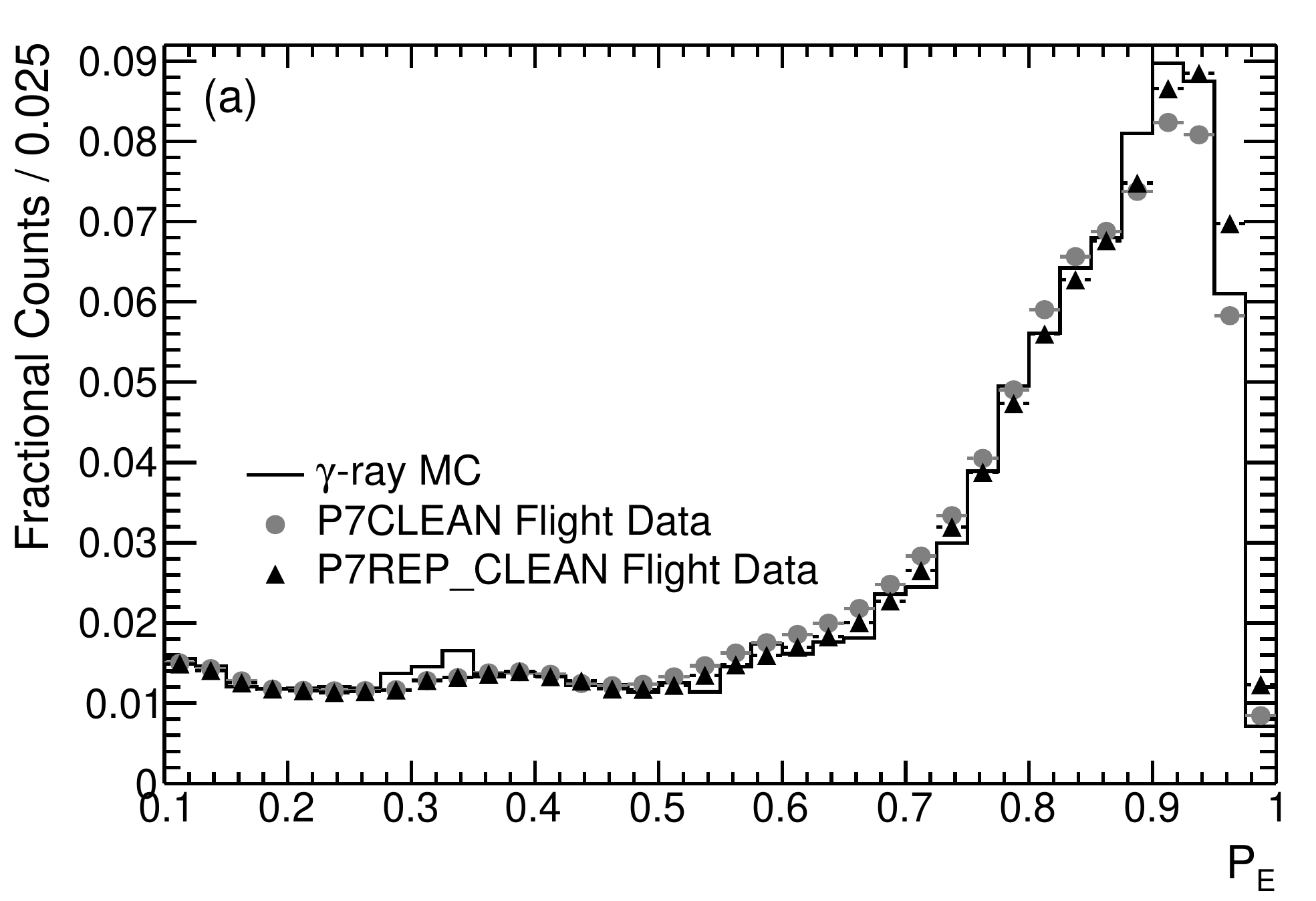}{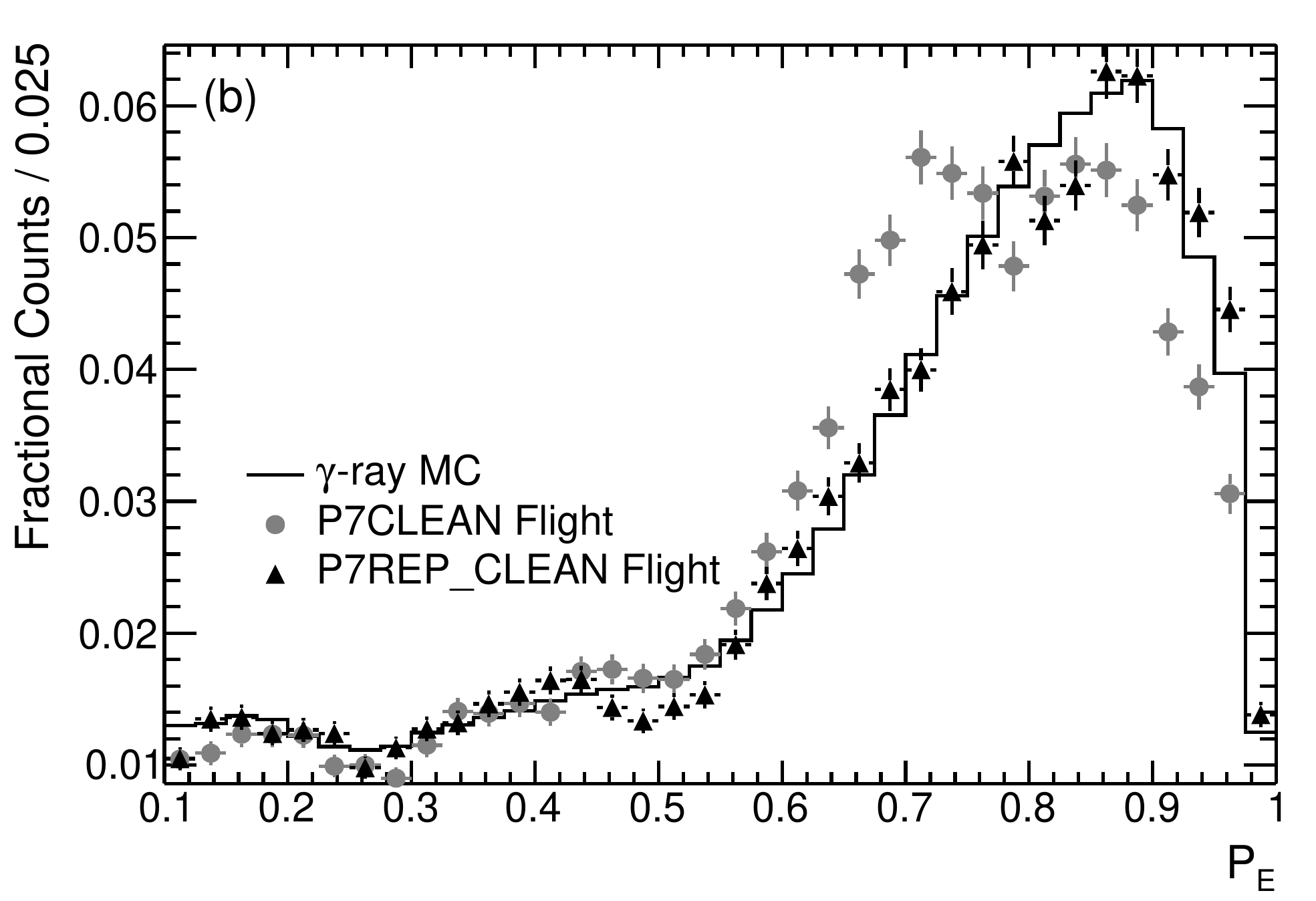}
{\caption{\label{fig:ProbDist_DataVsMC}\probE\ distributions for
    flight and all-sky MC datasets (see text) over (a) the entire energy range
    (2.6--541~GeV), and (b) the energy range  (52--541~GeV).}}

With the addition of \probE\ as a parameter for the energy dispersion we are not only testing that a possible line signal effectively follows the energy dispersion, but also that well-reconstructed events are clustered closer to the peak energy.  This contributes significant additional information to the likelihood fitting; to quantify the improvement we compared the effect on sensitivity from modeling the energy dispersion as simply a function of energy (``1D'' model, $D(\Ereco;\Egamma)$, as was done in~\cite{REF:2012.LineSearch}) to using a model with \probE\ (``2D'' model, $D(\Ereco;\Egamma,\probE)$) using simulations both with and without a spectral line.  Including the extra information, \probE, the statistical power is increased by $15\%$ on average.

%% file: Fitting.tex
\section{FITTING}\label{sec:fitting}

\subsection{Fitting method}\label{subsec:fitting_method}

We searched for spectral lines by performing maximum likelihood fits in sliding energy intervals in the five ROIs described in
\Secref{sec:method_roi}.   Specifically, we fit the count spectra in the energy domain, integrating over each ROI.  Given model uncertainties and
the relatively uniform coverage of the sky by the LAT, we made several assumptions to simplify the fitting procedure, which are outlined in
\Appref{app:Likelihood_Formalism}\footnote{Since we are using the \irf{P7REP\_CLEAN} event class and combining front- and back-converting events, we suppress the
\newText{event selection (\sVec)} relative to the notation used in \Appref{app:Likelihood_Formalism}.}.   We used the \textit{RooFit} toolkit~\cite{REF:RooFit.2003}, (version 3.12) to implement the models and perform the likelihood minimization.

In general, we performed unbinned maximum likelihood fits; however, because of the large number of events at the lower end of our energy range, 
we performed binned fits for energies $< 25$~GeV to reduce the required computation time.   When performing binned fits, we used 60 bins across 
the fit energy range.  Since the bin width is small compared to the instrument resolution, we lose very little information.  \newText{We confirmed that binned 
fits yield results nearly identical to unbinned fits by simulating 1000 pseudo-experiments both with and without a signal and fitting for a line at 20~GeV.}

We fit in narrow, approximately half-decade energy ranges, and therefore approximated the background spectrum, $\CBkg(\Ereco)$, as a single power law with an index \GamBkg\ that was allowed to float in the fit. We also incorporated the energy dependence of the exposure averaged across each ROI into the background component\footnote{Any variation across the energy interval in the exposure would introduce an artificial shaping to the background spectrum.} by means of an energy-dependent \newText{exposure} correction $\eta(\Ereco)$:

\begin{equation}\label{eq:BkgPDF_1D}
\CBkg(\Ereco|\GamBkg,\nBkg) =  \nBkg \left(\frac{\Ereco}{\Epivot}\right)^{-\GamBkg} \eta(\Ereco).
\end{equation}

\noindent where \Epivot\ is a reference energy (we used $\Epivot=1$ MeV), \nBkg\ is the total number of background events, \newText{which is a
free parameter in the fit, and $\eta(\Ereco)$ is given by  \Eqref{eq:Expsosure_Correction_1D}, which includes a normalization constraint that is defined by \Eqref{eq:Normalization_Bkg}.}   Since both the background and exposure vary smoothly and slowly across our fit ranges, 
we did not explicitly convolve the above equations with the energy dispersion to derive the expected models for the observed 
energy, i.e.,~we assumed  $\Ereco=\Etrue$ in \Eqref{eq:BkgPDF_1D}.

Our complete counts model to fit for a line at \Egamma\ is:

\begin{equation}\label{eq:SigPlusBkgCnts}
C(\Ereco,\probE|\vec{\alpha}) = 
\nSig \Deff(\Ereco;\Egamma,\probE) \wROI(\probE)  + \nBkg
\left(\frac{\Ereco}{\Epivot}\right)^{-\GamBkg} \eta(\Ereco) \wROI(\probE) \,,
\end{equation}

\noindent where the model parameters $\vec{\alpha}$ are \Egamma,
\GamBkg\, \nSig\ and \nBkg.   Note that we fit for \nSig\ independent of any DM model assumption; we then assumed a specific DM profile and calculated the J-factor in the ROI in order to solve for the annihilation cross-section or decay lifetime given the magnitude of the exposure in that ROI (see \Secref{sec:results}).

Since we incorporated \probE\ in the signal model, we included the distributions of \probE, $w(\probE)$.  For each fit in a specific ROI and energy interval, we took the \probE\ distributions for both signal and background from all of the data in the ROI and energy range; i.e., $\wBkg(\probE) = \wSig(\probE) = \wROI(\probE)$.  The small effect from this approximation is discussed in \Secref{sec:systematic_errors_edispmodel}. 

The energy interval for a fit at \Egamma\ in this search was $\pm6\sigE(\Egamma)$, where \sigE\ is the on-axis LAT energy resolution at the fit energy. (Specifically, \sigE\ is the half-width of the $\pm 34\%$ containment about the peak value of the energy dispersion for on-axis events.) The interval was broadened from the previous LAT analysis~\cite{REF:2012.LineSearch} to reduce the statistical uncertainty of \GamBkg.  This consequently reduced the uncertainty of \nSig\ because the maximum likelihood values of the parameters are correlated in the fits.  As discussed in~\cite{REF:Weniger:2012tx}, the significance of the fit has a slight dependence on interval size. However, for energy ranges wider than $\sim12\sigE$, the change in significance is small compared to the expected statistical variation. Additionally, fitting in wider intervals may reduce the validity of approximating the background as a power law. However, we do not find that this approximation induces a large systematic effect (see \Secref{sec:GalaxySmoothness}).

Each fit was performed at a specific energy \Egamma\ as opposed to letting the line energy float in the fit.  The spacing between adjacent fit energies is half the energy resolution.   Simulations show that with this choice, the loss of signal for potential lines offset with respect to our search grid is small;  at worst we underfit \nSig\ by less than 10\%.   \newText{We constrained \nSig\ to be positive to avoid unphysical measurements as well as negative likelihoods.}

We calculate the local significance by taking the square-root of the Test Statistic ($\slocal=\sqrt{TS}$), which is defined as twice the difference in the log-likelihood between the maximum likelihood hypothesis and the null hypothesis:

\begin{equation}\label{eq:TS}
TS=2\textrm{ln}\frac{\mathcal{L}(\nSig = n_{\rm sig,best})}{\mathcal{L}(\nSig=0)}.
\end{equation}

\noindent We expect at least 10 (and usually many more) effective background events (see
\Secref{sec:systematic_errors_fractional_residual}) for each energy range and ROI considered in the fits so the Gaussian
approximation for application of Chernoff's \cite{REF:Chernoff} theorem \newText{to predict a $\chi^2$
distribution (for one bounded degree of freedom) of $TS$ is well justified.}  

\subsection{Global significance}\label{subsec:LEE}

We fit lines for 88 different $\Egamma$ values in R16, R41, R90, and R180 and 44 in R3 (where we only fit for $\Egamma > 30$~GeV) for a total of 396 fits. Given this number of trials, it is 
reasonable to expect some of the fits to indicate apparently-significant values for the number of signal events even if the underlying data are
purely background.  If our trials involved independent data samples, we could translate a local p-value to a global (i.e., post-trial) p-value using $\pglobal = 1 - (1 - \plocal)^{396}$.    

However, our samples were not independent.  In fact, converting between the $TS$ and global significance, \sglobal\ (i.e., \pglobal\ expressed as a significance relative to the standard deviation of a Gaussian distribution, $\sigma$), is complicated because:
\begin{fermienumerate}
\item{The energy ranges overlap, meaning the fits were not all independent.}
\item{The ROIs are nested in those of larger radii (see \Secref{sec:method_roi}).}
\end{fermienumerate}

To estimate \sglobal, we simulated 1000 background-only realizations of our search.  For each realization, we generated five independent samples corresponding to the non-overlapping parts of our five ROIs (i.e., we generated samples representing R3, R16 without R3, R41 without R16, and so on).  For each sample, we simulated background-only events with $\GamBkg = 2.4$ and the exposure corrections from R3.  For simplicity, and to reduce the computational time required, we omitted \probE\ from the model used for these realizations.  We merged the independent samples to obtain simulated datasets matching our ROIs, with the correct amount of overlapping events.

For each realization, we performed all 396 fits for a line signal at the various energies in all of the ROIs and extracted the 
largest \newText{\slocal} value obtained by any of the fits \newText{(\smax)}.  Empirically, we found that the distribution of \newText{\smax} 
values for each realization was well modeled by the expected distribution for \newText{\ntrial\ trials where the $TS$ distribution
follows a $\chi^2$ distribution with one bounded degree of freedom (see \Figref{TF}~a):} 

\begin{equation}
f(\smax) = \frac{\ntrial}{2}(\delta(\smax) + \frac{e^{-\smax^{2}/2.}}{\sqrt{2\pi}}) (1 - \frac{1}{2}P_{\chi^2}(\smax^{2},1))^{(\ntrial-1)}.
\end{equation}

\noindent The best fit number of independent trials was \newText{$n_{\rm t} = 198 \pm 6$}.  Therefore, we estimate that our
search consists of \newText{198} effective independent trials and calculate the relation to convert from \slocal\ to \sglobal\ accordingly \newText{(see \Figref{TF}~b)}.

\twopanel{ht}{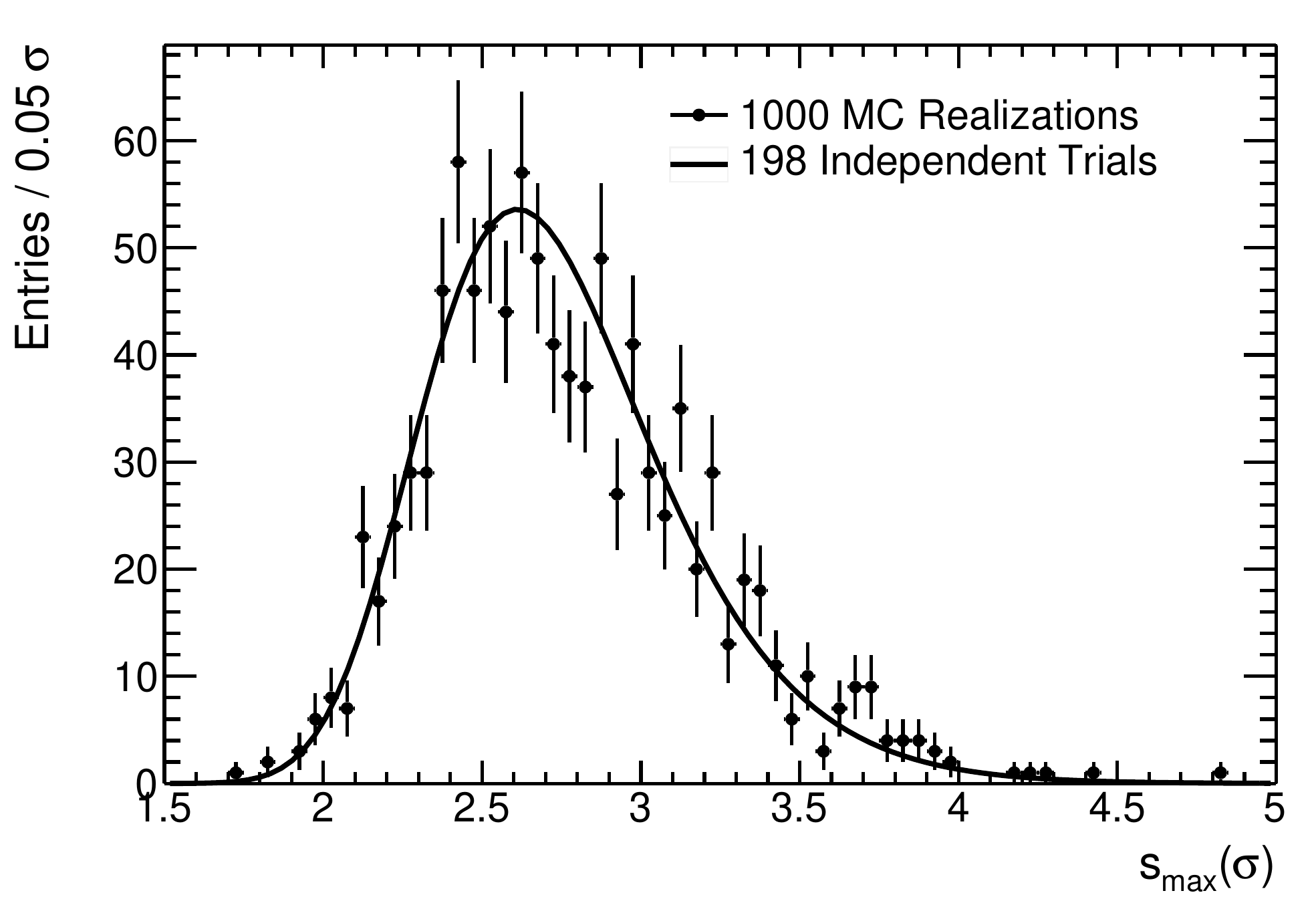}{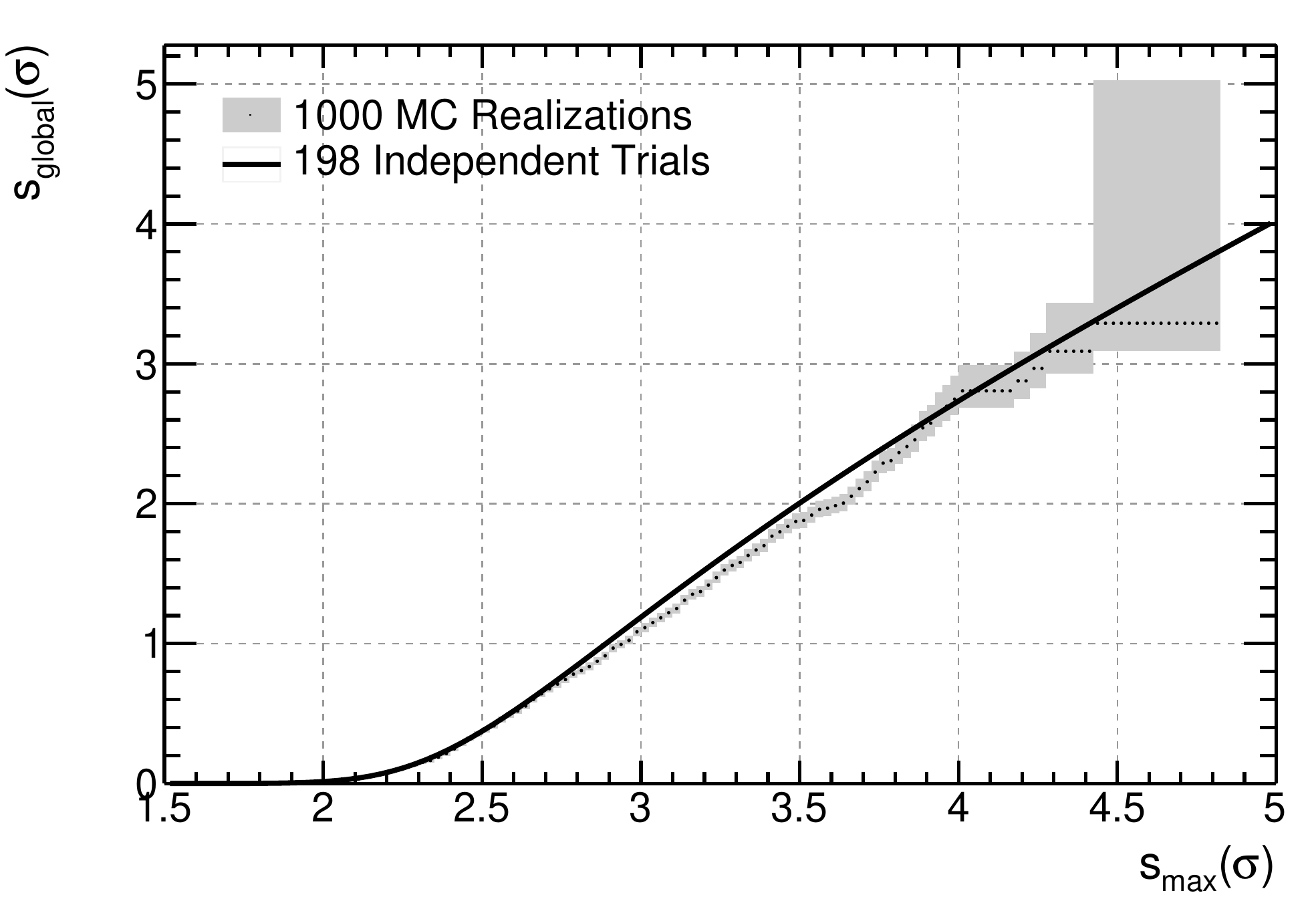}{\caption{\label{fig:TF}\newText{Derivation of global significance versus local significance given our scan over energy in $0.5\sigE$ steps and five ROIs: (a) distribution of the largest \slocal\ values obtained in any ROI at any energy from 1000 MC realizations (points) and the best-fit independent trials curve, (b) the corresponding \slocal\ to \sglobal\ transformation.}}}

We also extracted the largest $TS$ value obtained by any of fits in each ROI and fit for the number of independent trials for that ROI. In each case, we found that the best fit $n_{\rm t,ROI}$ was about \newText{60\%} of the number of trials actually performed, \newText{or slightly more that one trial for each step of $\sigE$.}

Summing the best fit number of independent trials from the five ROIs gives \newText{243}, while empirically we found \newText{$n_{\rm t} = 198$}.  This suggests that the data sets for the ROIs are largely independent, i.e., the overlap between the ROIs only reduces the effective number of trials by a factor of \newText{$0.81$}.  This is reasonable given that we gain a factor of $\sim 6$ events going from R3 to R16, $\sim 3.2$ going from R16 to R41, $\sim 2.2$ going from R41 to R90 and $\sim 1.6$ going from  R90 to R180.

\newText{Finally, we note that this conversion between \slocal\ and \sglobal\ is only applicable to the specific search using $0.5\sigE$ energy steps and five nested ROIs.  Accordingly we do not quote global signficances for fits made on control samples or with other event selections in the course of studying potential systematic biases.}

%% file: Systematic_Errors.tex
\section{SYSTEMATIC UNCERTAINTIES}\label{sec:systematics}
In this section we introduce and summarize systematic uncertainties related to our search for \gammaRayHyph\ lines.   Detailed discussions of each issue can be found in \Appref{app:systematics_details}. A discussion of systematic studies performed near $\Egamma\approx130$ GeV is postponed until \Secref{sec:133GeV_Feature}.

We consider three classes of systematic uncertainties:
\begin{fermienumerate}
\item{Uncertainties that enter in the conversion between the fit number of signal counts, \nSig, and the inferred fluxes. These uncertainties induce a corresponding uncertainty in the estimated model fluxes and upper limits on those fluxes, but do not affect fit significances.  We quantify these in terms of the relative uncertainty of the exposure: $\delta \Exposure / \Exposure$.  These are discussed in \Secref{subsec:syst_exposure}}.
\item{Uncertainties that would scale the fit estimates of the number of signal counts (i.e., affect fit significances and upper limits) but would not otherwise induce or mask a signal. These primarily consist of errors in signal model parametrization.   We quantify these in terms of the relative uncertainty of the number of signal counts: $\delta \nSig / \nSig$.   These are discussed in \Secref{sec:systematic_errors_unmodeled_features}--\Secref{sec:systematic_errors_edispmodel}.}
\item{Uncertainties that could mask a true signal, or induce a false signal.  We quantify these in terms of the induced fractional signal, $f$ (discussed in \Secref{sec:systematic_errors_fractional_residual}).   These uncertainties are discussed in \Secref{sec:systematic_errors_backgrounds}--\Secref{sec:systematic_errors_spectral_smoothness}.}
\end{fermienumerate}

\subsection{Induced fractional signal}\label{sec:systematic_errors_fractional_residual}

Many types of systematic uncertainties that could affect this analysis induce narrow spectral features of a fixed fractional size, which we call ``induced fractional signals''.   For example, unmodeled energy-dependent variations in the effective area at the 10\% level would induce features at the same level in the counts spectrum.   Therefore, it is useful to consider the signal-to-background ratio of any feature in addition to the statistical significance. 

When quantifying the signal-to-background ratio of narrow features, we are more concerned with the background under the signal peak than with the total background in the fit energy range.  Therefore, it is useful to consider the ``effective background'' $b_{\rm eff}$, which can be calculated in terms of the signal and background probability density functions \FSig\ and \FBkg (which are just \CSig\ and \CBkg\ normalized to unit values, see \Appref{app:Likelihood_Formalism}) and the total number of events in the fit range, N: 

\begin{equation}
  b_{\rm eff} = N \int \frac{\FSig(\Ereco)\FBkg(\Ereco)}{\FSig(\Ereco)+\FBkg(\Ereco)} d\Ereco.
\end{equation}

\noindent The integral is performed over the fit energy interval. 

The $TS$ is closely related to $b_{\rm eff}$, and we find that following
relation holds to within $5\%$ for fits to both flight data and MC simulations:

\begin{equation}
  b_{\rm eff} \simeq \frac{\nSig^{2}}{TS}.
\end{equation} 

As stated above, it is useful to report the 
magnitude of potential systematic uncertainty in terms of ``fractional signal'' $f$, i.e., the ratio of signal 
counts to effective background counts:

\begin{equation}\label{eq:FracSig}
  f = \frac{\nSig}{b_{\rm eff}} \simeq \frac{TS}{\nSig}.
\end{equation} 

\noindent The most practical aspect of this formulation is that it allows us to quickly and easily convert 
between systematically induced fractional signal and $TS$ for a given search region and energy.

Furthermore, from the above equations, we can see that for a given fractional signal, the local significance increases as
$\slocal\propto \sqrt{\makebox[0pt]{\phantom{\nSig}}TS} \propto \sqrt{\makebox[0pt]{\phantom{TS}}\nSig}$. 
Therefore, given adequate statistics, a small fractional signal can 
become highly statistically significant.   

\subsection{Summary of systematic uncertainties}\label{sec:systematic_errors_summary}

\Tableref{FracSig_Summary} summarizes the systematic effects discussed in \Appref{app:systematics_details}.   In \Tabref{Syst_Error_By_ROI} 
we have grouped the effects on $\delta\Exposure / \Exposure$, $\delta \nSig / \nSig$ and $\delta f$ for each ROI.

\begin{table}[ht]
\caption{\label{tab:FracSig_Summary} Summary of systematic effects.   As stated in the text, we quote either the relative uncertainty of the
exposure ($\delta\Exposure / \Exposure$), the relative uncertainty of the number of signal events ($\delta \nSig / \nSig$) or the uncertainty of the induced
fractional signal ($\delta f$).   We give representative values when the magnitude of the effect depends on energy, or varies between ROIs.}
\begin{tabular}{ccrlc}
\hline\hline
Systematic & \multicolumn{3}{c}{Effect} & Section \\
\hline
Effective area scale & & $\delta\Exposure / \Exposure =$ & $\pm 0.1$ & \ref{subsec:syst_exposure} \\
Averaging exposure over ROI & (R3) & $|\delta\Exposure / \Exposure| <$ & $0.01$ & \ref{subsec:syst_exposure} \\ 
 & (R180, $\Egamma = 300$~GeV) & $\delta\Exposure / \Exposure =$ & $\pm 0.13$ & \ref{subsec:syst_exposure} \\
\Egamma\ grid spacing & & $\delta \nSig / \nSig =$ & $\SPSB{+0.0}{-0.1}$ &  \ref{subsec:fitting_method} \\
Energy resolution &  &$\delta \nSig / \nSig =$ & $\pm 0.07$ &  \ref{sec:systematic_errors_unmodeled_features} \\
Broadening from Z width & ($\Egamma = 68$~GeV) & $\delta \nSig / \nSig =$ & $-0.07$ &  \ref{sec:systematic_errors_decay_width} \\
\probE\ distribution variation &  & $\delta \nSig / \nSig =$ & $\pm 0.01$ & \ref{sec:systematic_errors_edispmodel} \\
Energy dispersion model $\theta$-variation &  & $\delta \nSig / \nSig =$ & $\pm 0.02$ & \ref{sec:systematic_errors_edispmodel} \\
CR contamination & (R3) & $|\delta f| <$ & $0.005$ & \ref{sec:systematic_errors_backgrounds} \\
 & (R180) & $\delta f =$ & $\pm 0.014$  & \ref{sec:systematic_errors_backgrounds} \\
Point-source contamination &  & $|\delta f| <$ & $0.005$ & \ref{sec:systematic_errors_point_sources} \\
Effective area variations & ($\Egamma = 5$~GeV) & $\delta f =$ & $\pm 0.005$ & \ref{sec:LimbSmoothness} \\
 & ($\Egamma > 100$~GeV) & $\delta f =$ & $\pm 0.025$ & \ref{sec:LimbSmoothness} \\
Astrophysical background modeling & (R180, $\Egamma = 30$~GeV) & $\delta f =$ & $\pm 0.005$ & \ref{sec:GalaxySmoothness} \\
 & (R180, $\Egamma > 100$~GeV) & $\delta f =$ & $\pm 0.011$ & \ref{sec:GalaxySmoothness}\\
 & (R3) & $\delta f =$ & $\pm 0.019$ & \ref{subsec:syst_background_model} \\
\hline\hline
\end{tabular}
\end{table}

\begin{table}[ht]
\caption{\label{tab:Syst_Error_By_ROI} Total magnitude of systematic effects, by ROI and Energy.   We obtained these estimates by adding in
quadrature the magnitudes of all the potential uncertainties on $\delta\Exposure / \Exposure$, $\delta \nSig / \nSig$ and $\delta f$ for each ROI.}
\begin{tabular}{clccccc}
\hline\hline
Quantity & Energy & R3 & R16 & R41 & R90 & R180 \\
\hline
$\delta\Exposure / \Exposure $ & 5~GeV & $\pm$0.10 & $\pm$0.10 & $\pm$0.11 & $\pm$0.12 & $\pm$0.14 \\
$\delta\Exposure / \Exposure $ & 300~GeV & $\pm$0.10 & $\pm$0.10 & $\pm$0.12 & $\pm$0.13 & $\pm$0.16 \\
$\delta \nSig / \nSig$ & All & $\SPSB{+0.07} {-0.12}$ & $\SPSB{+0.07} {-0.12}$ & $\SPSB{+0.07} {-0.12}$ & $\SPSB{+0.07} {-0.12}$ & $\SPSB{+0.07} {-0.12}$ \\
$\delta f$ & 5~GeV & $\pm$0.020 & $\pm$0.020 & $\pm$0.008 & $\pm$0.008 & $\pm$0.008 \\
$\delta f$ & 50~GeV & $\pm$0.024 & $\pm$0.024 & $\pm$0.015 & $\pm$0.015 & $\pm$0.015 \\
$\delta f$ & 300~GeV & $\pm$0.032 & $\pm$0.032 & $\pm$0.035 & $\pm$0.035 & $\pm$0.035 \\
\hline\hline
\end{tabular}
\end{table}

The systematic uncertainties related to the exposure, when summed in quadrature, can reach up to $\delta \Exposure / \Exposure = 0.16$ for the R180 and R90 ROIs. However, as stated earlier, they do not affect the signal significance.  Furthermore, they only have a minor impact on the limits on $\Phi_{\gamma\gamma}$ and $<\sigma v>$, as they are less than $40\%$ of the expected statistical variations in limits, which are typically $40-50\%$.  

The uncertainties of the energy dispersion modeling could cause us to underestimate a true signal, or inflate a statistical fluctuation. These range over $-0.12 < \delta \nSig / \nSig < 0.07$. In other words, we might estimate a true $5\sigma$ signal to be only $4.4\sigma$, or inflate a $3\sigma$ fluctuation to be $3.2\sigma$. These uncertainties also only have minor impact on the analysis, as even for $5\sigma$ signals they result in systematic errors that are less than the expected statistical fluctuations, \newText{which are about $1\sigma$}. 

Uncertainties that can induce or mask a signal can be more problematic. In \Tabref{Syst_Error_By_ROI} these range in magnitude from $\delta f = 0.008$ at low energies up to
$\delta f = 0.035$ at high energies.  However, because of increased statistics in the larger ROIs at low energies, even a $f = 0.01$ induced signal
can become highly statistically significant.   We will discuss this question further in \Secref{sec:results}. \newText{In summary, at energies up to 100~GeV the dominant source of potential systematic bias for the smaller ROIs (R3, R16) is the modeling of the astrophysical backgrounds as a power law, while for the larger ROIs it is CR contamination.  Above 100~GeV, because of the limited statistics of the Limb control sample, uncertainties of potential features in the effective area dominate the systematic uncertainties.}

%% file: Results.tex
\section{FITTING RESULTS AND UPPER LIMITS}\label{sec:results}

We have performed a scan for spectral lines from 5--300~GeV in the five ROIs described in \Secref{sec:method_roi} and find no
globally significant lines. \Figureref{Signif_vs_En} shows the local fit significance for each of the fit energies and all five
ROIs; all of the fits are below 2$\sigma$ global significance.  \newText{As shown in \Figref{Signif_MCvsData}, the distribution
of the \slocal\ values from our line search is well modeled by the null hypothesis expectation according to Chernoff's
theorem~\cite{REF:Chernoff}.}

\onepanel{ht!}{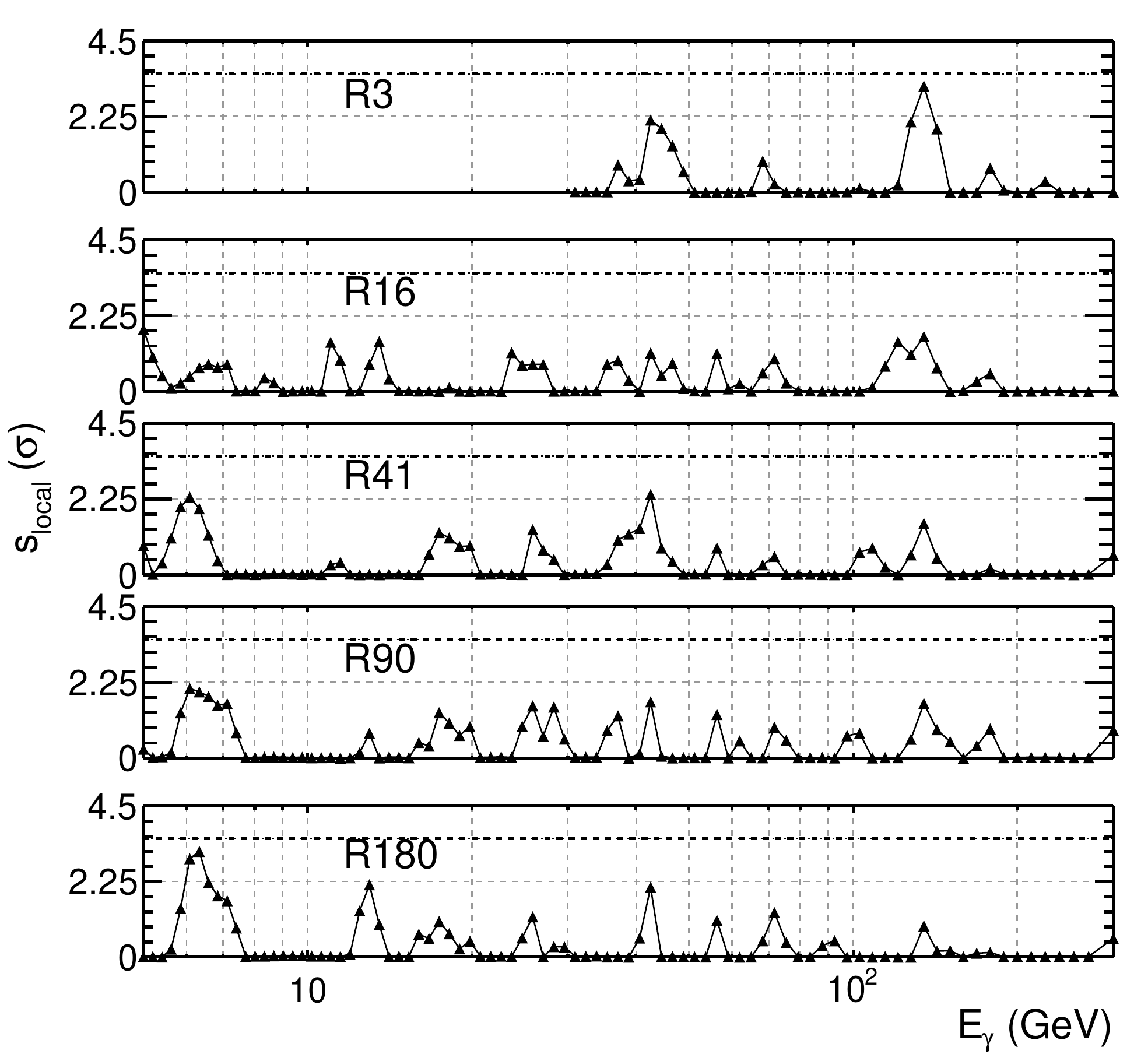}{\caption{\label{fig:Signif_vs_En}Local fit significance vs. line energy in all five
ROIs.  Note that \nSig\ was required to be non-negative.  The dashed line at the top of the plot indicates the local
significance corresponding to the $2 \sigma$ global significance derived with the method described in \Secref{subsec:LEE}.}}

\onepanel{ht!}{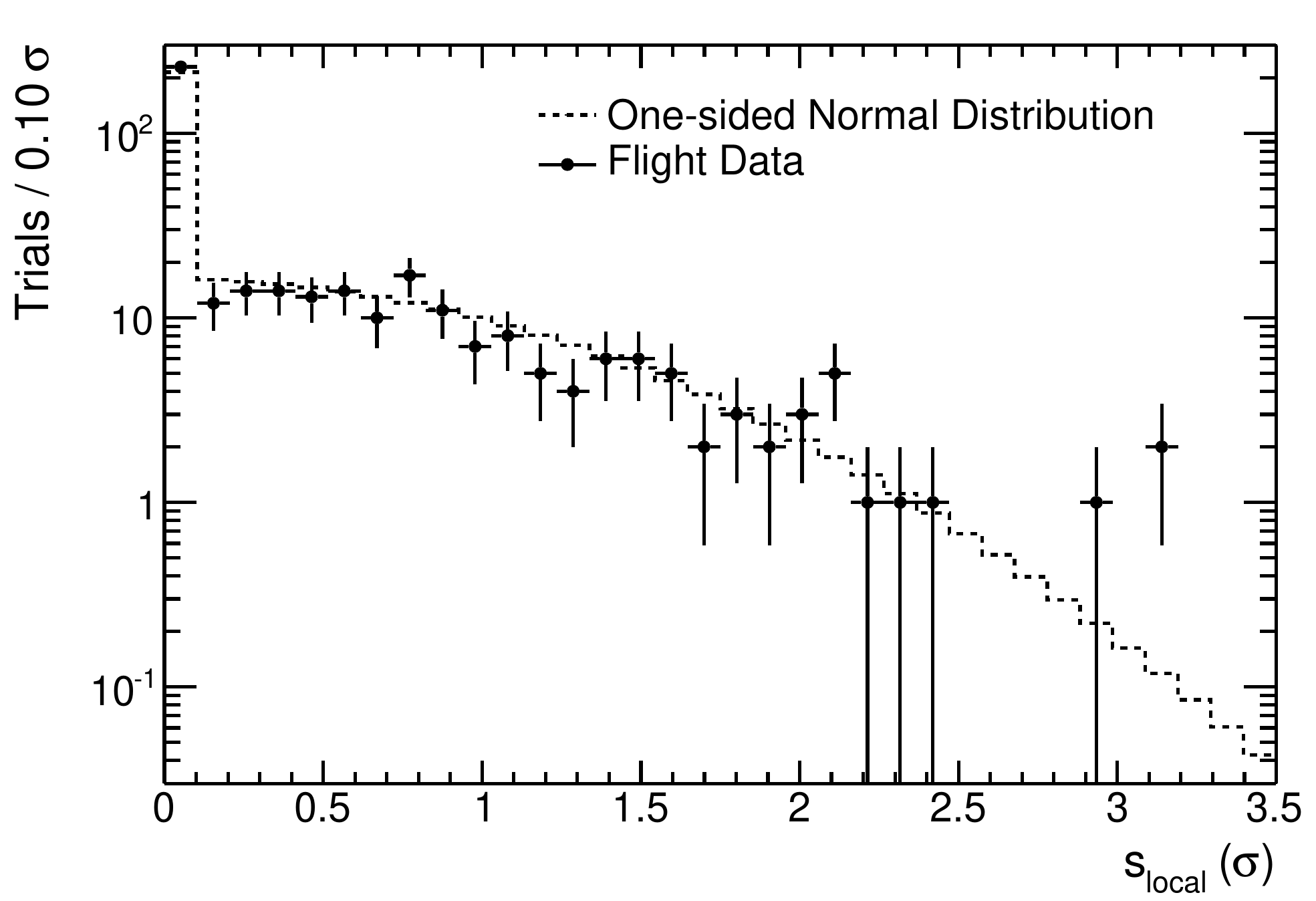}{\caption{\label{fig:Signif_MCvsData}\newText{Local fit significance distribution from
our line search.  The dashed line shows the expected distribution from the null hypothesis.  Note that \nSig\ was bounded to be
positive.}}}

The two most statistically significant fits were in R180 at 6.3~GeV, with $\slocal = 3.1\sigma$ and $f=0.010 \pm0.002$, and in
R3 at 135~GeV, with $\slocal = 3.2\sigma$ (corresponding to $\sglobal=1.5\sigma$) and $f=0.58 \pm 0.18$, where $f$ is the
effective signal fraction at the line energy (\Eqref{eq:FracSig}).  Although the fit at 6.3~GeV in R180 has a relatively large
$TS$ value, the signal fraction is similar to the expected systematic uncertainty of $\delta f = \pm0.008$ (see
\Tabref{Syst_Error_By_ROI}) for R180 at that energy.   \newText{A fine scan ($0.1\sigE$ steps) near 135~GeV in R3 found the
largest  significance at 133~GeV, with $s_{local}=3.3\sigma$.}   We discuss the results near 133~GeV in considerably more
detail in \Secref{sec:133GeV_Feature}.

Since no globally significant lines were detected, we have derived 95\% confidence level (CL) upper limits on the
\gammaRayHyph~flux from spectral lines ($\Phi_{\gamma\gamma}$).  \newText{We set upper limits on $\nSig(\Egamma)$ at the point
where the log-likelihood changes by 1.36 (2.71/2) with respect to the maximum.  Then,} using the magnitude of the averaged
exposure in each ROI at the fit line energy ($\Exposure_{ROI}(\Egamma)$), we can convert the 95\% CL upper limit on
$\nSig(\Egamma)$ to the 95\% CL upper limit on $\Phi_{\gamma\gamma}(\Egamma)$ using 

\begin{equation}\label{equ:FluxToCnt} \Phi_{\gamma\gamma}(\Egamma)=\frac{\nSig(\Egamma)}{\Exposure_{ROI}(\Egamma)}\,.
\end{equation}

\noindent Note that we solve for the $\Phi_{\gamma\gamma}(\Egamma)$ limits generally. If $\Phi_{\gamma\gamma}(\Egamma)$ is
associated with DM annihilation or decay, the corresponding annihilation cross-section or decay lifetime can be solved for
using specific DM model parameters  (e.g., J-factors).

\Figureref{R16_FluxUL} shows the flux upper limits in the R16 (Einasto-optimized) ROI.  Also shown are the expected limits and
expected 68\% and 95\% containment \newText{bands} derived from 1000 single-power law (no DM) MC simulations with
$\GamBkg=2.4$.  Therefore, these containment bands represent the expected statistical variation of a power-law distribution
normalized to the number of events in the dataset. 

\onepanel{ht}{\colSw{Figure_9}}{\caption{\label{fig:R16_FluxUL}95\% CL $\Phi_{\gamma\gamma}$ in the R16 ROI (black). 
Yellow (green) bands show the 68\% (95\%) expected containment derived from 1000 single-power law (no DM) MC simulations.  The
dashed lines show the median expected limits from those simulations.}}

Using \Eqref{eq:annSigTot} with $\frac{dN_{\gamma}}{dE}(\Egamma) = 2\delta(\Egamma-\Ereco)$ and $\Egamma = m_{\chi}$, we solve
for the corresponding upper limits on $\langle\sigma v\rangle_{\gamma\gamma}$, which are shown in \Figref{SigmaVUL} for the R3,
R16, R41, and R90 ROIs for contracted NFW, Einasto, NFW, and Isothermal profiles, respectively.  \newText{When directly
comparable, the upper limits on $\langle\sigma v\rangle_{\gamma\gamma}$ derived by~\citet{REF:Weniger:2012tx} are also shown.
Note that the contracted NFW limits are not compared since significantly different ROIs were used in this search compared to
the search in~\cite{REF:Weniger:2012tx}.  Also, the isothermal limits are not compared since different values of $r_s$ were
used.}

\fourpanel{ht}{\colSw{Figure_10a}}{\colSw{Figure_10b}}{\colSw{Figure_10c}}{\colSw{Figure_10d}}{\caption{\label{fig:SigmaVUL}95\%
CL $\langle\sigma v\rangle_{\gamma\gamma}$ upper limits for each DM profile considered in the corresponding optimized ROI. 
Yellow (green) bands show the 68\% (95\%) expected containment derived from 1000 single-power law (no DM) MC simulations.  The
dashed lines show the median expected limits from those simulations.  \newText{The solid gray line shows the limits derived
by~\citet{REF:Weniger:2012tx} (an independent search for spectral lines from 20$-$300~GeV) when comparable ROIs and identical
DM density profiles were used.}}} 

R180 is optimized for searches for spectral lines from WIMP decays (e.g., $\chi\rightarrow\gamma\nu$).  The flux upper limits
are related to the  lifetime ($\tau_{\gamma\nu}$) lower limits via \Eqref{eq:decSigTot} with $\frac{dN_{\gamma}}{dE}(\Egamma) =
\delta(\Egamma-\Ereco)$ and $m_{\chi}=2\Egamma$,  which are shown in \Figref{R180_TauLL}.

\onepanel{ht}{\colSw{Figure_11}}{\caption{\label{fig:R180_TauLL}95\% CL $\tau_{\gamma\nu}$ lower limits in R180 for an
NFW profile.  Yellow (green) bands show the 68\% (95\%) expected containment derived from 1000 single-power law (no DM) MC
simulations.  The dashed lines show the median expected limits from those simulations.}} 

We present the flux upper limits in all five ROIs and the relevant DM annihilation or decay limits explicitly in
\Appref{app:results}. Recall that we limited our search to energies greater than 30 GeV in R3 (see \Secref{sec:method_roi}). 

The limits presented do not include systematic errors.   As stated in \Secref{sec:systematic_errors_summary} the uncertainties
of the exposure (~$|\delta \Exposure / \Exposure| < 0.16$~) and the energy dispersion modeling (~$\delta \nSig / \nSig =
\SPSB{+0.06} {-0.12}$~) contribute negligibly to the limits when considered in quadrature with the statistical uncertainties.  
On the other hand, the inferred uncertainties of $\delta f$ from \Tabref{Syst_Error_By_ROI} can become significantly larger
than the statistical uncertainties at lower energies and for the larger ROIs.   In fact, the uncertainty of $\delta f$ from
\Tabref{Syst_Error_By_ROI} equals the expected statistical uncertainty at 10~GeV (for R16 and R41), 30~GeV (for R90)  and
70~GeV (for R180).   Empirically, the limits presented in \Figrefs{SigmaVUL}{R180_TauLL} generally lie within the expected
statistical variations, indicating that the systematic uncertainties are not dominating the statistical uncertainties.

%% file: 135Feature.tex
\section{THE LINE-LIKE FEATURE NEAR 133~GeV}\label{sec:133GeV_Feature}

\newText{The most significant fit from our search for spectral lines is for $E_{\gamma}=135$~GeV in our smallest ROI, R3 (see
\Secref{sec:results}).  A fine scan ($0.1\sigE$ steps) around this energy in R3 found the largest significance at 133~GeV, with
$s_{local}=3.3\sigma$.   The finer scan reduces the potential negative bias from the grid spacing to $\delta \nSig / \nSig < 0.02$,
but contributes to the trials factor. Using the procedure described in \Secref{subsec:LEE}, we estimate that if we had used
$0.1\sigE$ steps everywhere the effective number of independent trials would be $\ntrial=295\pm9$,  with which we extract $s_{global}=1.5\sigma$.}  

\newText{This is the same feature that has been reported in the un-reprocessed \pseven\ data at 130~GeV~\cite{Bringmann:2012vr,REF:Weniger:2012tx,Tempel:2012ey,Su:2012ft}.}  The feature has shifted from 130~GeV to 133~GeV \newText{in the reprocessed data}, as expected from the application of improved calibrations (see \Appref{app:pass_7_rep}).  In \newText{the rest of} this section, we discuss the 133~GeV feature in detail.  

\subsection{Evolution of 133~GeV feature with different datasets and signal models}\label{subsec:LineEvolution}

We studied how using reprocessed data and the 2D energy dispersion model (see \Secref{sec:method_2DPDF}) affects the significance of the observed feature in the two smallest ROIs (i.e., where the significances were the greatest): R3, optimized for a contracted NFW profile and R16, optimized for the Einasto profile. Recall that for the R16 dataset, we removed events near bright 2FGL sources (see \Secref{subsec:method_event}).  However, this masking only removes four events near 133~GeV within 3$^{\circ}$ of the Galactic center.

\newText{In order to better compare our results with the works referenced above we fit the \irf{P7CLEAN}\ (un-reprocessed) data
in these ROIs with the 1D energy dispersion model that does not incorporate parametrization  with $P_E$.} 
The local significances for fits at 130~GeV in R3 and R16 are $4.5\sigma$ and $3.9\sigma$, respectively
(see \Figref{line-comparisons-R3}~(a) and \Figref{line-comparisons-R16}~(a)).   \newText{Since these fits were motivated by results 
outside of our search, we cannot estimate an effective trials factor, and do not quote global significances for these fits.}

Using the 1D energy dispersion model and
fitting the \irf{P7REP\_CLEAN}\ at 133~GeV, we found local significances of $4.1\sigma$ (R3) and $2.2\sigma$ (R16) (see \Figref{line-comparisons-R3}~(b) and \Figref{line-comparisons-R16}~(b)).  It is worth noting that 70--80\% of events in the \irf{P7CLEAN} dataset are also in the \irf{P7REP\_CLEAN} dataset, depending on energy.  Therefore, small differences in \slocal\ are expected when evaluated with the \irf{P7CLEAN} or \irf{P7REP\_CLEAN} datasets.   We note in passing that the unmodeled, slight, smearing caused by the time-dependent shift in the absolute energy scale in the un-reprocessed data degraded the energy resolution by less than 5\% relative to the performance for \irf{P7REP\_CLEAN}.

Finally, when we used the 2D signal model, we found that the fits at 133~GeV have local significances of $3.3\sigma$ (R3) and $1.6\sigma$ (R16) (see \Figref{line-comparisons-R3}~(c) and \Figref{line-comparisons-R16}~(c)). 

Fitting the \irf{P7REP\_CLEAN} dataset with the 2D energy dispersion model causes \slocal\ to decrease by $20\%$ in R3 and $27\%$ in R16 compared to fitting with the 1D model. Simulations predict that \slocal\ should increase, on average, by $15\%$ in this case. A decrease by 20\% or more occurred in 2\% of
the simulations.  The decrease in significance with the 2D model implies that the clustering of events around the peak energy as a function of \probE\ in the 
flight data does not match variations in instrument performance well; this somewhat disfavors the interpretation of the 133~GeV feature as a DM line. 

\threepanelvert{ht!}{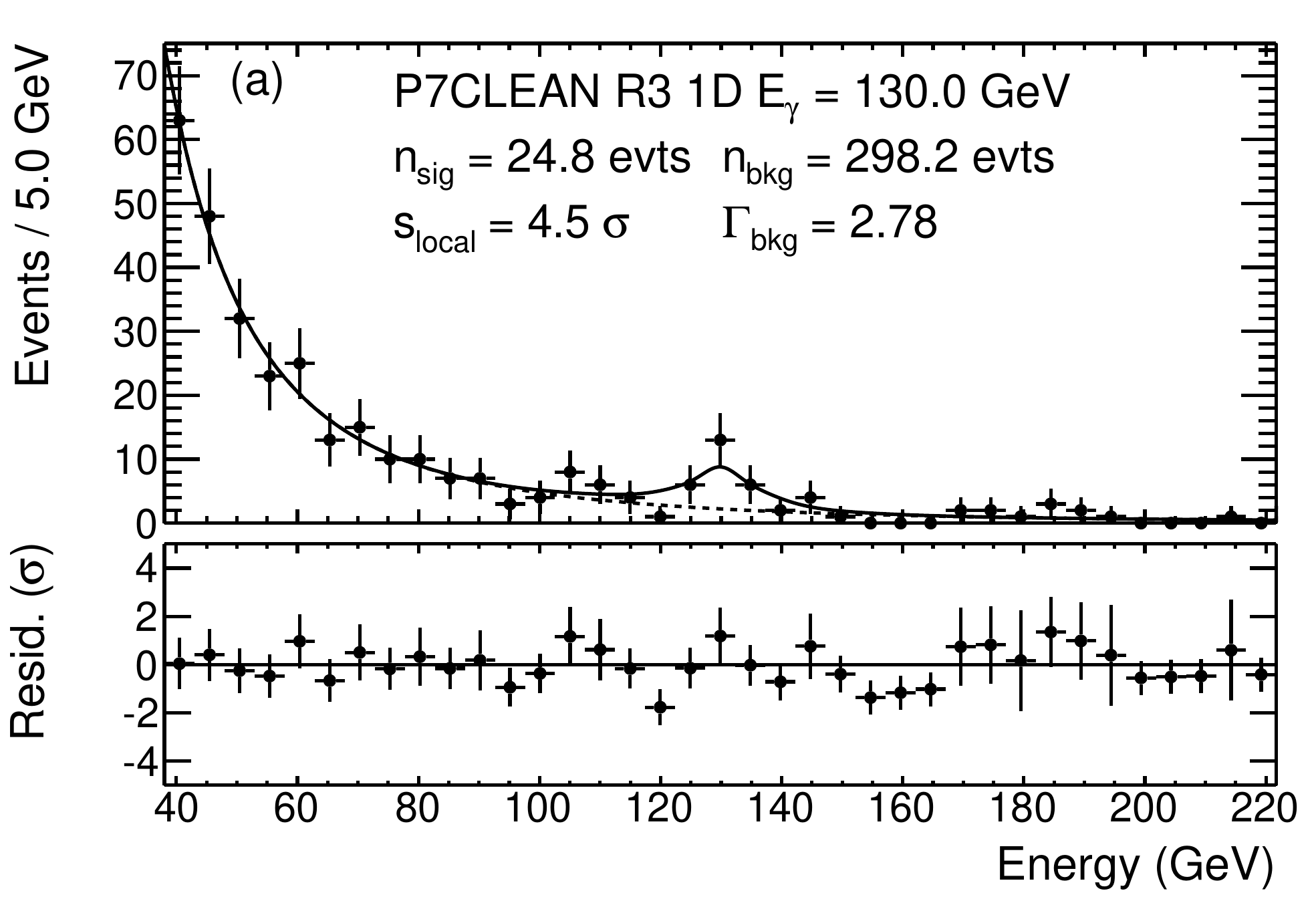}{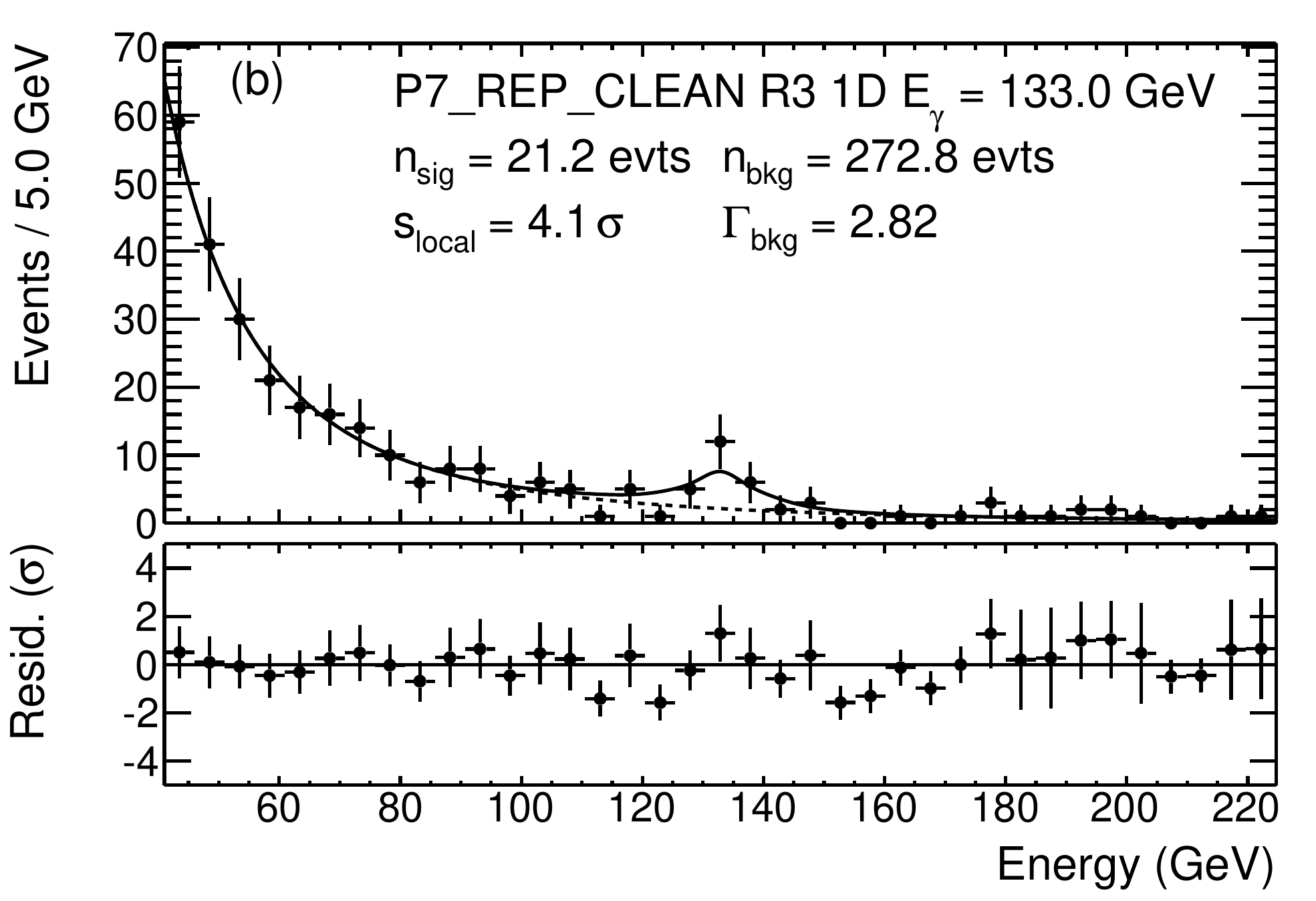}{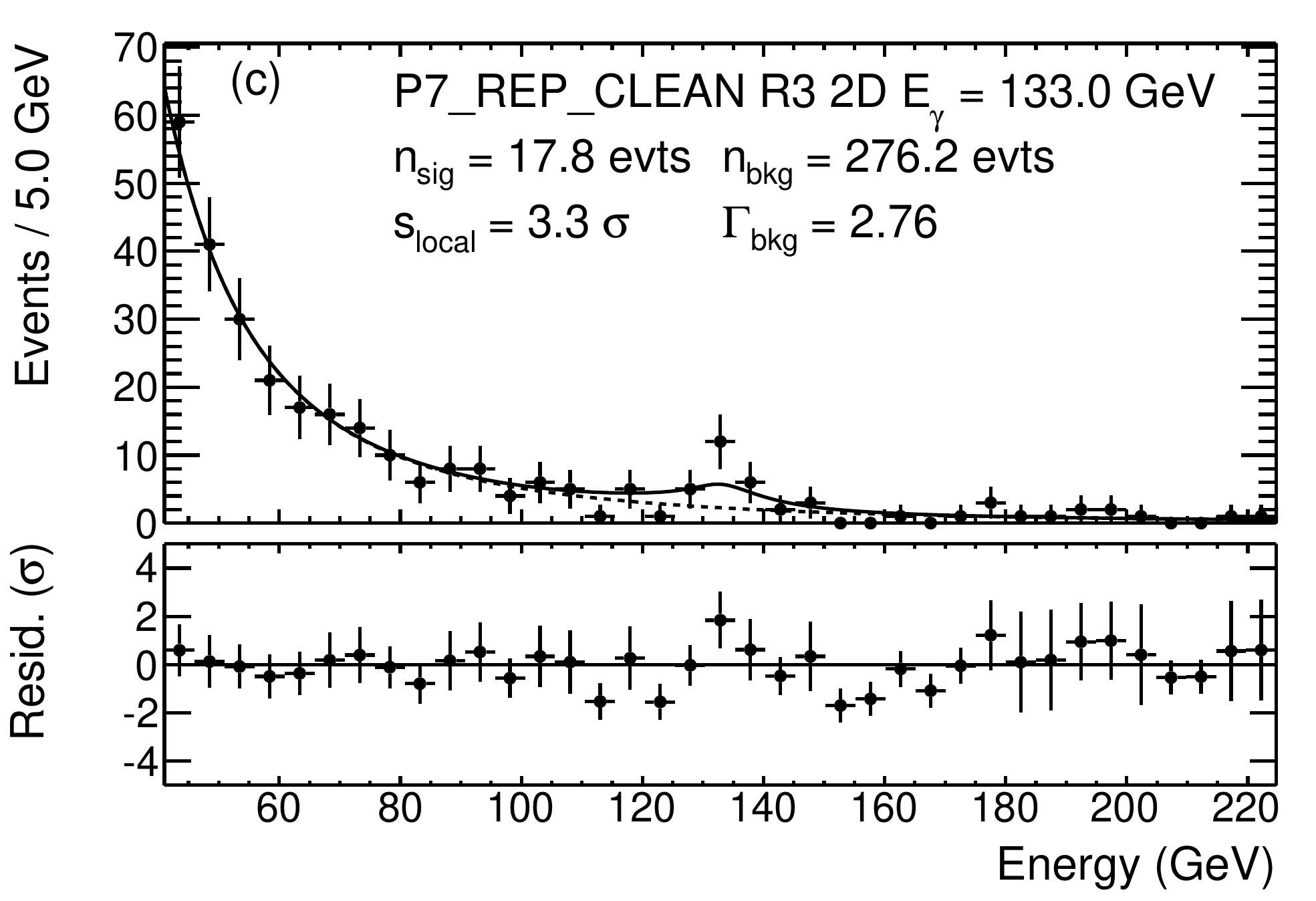}
{\caption{\label{fig:line-comparisons-R3} 
Fits for a line near 130~GeV in R3:
(a) at 130~GeV in the \irf{P7CLEAN}\  data using the 1D energy dispersion model (see \Secref{sec:method_2DPDF});
(b) at 133~GeV in the \irf{P7REP\_CLEAN}\ data again using the 1D model;
(c) same as (b), but using the 2D energy dispersion model  (see \Secref{sec:method_2DPDF}). \newText{The solid curve shows the average model weighted using the \probE\ distribution of the fitted events.}  Note that these fits were unbinned;
the binning here is for visualization purposes, and also that the x-axis binning in (a) is offset by 3~GeV relative to (b) and (c).}}

\threepanelvert{ht!}{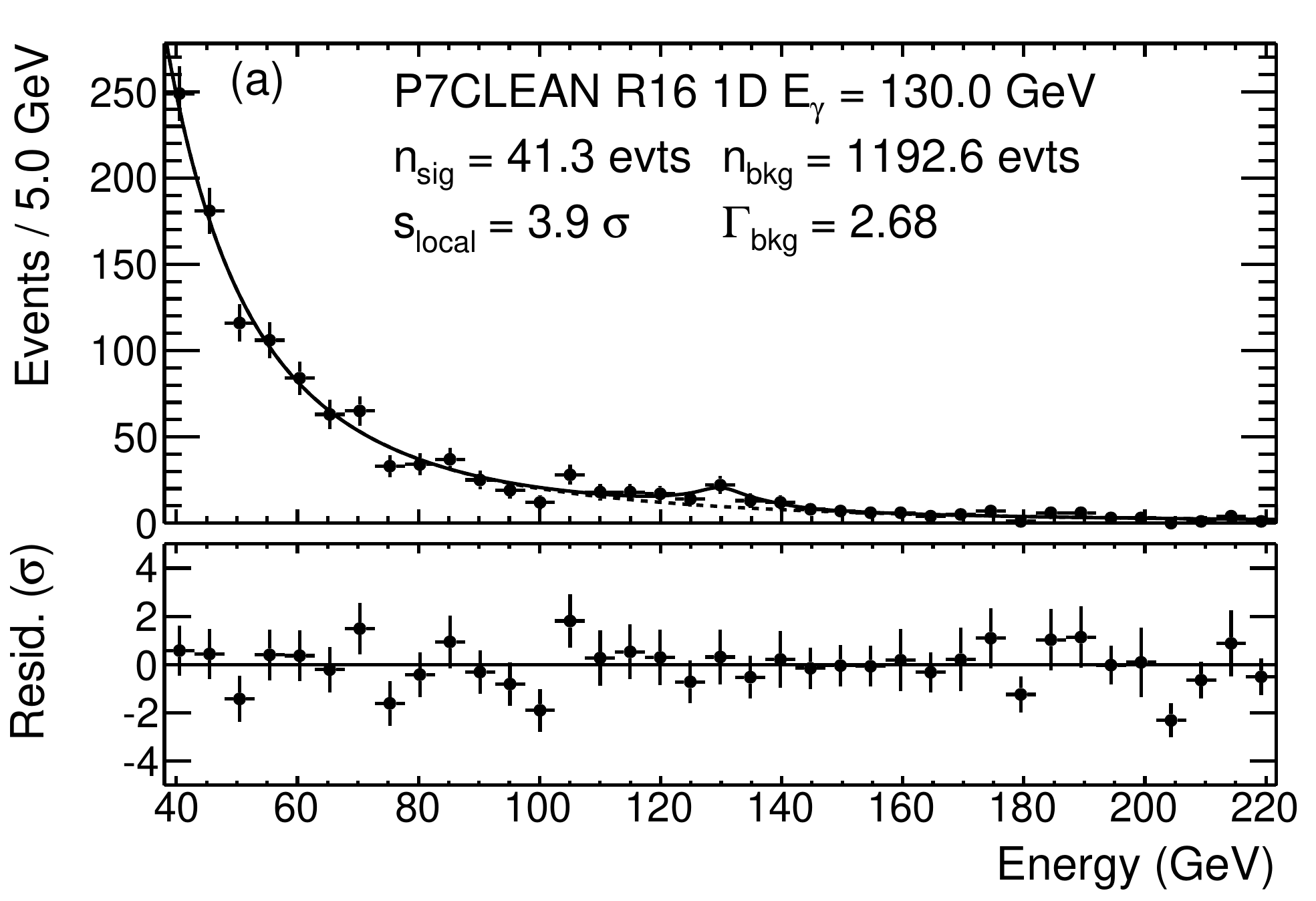}{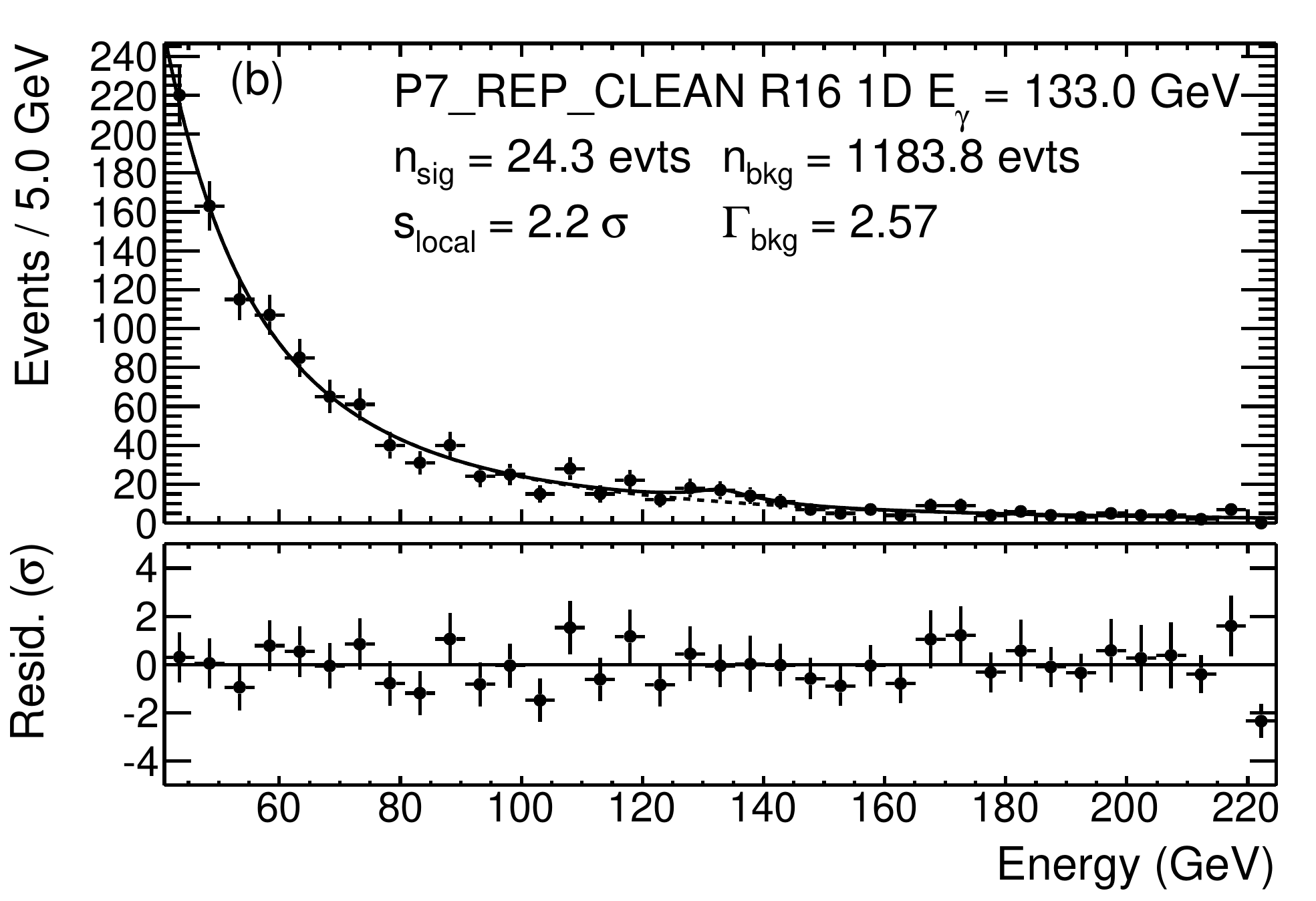}{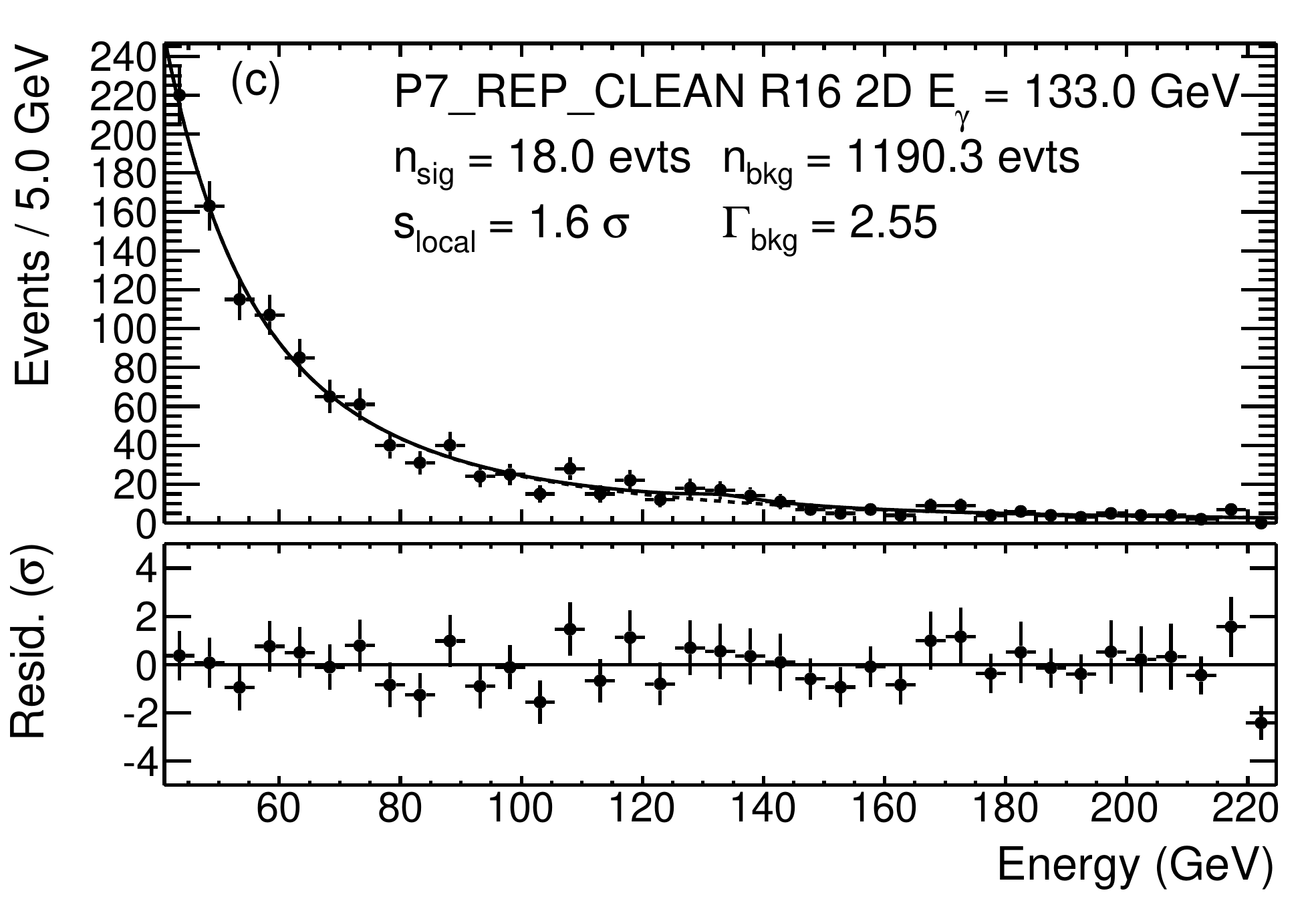}
{\caption{\label{fig:line-comparisons-R16} 
Fit for a line near 130~GeV in R16:  
(a) at 130~GeV in the \irf{P7CLEAN}\  data using the 1D energy dispersion model (see \Secref{sec:method_2DPDF});
(b) at 133~GeV in the \irf{P7REP\_CLEAN}\ data again using the 1D model;
(c) same as (b), but using the 2D energy dispersion model  (see \Secref{sec:method_2DPDF}). \newText{The solid curve shows the average model weighted using the \probE\ distribution of the fitted events.} 
Note that these fits were unbinned;
the binning here is for visualization purposes, and also that the x-axis binning in (a) is offset by 3~GeV relative to (b) and (c).}}

\newText{To test if the feature persists with additional data, we} also extracted a \irf{P7REP\_CLEAN} dataset in R3 that includes data through 12 December 2012 and fit at 133~GeV.  \Figureref{133_4.4yr} shows the fit results to this 4.4-year dataset using both the 1D and 2D energy dispersion models.  The local significance for the 1D energy dispersion model is $3.7\sigma$, and $2.9\sigma$ for the 2D energy dispersion model.  The significance decreased by $\sim 10\%$ with the 4.4-year dataset relative to the 3.7 year dataset.   \newText{This is well within the expected statistical fluctuations for either the signal or null hypothesis.  Here again we do not quote global significances for these fits as the studies were performed outside the context of our original search and it is difficult to estimate a trials factor.}

\clearpage

\twopanel{!ht}{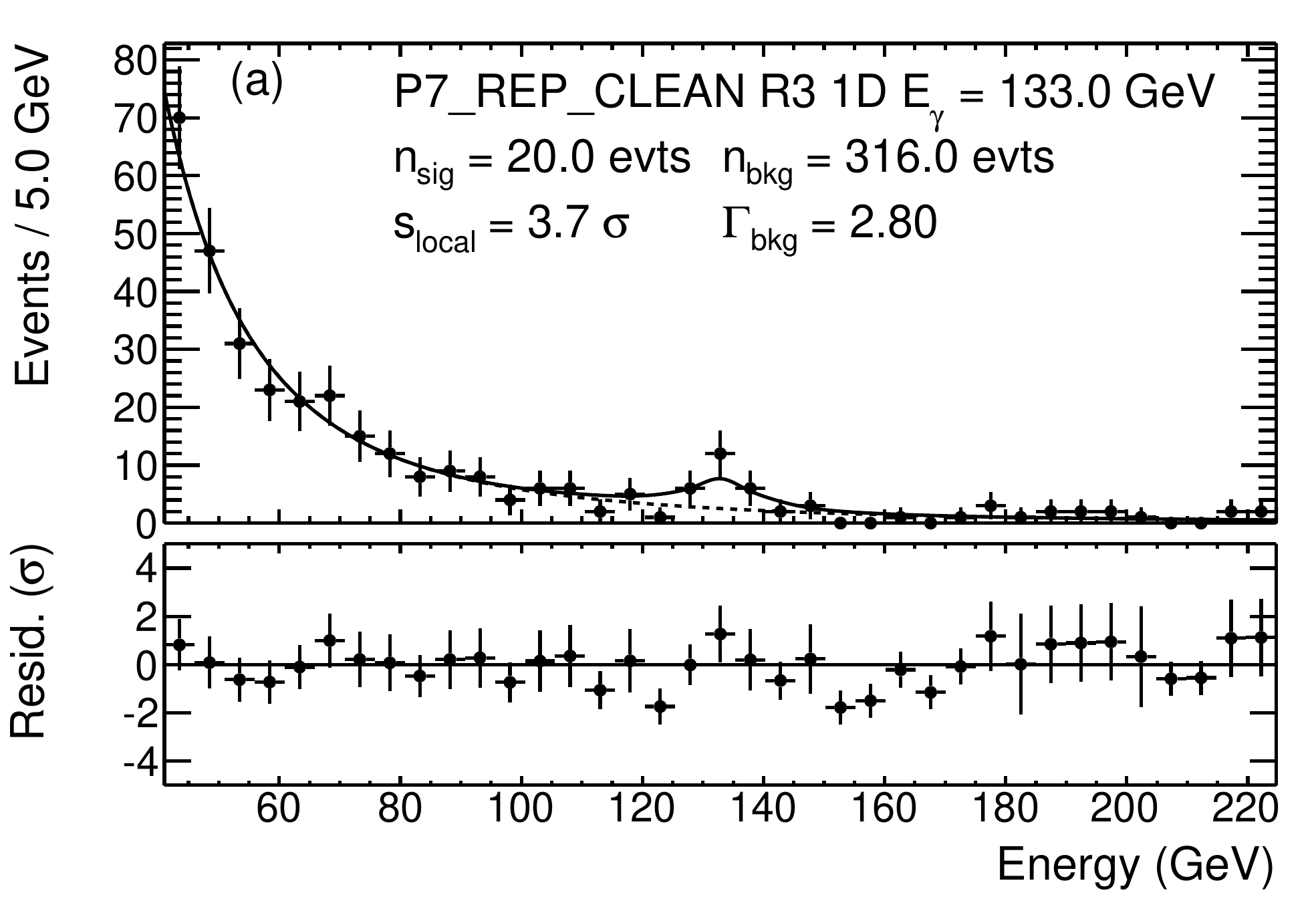}{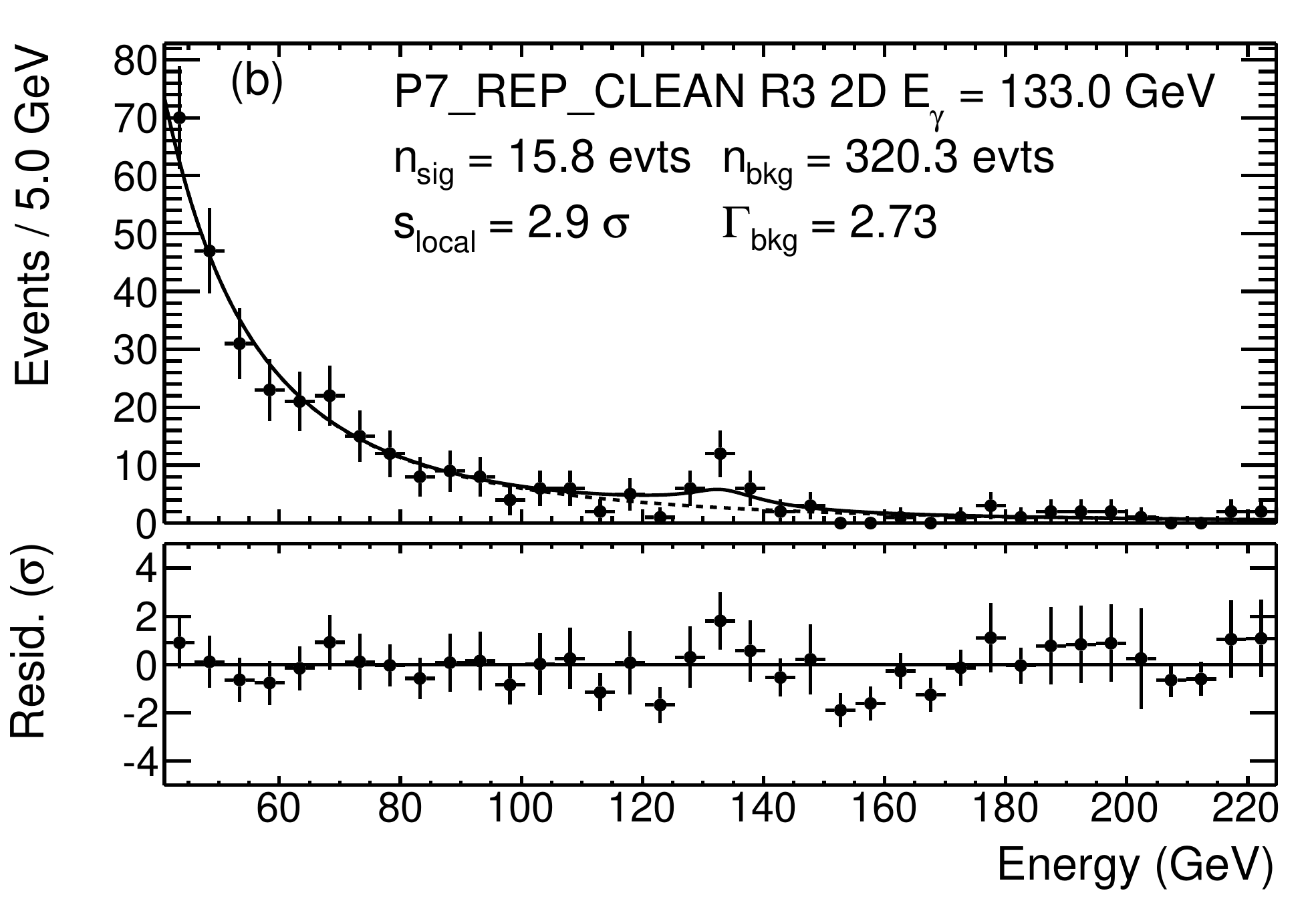}
{\caption{\label{fig:133_4.4yr} Fit for a line-signal signal at 133~GeV in R3 using a 4.4 year \irf{P7REP\_CLEAN} dataset and (a) the 1D energy dispersion model; (b) the 2D energy dispersion model. \newText{The solid curve shows the average model weighted using the \probE\ distribution of the fitted events.  Note that these fits were unbinned;
the binning here is for visualization purposes.}}}

\subsection{Width of the feature near 133~GeV.}\label{subsec:Width_133GeV_Feature}

We note that the 2D model predicts a slightly broader energy distribution than the 1D model.  As discussed in \Secref{subsec:fitting_method}, the 2D model depends on the \probE\ distribution in the data.  In fact, by inspection, the feature in the flight data appears to be narrower than both the 1D and 2D models, e.g., \Figref{line-comparisons-R3}~(b and c). To quantify this, we scaled the standard deviations of each of the three Gaussian functions that together are used to model the energy dispersion in each \probE\ bin in the 2D model by a common scale factor ($s_{\sigma}$), while also scaling the means to preserve the overall shape of the model.  \newText{This adds another degree of freedom to the fit with signal relative to the background-only fit (n=2).}  We then refit at 133~GeV in the R3 ROI; the best fit value was $s_{\sigma}=0.32^{+0.11}_{-0.07}$ as shown in \Figref{133Width}. The fit with $s_{\sigma}$ increases \newText{the $TS$ by 9.4 relative to the fit without $s_{\sigma}$}.  In the case of scaling the 1D model, the fit returns $s_{\sigma}=0.44^{+0.16}_{-0.13}$ ($\Delta TS=5.3$). \newText{Pre-launch beam tests indicated that the uncertainty of the energy resolution is less than 10\% of the measured resolution up to the maximum accessible beam energy of 280~GeV~\cite{REF:2012.P7Perf}.  Therefore, we conclude that the feature in the data is narrower than our expected energy resolution by a factor of 2$-$3, and is inconsistent with the expected resolution at the 2$-$3$\sigma$ level.}

\onepanel{!ht}{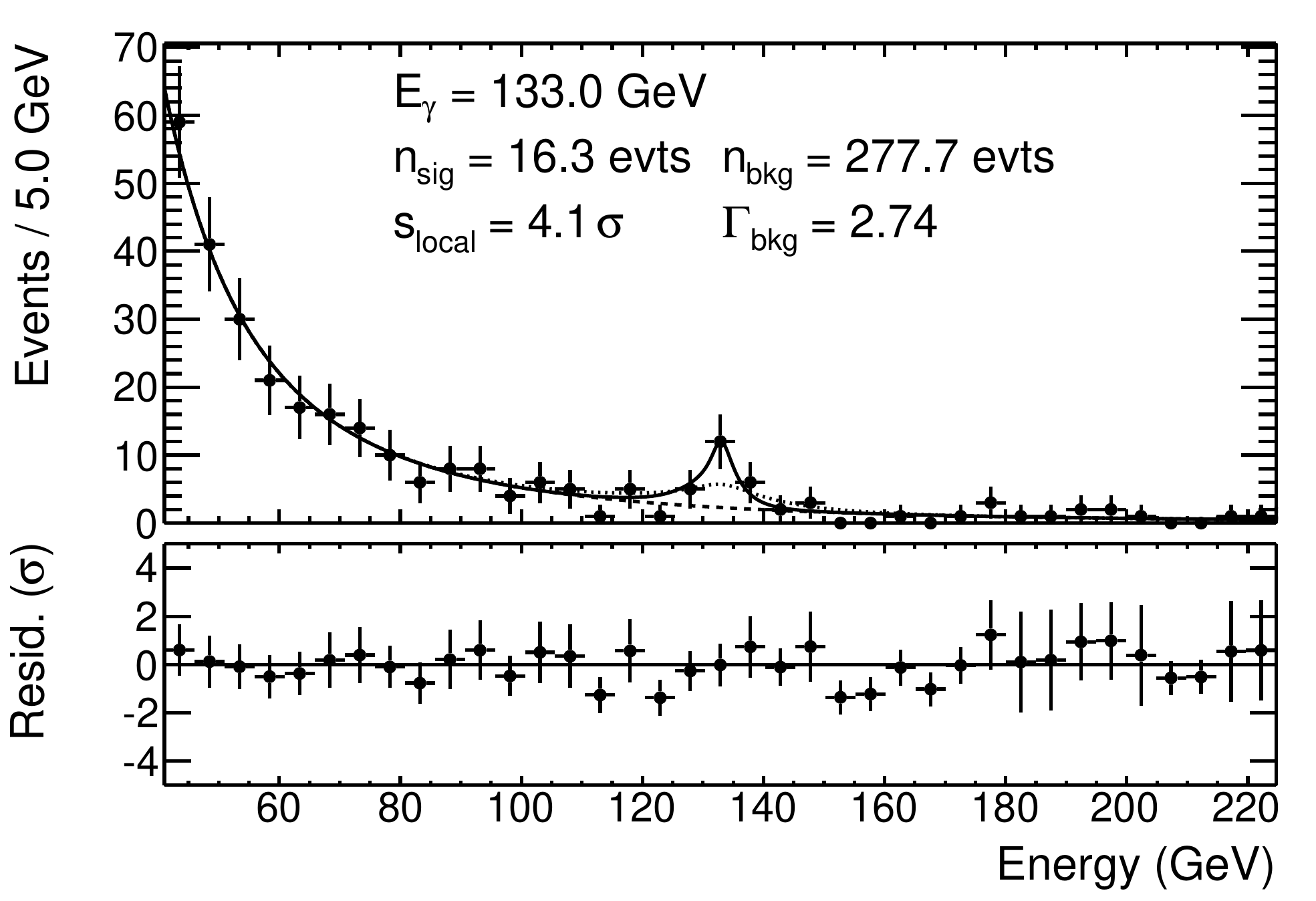}
{\caption{\label{fig:133Width} Fit to a \gammaRayHyph\ line at 133~GeV in the 3.7 year \irf{P7REP\_CLEAN}\ R3 data using the 2D model including a scale factor for the width of the energy dispersion. \newText{ The solid curves show the average model weighted using the \probE\ distribution of the fitted events.} The best fit width of the energy resolution was $s_{\sigma}=0.32^{+0.30}_{-0.13}~(95\%~{\rm CL})$ of that predicted from MC simulations. The dotted line shows the best-fit curve with $s_{\sigma}$ fixed to 1.0. Note that when $s_{\sigma}$ is allowed to vary the signal model includes two more degrees of freedom than the null hypothesis, so \slocal\ is less than $\sqrt{TS}$.  \newText{Also, note that these fits were unbinned;
the binning here is for visualization purposes.}}}

\subsection{133~GeV feature in the control regions}\label{subsec:133ControlRegions}

We examine two control datasets that are expected to contain little or no DM.  The first was the Limb dataset (described in \Secref{subsec:method_event}), while the second was a region centered on the Galactic plane but excluding the Galactic center, which we call the inverse ROI.  The inverse ROI contains a variety of \gammaRayHyph\ sources, but provides good statistics and a reasonable sample of the astrophysical backgrounds that we might expect from the Galactic center.   See \Tabref{event_samples} for event selection details. 

\subsubsection{The Earth Limb}\label{subsec:133GeV_Feature_Earth_Limb}

\Figureref{Limb6Sigma133GeVFit} shows the fit using our 2D energy dispersion model (see \Secref{sec:method_2DPDF}) at 133~GeV to the Limb data, 
which indicates a $2.0\sigma$ excess.  We calculated the fractional size of the signal using \Eqref{eq:FracSig} to be $f(133~{\rm GeV})_{\rm Limb}=0.14 \pm 0.07$.   
The gamma-ray spectrum of the Limb is expected to be featureless.  Therefore, the appearance of a line-like feature in the Limb at the same energy as the feature seen in the Galactic center suggests that some of the 133~GeV Galactic center feature may be due to a systematic effect.  We do note that the fractional size of the feature in the Limb is smaller than observed in the smallest ROIs around the Galactic center: $f(133~{\rm GeV})_{\rm R3}=0.61 \pm 0.19$.   We also note that significance of the feature in the Limb is somewhat reduced in the \irf{P7REP\_CLEAN} dataset relative to the \irf{P7CLEAN} dataset, where $f(130~{\rm GeV})_{\rm Limb}=0.18$.

\onepanel{ht!}{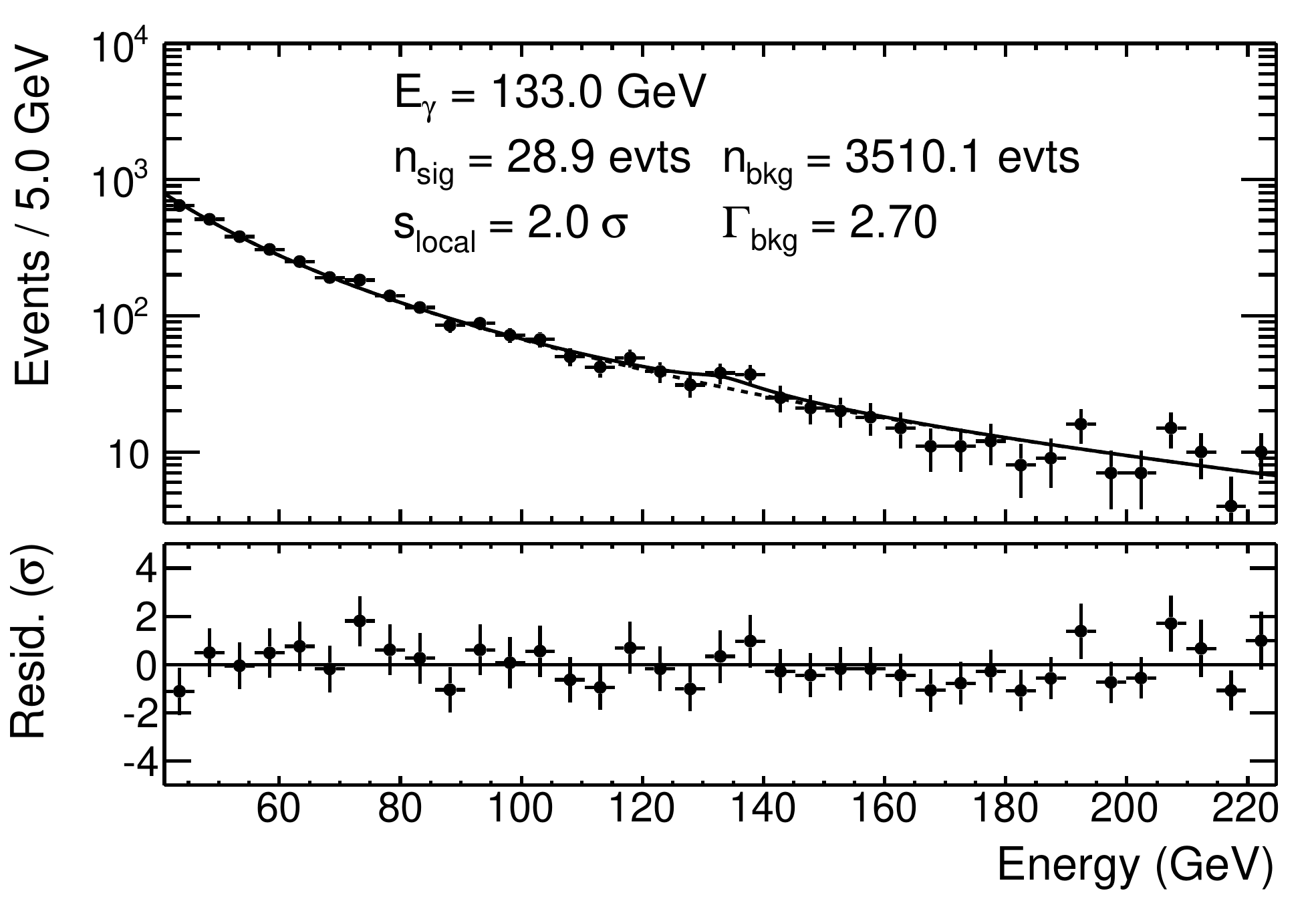}
{\caption{\label{fig:Limb6Sigma133GeVFit} Fit at 133~GeV line to the Limb data (\irf{P7REP\_CLEAN}) using the 2D energy dispersion model. \newText{The solid curve shows the average model weighted using the \probE\ distribution of the fitted events.  Note that these fits were unbinned;
the binning here is for visualization purposes.}}}

The Limb is bright enough to be seen in the \newText{least stringent \gammaRayHyph\ selection},
\irf{P7REP\_TRANSIENT}, which is meant to be used to study transient phenomena like \gammaRayHyph\ bursts.  The \irf{P7REP\_TRANSIENT}
event class has much higher rates of CR contamination than the \irf{P7REP\_CLEAN} class, ($\sim 10$~Hz compared to $< 0.1$~Hz), as it does not
include some of the more stringent criteria needed to achieve the $\mathcal{O}(10^5)$ CR rejection required for point-source analysis.  More details 
about the specific event selection criteria for the various event classes are available in Sec.~3.3 of reference~\cite{REF:2012.P7Perf}.  \newText{We note
for completeness that the \irf{P7REP\_TRANSIENT} Limb event sample does not show any feature at 133~GeV.}

\newText{We have used the background subtraction technique described in Sec.~5.3.1 of~\cite{REF:2012.P7Perf} on both the
\irf{P7REP\_TRANSIENT} and \irf{P7REP\_CLEAN} Limb samples to measure the \gammaRayHyph\ efficiency going from the
\irf{P7REP\_TRANSIENT} to \irf{P7REP\_CLEAN} selection as a function of energy.}  For this study, we used $111^{\circ} <
\theta_{\rm z} < 113^{\circ}$ for the signal region and $108.5^{\circ} < \theta_{z} < 109.4805^{\circ}$ and  $114.4701^{\circ}
< \theta_{\rm z} < 115.5^{\circ}$ for the background regions.  The specific angles were chosen such that the signal and
background regions contain the same solid angle.  The $\theta_{\rm z}$ distributions for the signal and background regions, as
well as the extracted efficiencies are shown in \Figref{LimbEfficiency}.  The predicted efficiency based on the
\irf{P7REP\_TRANSIENT} and \irf{P7REP\_CLEAN} IRFs and the observing profile for the Limb, is also shown for comparison.  
While the predicted efficiency is smooth and featureless, the flight data suggest dips in efficiency above and below 133~GeV.  The efficiency at $120$~GeV is $\sim 80\%$ of the MC prediction, and $\sim 60\%$ for the dip above 133~GeV.  

\newText{We performed MC simulation studies with background-only event samples, modified the exposure correction $\eta(E)$, 
based on the data to MC efficiency ratio from \Figref{LimbEfficiency}, and estimated the expected induced fractional signal when fitting with our 1D PDF. 
In all cases the average induced fractional signal was less than 3\%.  However, we do not have adequate statistics to measure the efficiency in finer energy 
bins and can not rule out narrower sub-structures contributing to an induced signal.  Also, we observed in these simulations a systematic bias in the fitted 
spectral index of $\delta\GamBkg\sim-0.05$, which in turn caused the fits to be more affected by upward 
statistical fluctuations near the fit energy and broadened the distribution of \nSig\ by $\sim20-30$\%.}
  
\newText{Therefore, although suggestive, we do not believe that the measured variations in \gammaRayHyph\ efficiency provide a complete explanation for the observed feature of $f(133~{\rm GeV})_{\rm Limb}=0.14$, which is well outside the range of induced signals seen in the Limb, which are typically less than $f = 0.05$, see \Secref{sec:LimbSmoothness}.}   The potential origin of the features observed in the Transient-to-Clean efficiency observed in the Limb data is discussed further in \Secref{subsec:133GeV_Feature_Events}.

In \Figref{LimbEfficiency}~(a), it is clear that in the \irf{P7REP\_CLEAN} selection the $\theta_{\rm z}$ background regions contain very few events;  in fact, the exposure for the Limb sample is over 400 times smaller than for the Celestial sample, therefore the expected cross-contamination of the Limb sample from a signal of $\sim 25$ events at the Galactic center would be less than a single event.  

\twopanel{!ht}{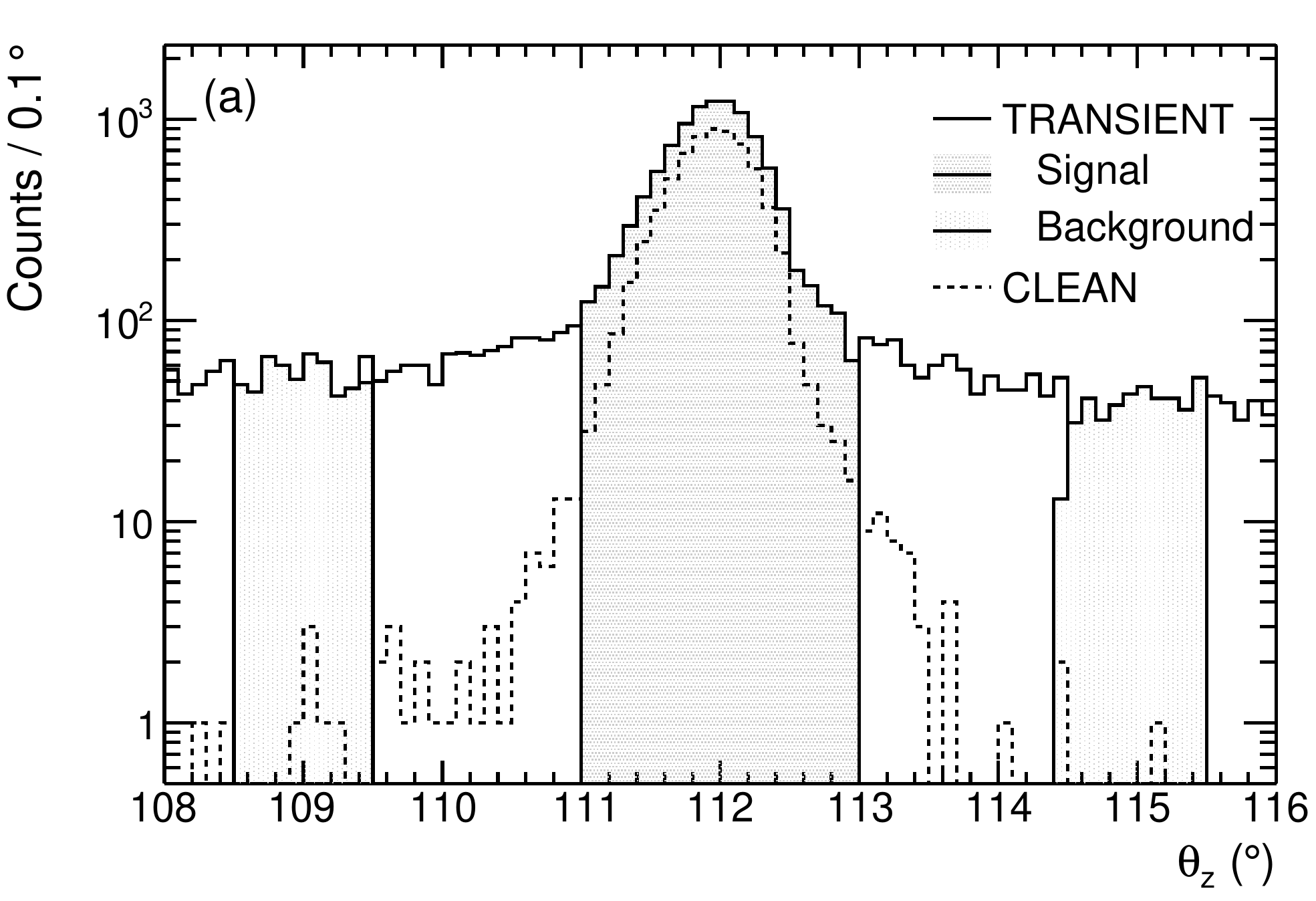}{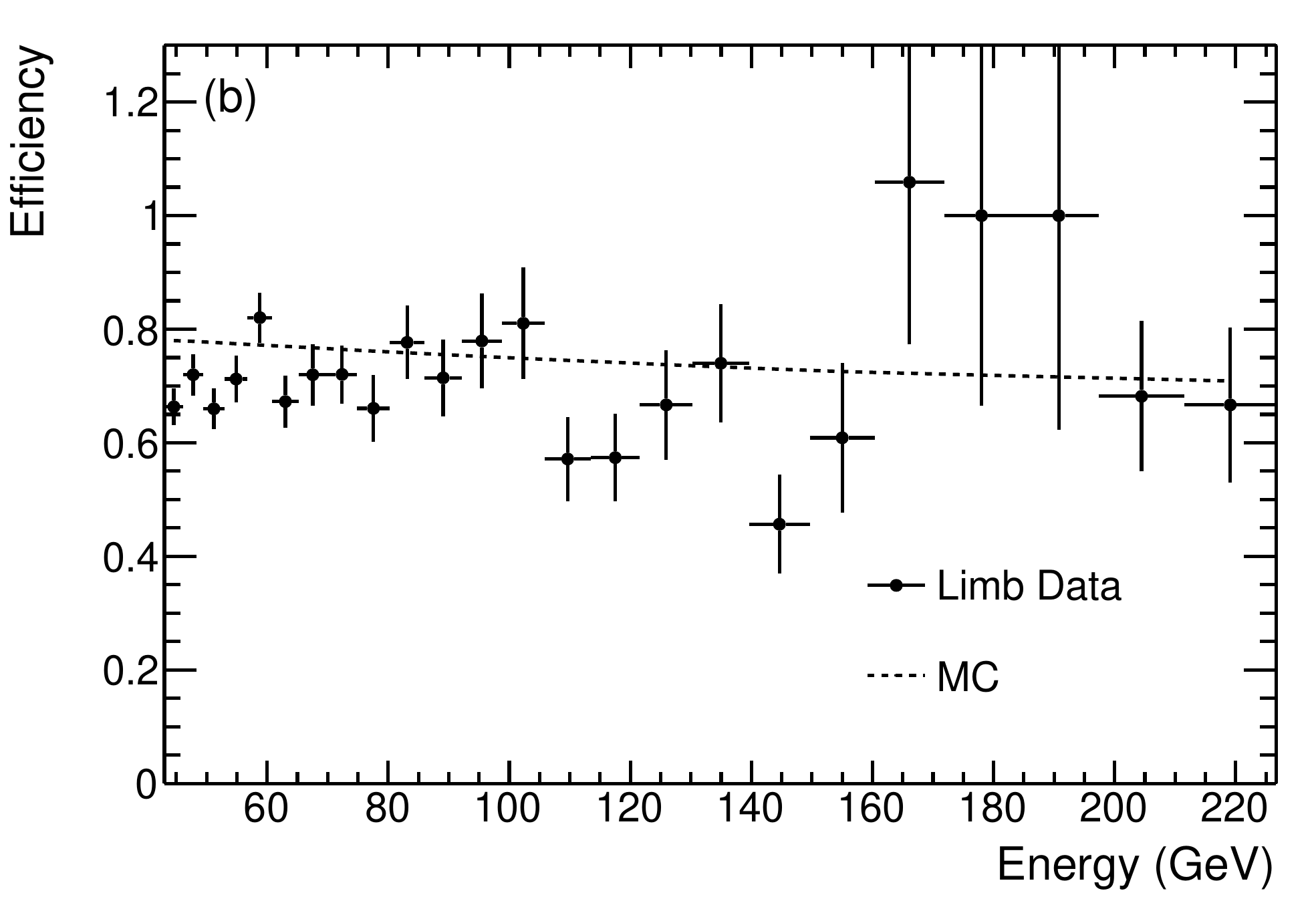}
{\caption{\label{fig:LimbEfficiency}
Measurement of the \irf{P7REP\_TRANSIENT}-to-\irf{P7REP\_CLEAN}\ efficiency using the Limb control sample:
(a) the distribution of $\theta_z$ for all events in the \irf{P7REP\_TRANSIENT} and  \irf{P7REP\_CLEAN} Limb samples for 2.6 GeV $< \Ereco <$ 541 GeV.
including signal and background regions;  (b) the \irf{P7REP\_TRANSIENT}-to-\irf{P7REP\_CLEAN} efficiency for Limb data and MC.  MC has been weighted to have the same livetime distribution with $\theta$ as the Limb data.}}

\subsubsection{The Inverse ROI}\label{subsec:133GeV_Feature_Inverse_ROI}

\newText{As a further control study we also searched for features elsewhere along the Galactic plane.}   We define the inverse ROI A to be events with $|b|<10^{\circ}$, excluding a $20^{\circ} \times 20^{\circ}$ square in the Galactic center in the Celestial dataset. In addition to A, we also examined inverse ROIs B and C, which are subsets of inverse ROI A with $|b|>1^{\circ}$ and $|b|<1^{\circ}$ respectively. \Figureref{InverseROIFit} shows the results of fits for lines at 133~GeV in the three inverse ROI regions. Regions A, B, and C show no indication of a line-like feature at 133~GeV with $\slocal>1.1\sigma$. We also scanned using $20^{\circ}\times20^{\circ}$ ROIs along the Galactic plane resulting in 17 independent fits.  \Figureref{InverseROIFit} (d) shows the  results from the fit at 133~GeV with the greatest statistical significance, where $\slocal=2.0\sigma$.  Thus we find no clear indication for a 133~GeV line feature in these inverse ROI control datasets.

\fourpanel{!ht}{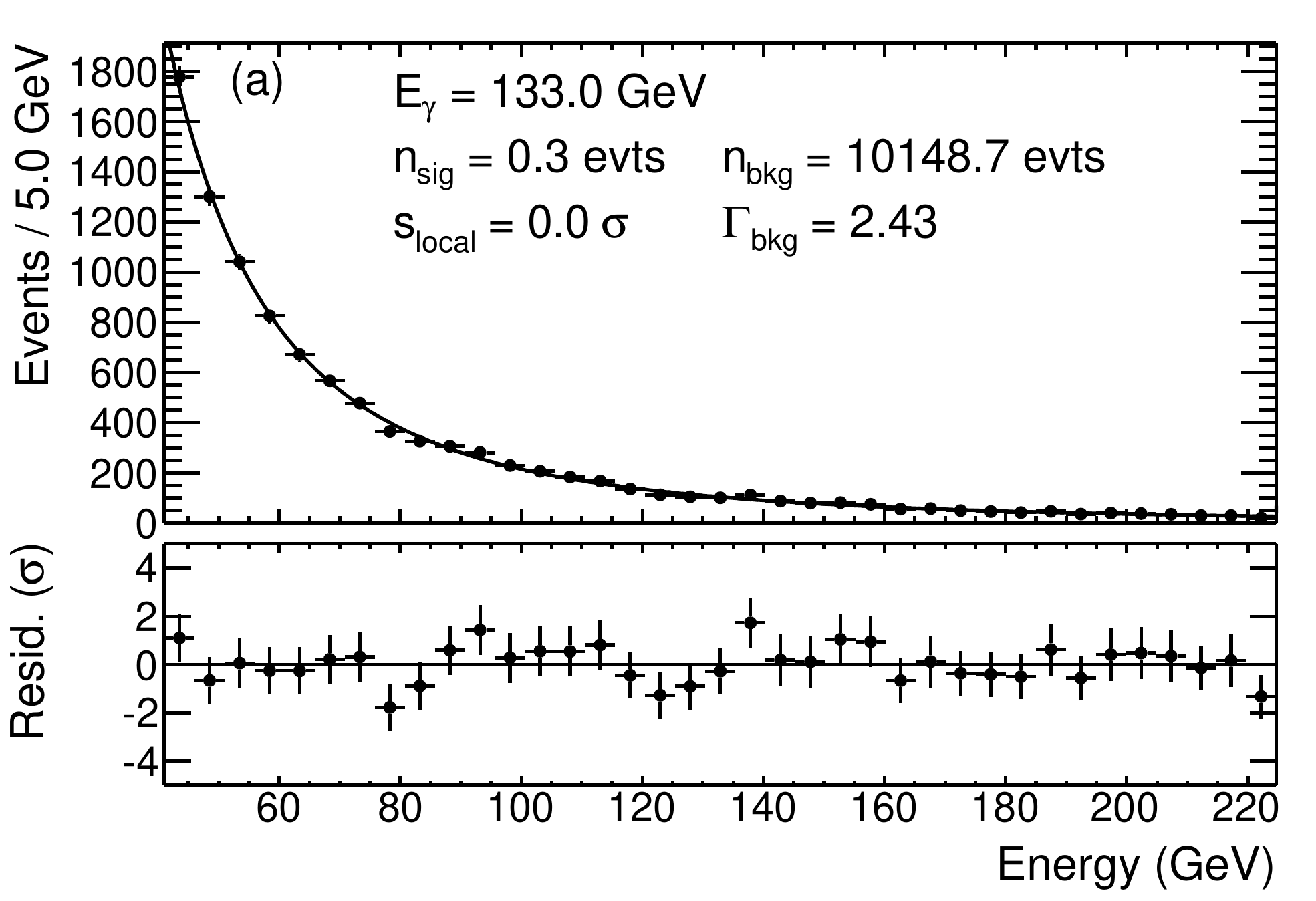}{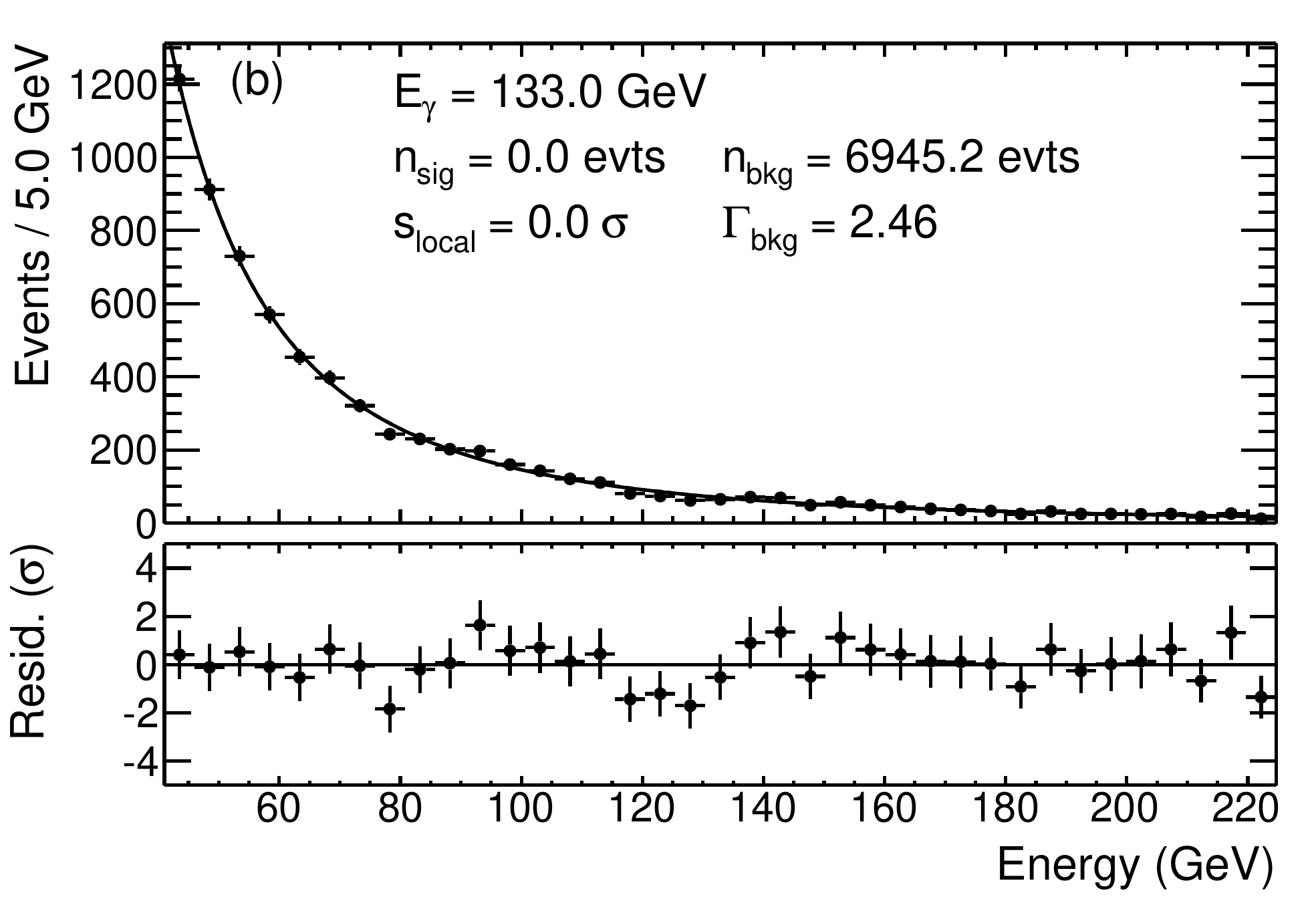}{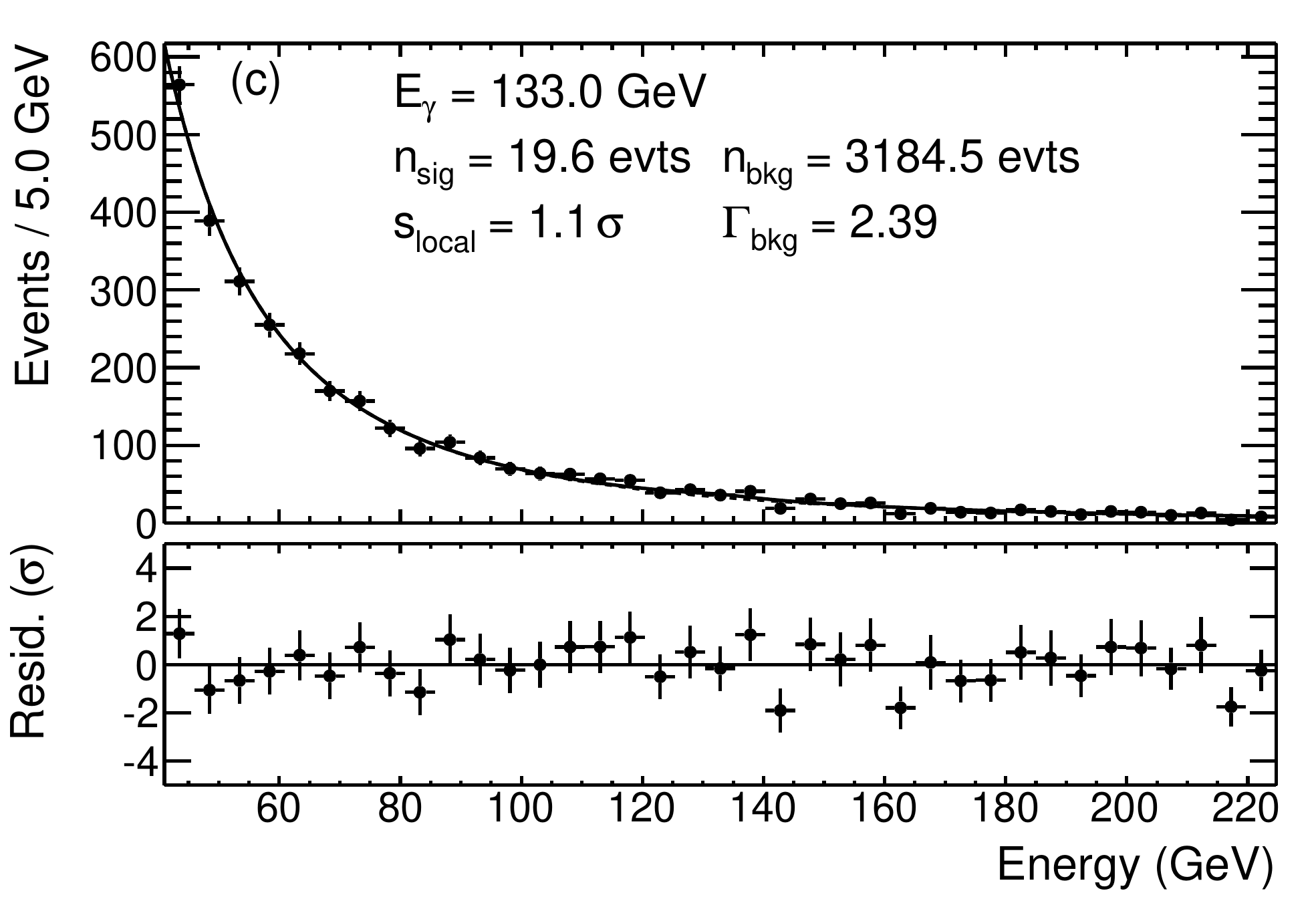}{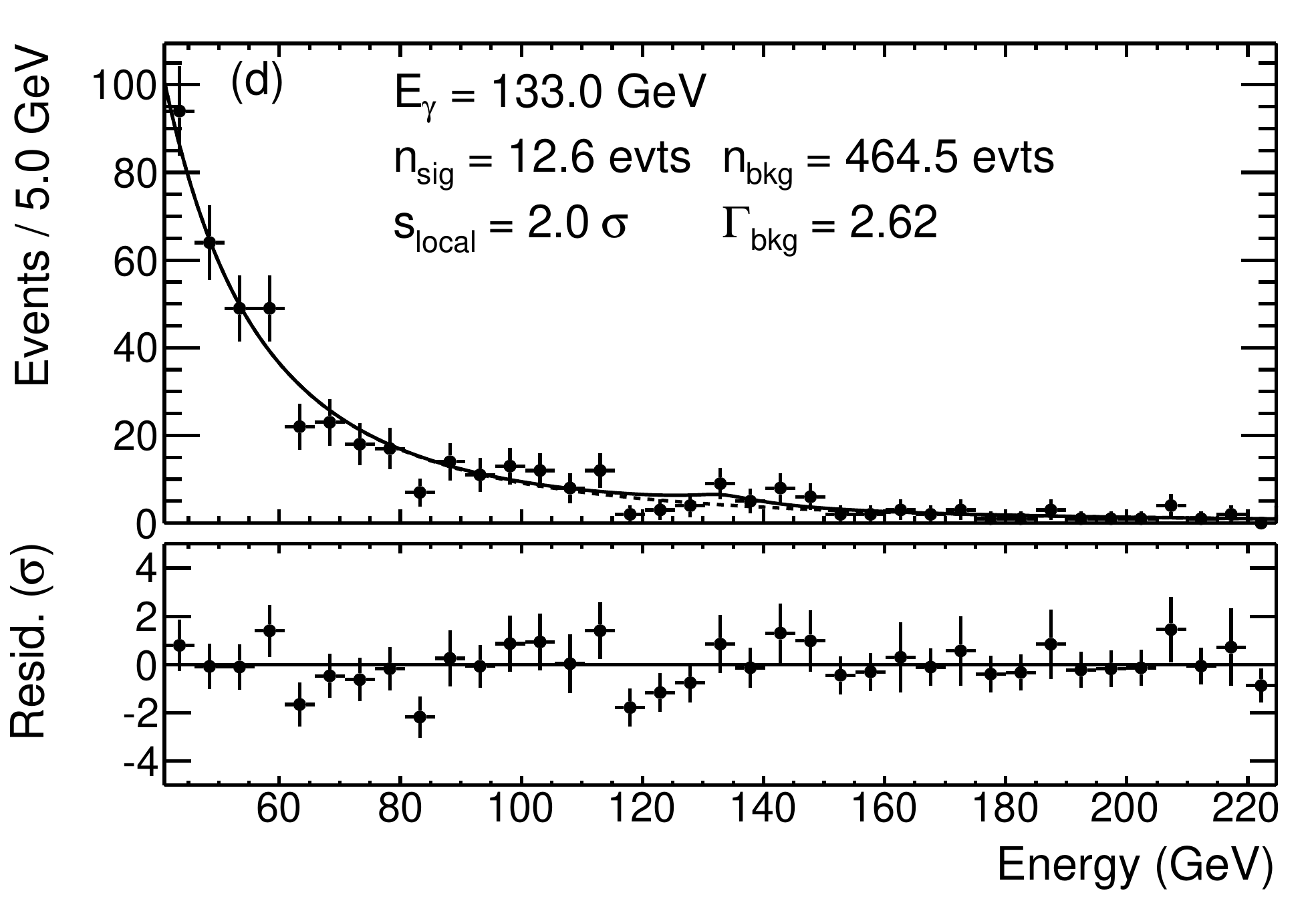}
{\caption{\label{fig:InverseROIFit}Fits for a 133~GeV line in inverse ROIs (\irf{P7REP\_CLEAN}) using the 2D energy dispersion model: (a) A, (b) B, (c) C, and (d) D.  See text for definitions of Regions A,B and C, region D is $|b|<10^{\circ}, 90^{\circ}<l<110^{\circ}$.  \newText{The solid curves show the average model weighted using the \probE\ distribution of the fitted events.  Note that these fits were unbinned;
the binning here is for visualization purposes.}}}

\subsection{Examination of the events contributing the 133~GeV feature}\label{subsec:133GeV_Feature_Events}

We have examined many aspects of the events contributing to the 133~GeV feature, and compared them to events at nearby energies as well as with MC simulations. Within the limited statistics available, the events contributing to the 133~GeV feature exhibit few particularly striking characteristics. The two most notable features are: 
\begin{fermienumerate}
\item{The consistency between the reconstructed direction as estimated by the tracker and the primary axis of the energy deposition in the calorimeter is somewhat worse in the flight data than in the MC simulations (\Figref{CAL_TKR_Variables}).  The disagreement was even greater before reprocessing the data with updated calorimeter calibration constants.  This disagreement is seen in several quantities that contribute strongly to determining \probE, so it is unsurprising that \probE\ tends to have slightly lower values in the flight data, or that the data-MC agreement of the \probE\ distribution has improved with the reprocessed data (see \Figref{ProbDist_DataVsMC}).   We also note
that, with the available statistics, the flight-data from R16 is consistent with the distribution from the entire sky.}
\item{The $\theta$ distribution of the events contributing to the 133~GeV feature is marginally different statistically than for events at other energies and the MC predictions.  This is discussed in more detail in \Secref{subsec:133GeV_Feature_Theta}.}
\end{fermienumerate}

\twopanel{!ht}{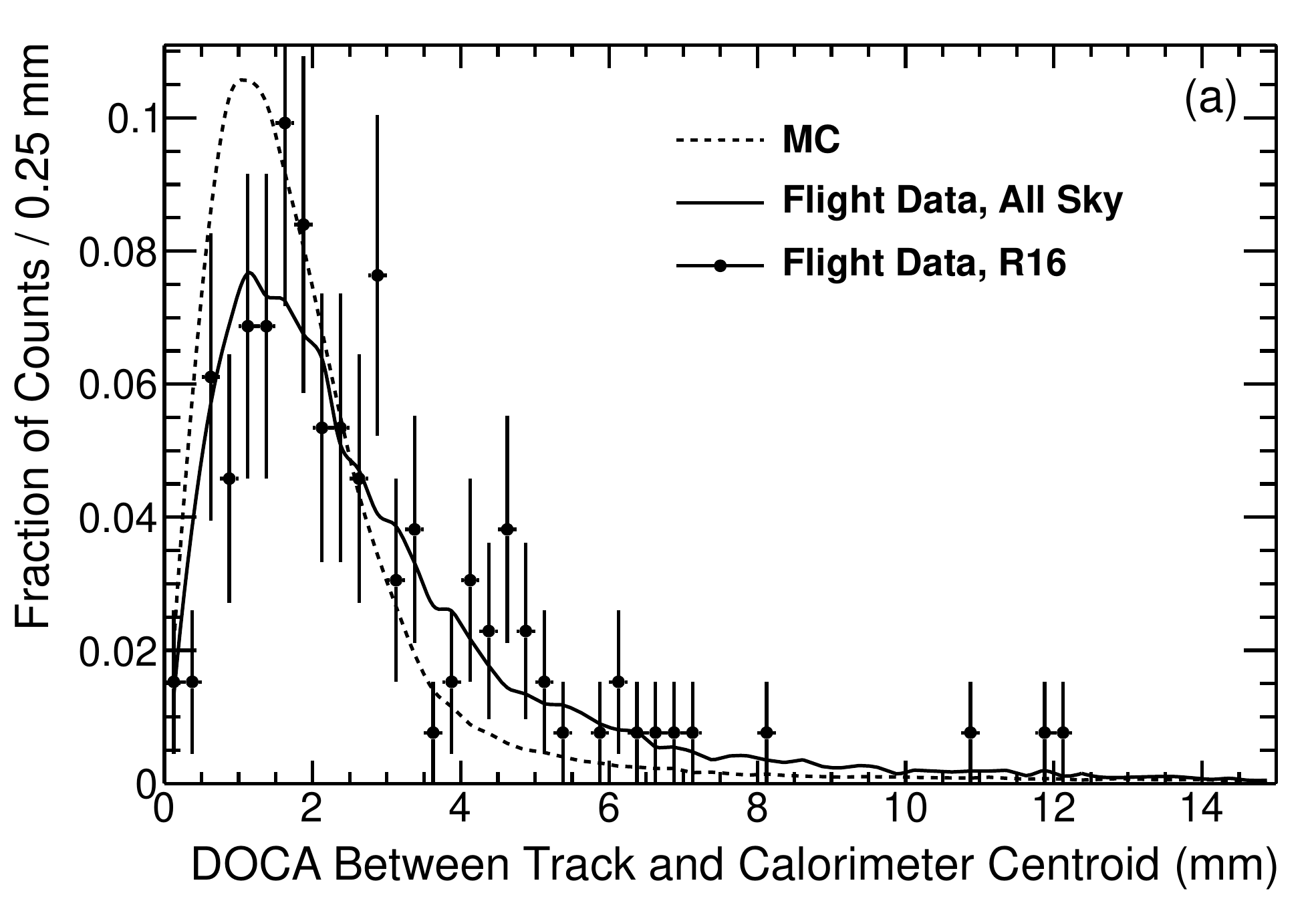}{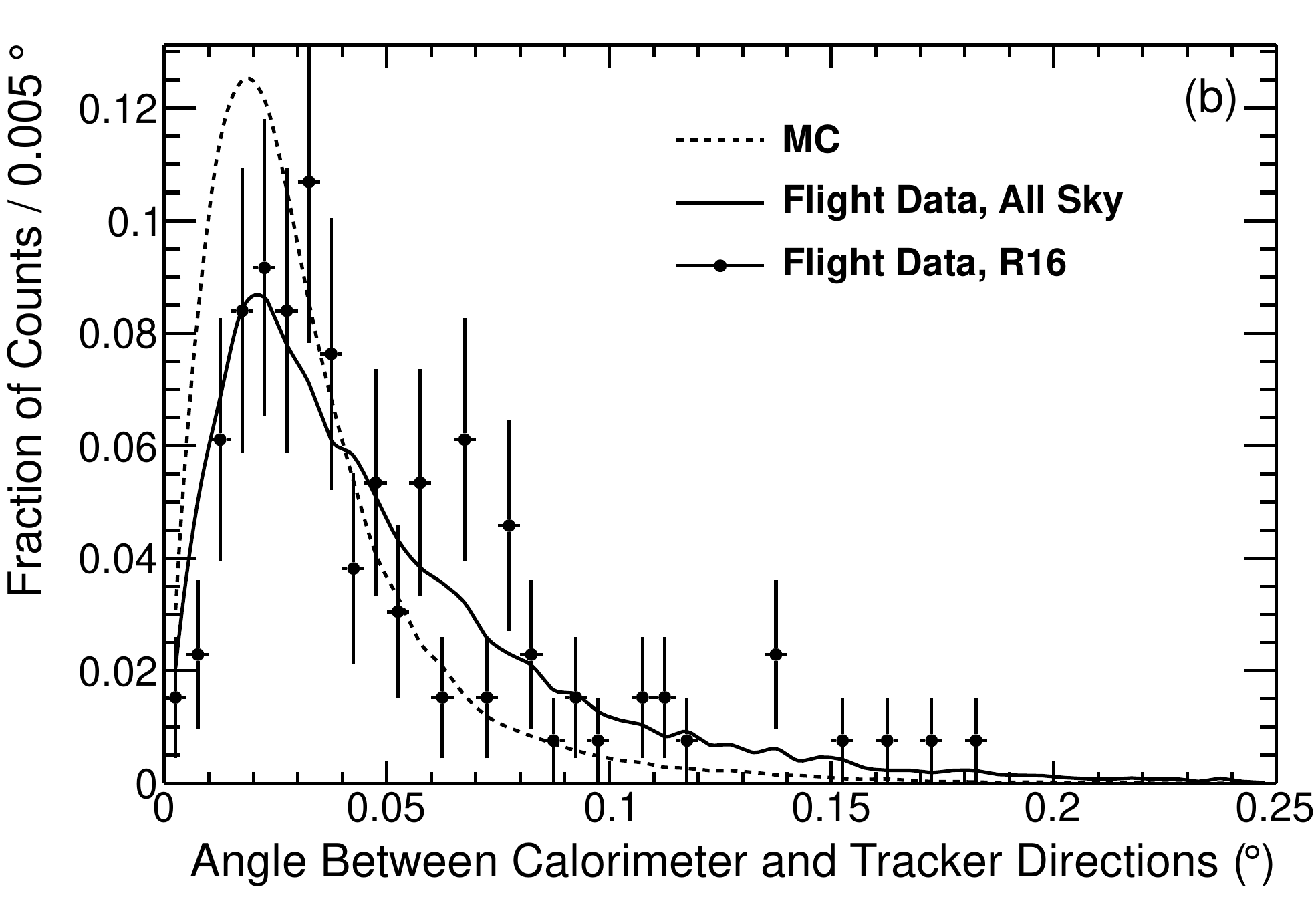}
{\caption{\label{fig:CAL_TKR_Variables}Two measures of the calorimeter-tracker event direction agreement for events with $125 \rm{GeV} < \Ereco < 145 \rm{GeV}$:  (a) the distance of closest approach (DOCA) between the extrapolation of the tracker direction and the centroid of the calorimeter energy deposition; (b) the angle between the tracker direction and the primary axis of the energy deposition in the calorimeter.  In both curves, the MC has been weighted to have the same livetime distribution with $\theta$ as the all-sky data, and the small ($< 0.005$) error bars on the All Sky sample have been suppressed for visual clarity.}}

\subsection{\texorpdfstring{$\theta$}{theta}-dependence of the 133~GeV feature}\label{subsec:133GeV_Feature_Theta}

Several authors have reported a $\theta$ dependence of the prominence of the spectral feature in both the Limb and Galactic center datasets, which is unexpected~\cite{Whiteson:2012hr,Hektor:2012ev,Finkbeiner:2012ez}.  Our results are broadly consistent with those previously reported; the feature appears with a larger statistical significance in data sets of events with smaller incident angles. 
To study this near the Galactic center, we fit for a line at 133~GeV in R16 in two $\theta$ ranges: $\theta<50^{\circ}$ and $\theta>50^{\circ}$.  This ROI was chosen for this study to have enough events to separately consider both $\theta$ ranges.  \Figureref{ThetaSlice} shows the fit results in both $\theta$ ranges.  There is no evidence of any feature at 133~GeV from events with $\theta>50^{\circ}$, while the fit using events with $\theta<50^{\circ}$ indicates a feature at 133~GeV with $\slocal=1.9\sigma$. Though there are fewer events with $\theta>50^{\circ}$, the observed fractional size from the events with $\theta<50^{\circ}$, $f(\Egamma=133~{\rm GeV})_{\theta<50^{\circ}}=0.18$, should scale to produce a feature with $1.0\sigma$ given the number of events with $\theta>50^{\circ}$; see \Eqref{eq:FracSig}. 

\twopanel{!ht}{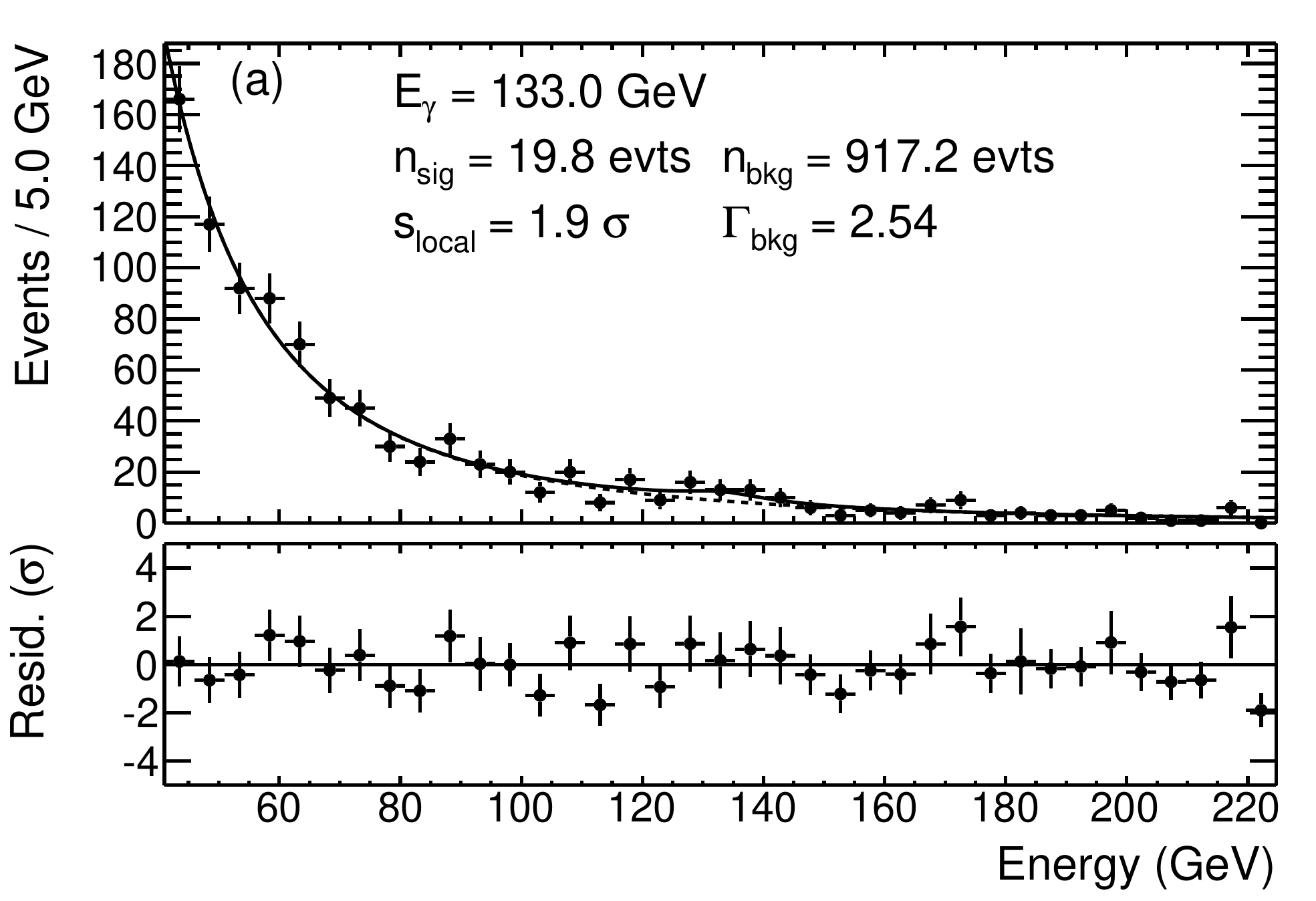}{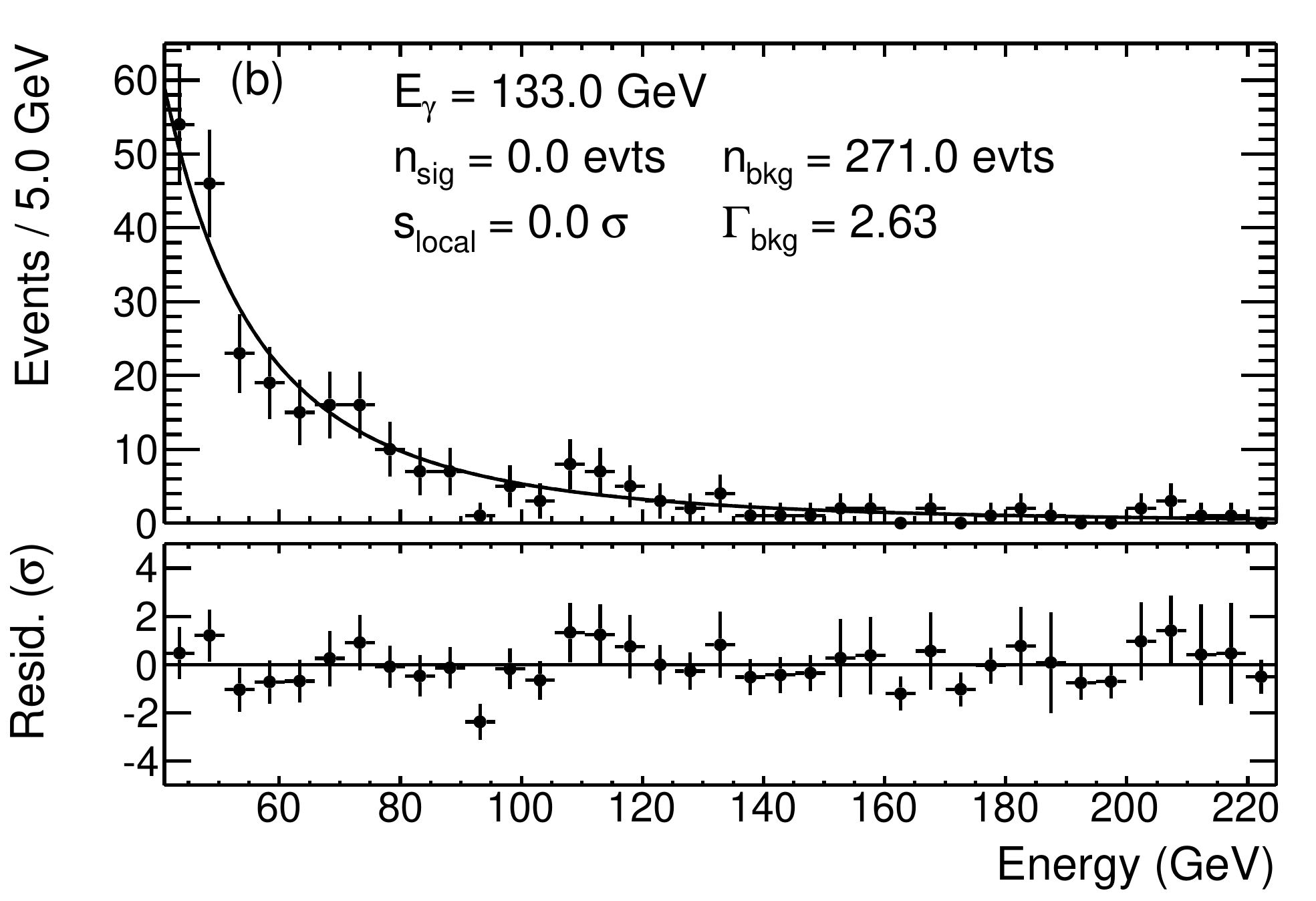}
{\caption{\label{fig:ThetaSlice}Fit at 133~GeV line in R16 (\irf{P7REP\_CLEAN}) using the 2D energy dispersion model: (a) for events with $\theta<50^{\circ}$, (b) for events with $\theta>50^{\circ}$.  \newText{The solid curves show the average model weighted using the \probE\ distribution of the fitted events.  Note that these fits were unbinned;
the binning here is for visualization purposes.}}}

Similarly, we split the Limb dataset into the same ranges of $\theta$.  For events with $\theta<50^{\circ}$ the significance is $\slocal=2.6\sigma$, while for the events with $\theta>50^{\circ}$ the significance is $\slocal=0.0\sigma$ (see \Figref{ThetaSlice_Limb}).

\twopanel{!ht}{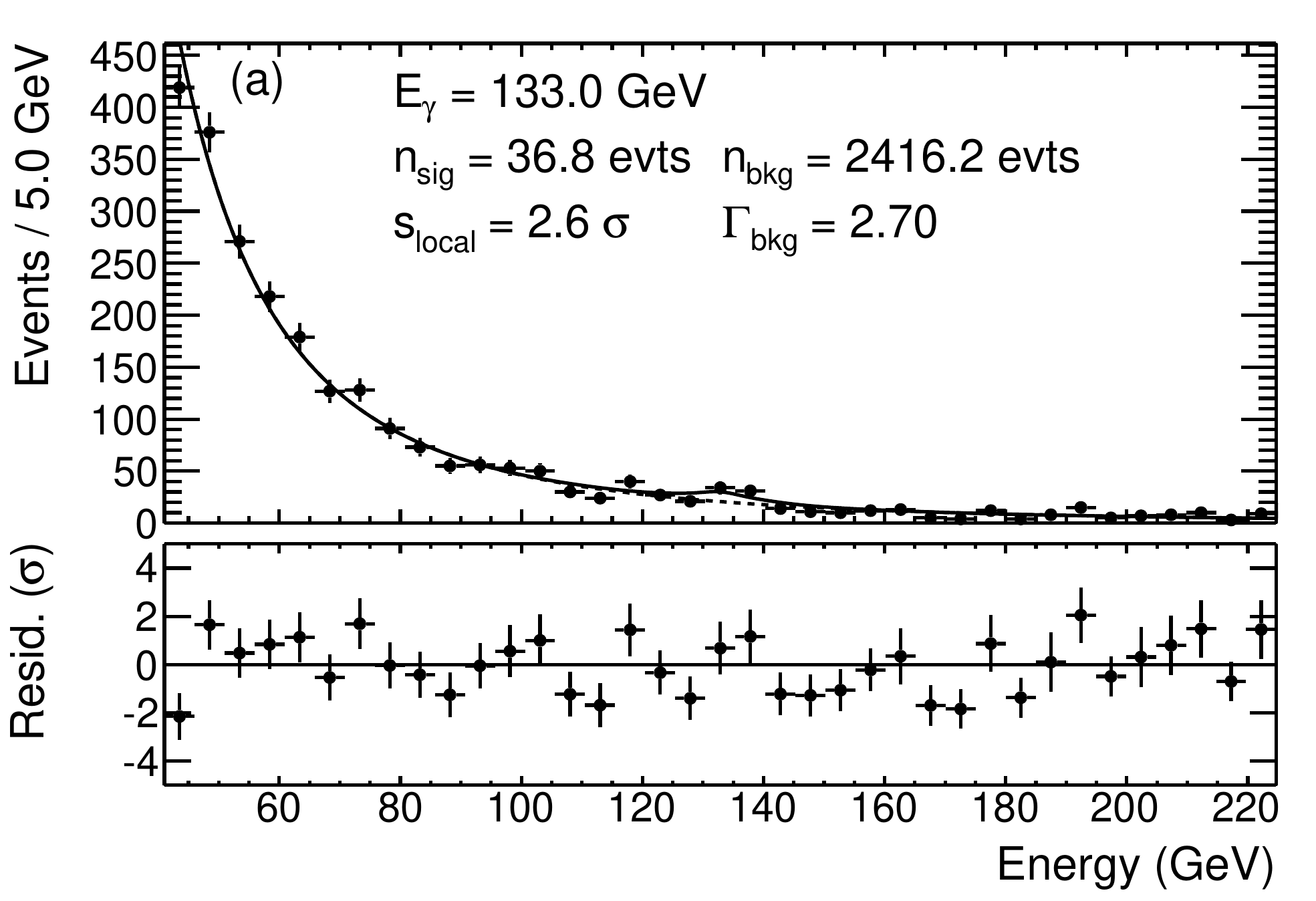}{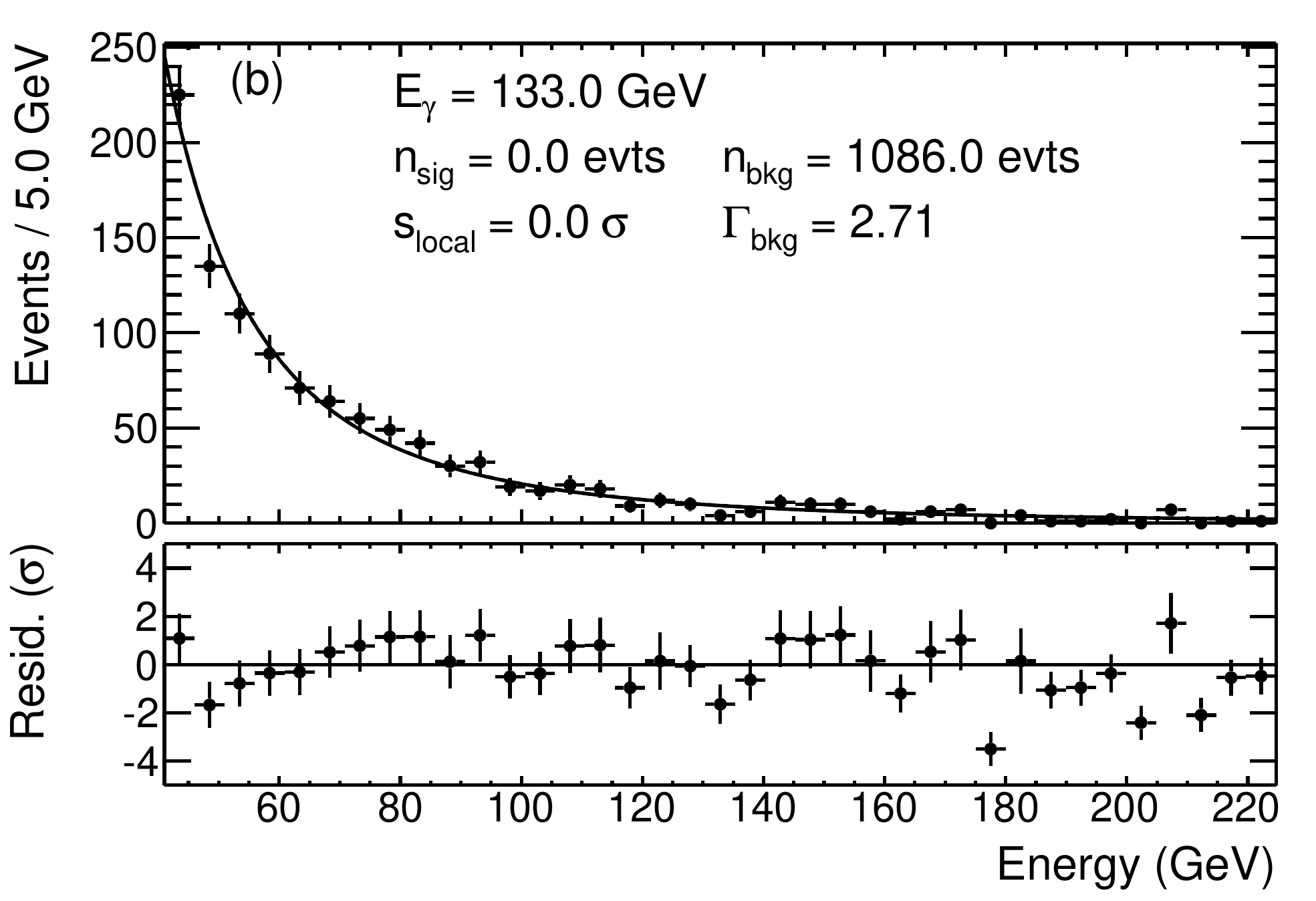}
{\caption{\label{fig:ThetaSlice_Limb}Fit at 133~GeV line in the Limb (\irf{P7REP\_CLEAN}) using the 2D energy dispersion model.: (a) for events with $\theta<50^{\circ}$, (b) for events with $\theta>50^{\circ}$. \newText{The solid curves show the average model weighted using the \probE\ distribution of the fitted events.  Note that these fits were unbinned;
the binning here is for visualization purposes.}}}

%% file: Discussion.tex
\section{DISCUSSION AND SUMMARY}\label{sec:discussion}

We have performed a search for \gammaRayHyph\ spectral lines from 5--300 GeV in five ROIs defined \textit{a priori} to optimize sensitivity for various DM density profiles.   This search was performed using 3.7~years of data that have been reprocessed using updated calorimeter
calibration constants, and the 2D energy dispersion model that includes information about the event-by-event energy reconstruction quality. 

We found no globally significant spectral line signals and present flux upper limits for monochromatic sources (see Tabs~\ref{tab:allLimits_lowerEn}--\ref{tab:allLimits_higherEn}).  For a particular DM density profile for the Milky Way, the flux upper limits can be translated to annihilation cross section upper limits or decay-lifetime lower limits. \Figureref{SigmaVUL} shows the 95\% CL upper limits on $\langle\sigma v\rangle_{\gamma\gamma}$ for the contracted NFW (R3), Einasto (R16), NFW (R41), and Isothermal (R90) DM density profiles for the ROIs that provide the best sensitivity. Also shown are the 95\% CL lower limits on the decay lifetime ($\tau_{\nu\gamma}$) for the R180 ROI assuming an NFW profile.  The cross section upper limits have been improved in some cases by a factor of several relative to the LAT Collaboration 2-year limits~\cite{REF:2012.LineSearch} and represent an extension of the search range from 7--200~GeV to 5--300~GeV.

Our new velocity-averaged cross section limits lie in the range $\langle \sigma v\rangle_{\gamma\gamma} \sim 10^{-29}-10^{-27}$cm$^{3}$s$^{-1}$, with the precise limit depending on the WIMP mass and the DM density profile assumed for the Milky Way; cuspier profiles and lower masses are constrained more strongly. The limits are a factor of $\sim 5-5000$ times below the canonical thermal relic cross section of $\langle \sigma v\rangle_{\rm WIMP} \sim 3\times 10^{-26}$cm$^{3}$s$^{-1}$ and therefore strongly constrain models in which DM particles can annihilate to Standard Model particles through tree-level diagrams. However, since DM is constrained to be electrically neutral to a very good approximation, WIMP interactions in most models produce monochromatic photons only through higher-order processes, \newText{the cross sections of which are typically suppressed by three or more orders of magnitude}.  This means that our limits do not disfavor the WIMP hypothesis in general.

Our two most significant fits occurred at 6~GeV in R180 and at 133 GeV in R3.  While the fit at 6~GeV in R180 has a relatively large $TS$ value, the signal fraction ($1\%$), was similar to the expected systematic uncertainty of $f \sim 0.008$ for R180 at that energy. 

Reports of a line-like feature in the Galactic Center using the public data have appeared in the literature~\cite{Bringmann:2012vr,REF:Weniger:2012tx,Su:2012ft}. The authors calculated the flux of the source producing the line-like feature to be $\sim2\times10^{-10}\rm{cm}^{-2}\rm{s}^{-1}$, which is not ruled out by our 95\% CL $\Phi_{\gamma\gamma}$ limits in R3 ($3.4\times10^{-10}\rm{cm}^{-2}\rm{s}^{-1}$ for $E_{\gamma}=135$~GeV, see~\Tabref{allLimits_higherEn}).  Additionally, these reported fluxes are similar to the mean value obtained from our fit at 133~GeV in R3 of $\Phi_{\gamma\gamma}^{R3}(E_{\gamma}=133~\rm{GeV})=1.9\times10^{-10}\rm{cm}^{-2}\rm{s}^{-1}$.

The fit at 133~GeV in R3 yields $\slocal = 3.3\sigma$ with $f(133~{\rm GeV})_{R3} = 0.61$, which is larger than any of the systematic effects summarized in \Secref{sec:systematics} (see \Tabref{FracSig_Summary}) and is larger than the feature seen at 133~GeV in the Limb: $f(133~{\rm GeV})_{\rm Limb}=0.14$.  Also, if the feature is due to an instrumental effect, one would have expected it to appear in the spectra of \gammaRays\ from the inverse ROI, which it does not. Therefore, the 133~GeV feature in R3 cannot be entirely explained in terms of known systematic effects. However, as discussed in \Secref{sec:133GeV_Feature}, the 133~GeV feature does have certain characteristics that disfavor
interpreting it as a DM signal. The fit significance reduces when using the 2D energy dispersion model, making the global significance of the feature $\sglobal = 1.5\sigma$. This decrease in significance is in large part due to the 133~GeV feature being much narrower than the LAT energy resolution, and not being present in events with $\theta>50^{\circ}$.  More data and study are needed to clarify the origin of this feature.

Two ongoing developments will help to resolve the question of the origin of the 133~GeV feature and also benefit future line searches:  
\begin{fermiitemize}
\item Beginning 2012 October, the LAT started collecting more data from the Limb through weekly 2-orbit pointed observations at the orbital pole. This change alone should increase the available Limb dataset by $\sim 40\%$ over the next year and will decrease the current statistical limitations at high energies ($>100~{\rm GeV}$) in the Limb.  Additional Limb data can also be collected during ``Target of Opportunity'' pointed observations if Limb tracing is implemented while the target is occulted by the Earth. These data will help to constrain the uncertainties from narrow features in the effective area, which are among the dominant source of potential systematic uncertainties that may induce a false signal.
\item \newText{Almost every aspect of the LAT event reconstruction and selection algorithms have been re-written in the new and upcoming \peight\ versions.}   Expected improvements most relevant to a line search are an increased effective area at all energies and an improved energy resolution, particularly at higher energies\cite{Atwood:2013rka}.  Furthermore, aside from any performance improvements, \newText{given the scope of the changes in \peight},  systematic biases associated with the event reconstruction and selection in \peight-based analyses are likely to be uncorrelated with similar biases in \pseven-based analyses, which will help clarify if the feature at 133~GeV is a systematically induced artifact. 
\end{fermiitemize}

%% file: Data_Reprocessing.tex
\section{\irf{PASS~7}\ DATA REPROCESSING}\label{app:pass_7_rep}

In 2012, the LAT Collaboration reprocessed all of the data from the mission to-date with updated calibrations for the instrument subsystems, but 
with the same reconstruction and event-level analysis algorithms as the previously released \pseven\ data.   

The primary goal of this reprocessing was to incorporate improved calibrations of the measurement of the light asymmetry between the ends of the calorimeter
crystals.  This asymmetry is used to derive position information that is critical to measuring the centroid and axis of the electromagnetic shower in the
calorimeter.   \newText{See Fig. 12 of \cite{REF:2012.P7Perf} for an illustration of the calorimeter shower, including the centroid and axis.}  Above a few GeV, both the centroid and axis of the electromagnetic shower are  useful in constraining the event reconstruction in the tracker, which would otherwise be degraded
because of the increased event complexity at these high energies caused by the \newText{back-scattering of} particles from the calorimeter back into the tracker.  
Specifically, using the calorimeter shower centroid as an additional constraint on the event direction can significantly 
reduce the tails of the \newText{PSF}.  Furthermore, the consistency between the tracker direction solution and both the calorimeter 
shower axis and centroid are powerful discriminators between \gammaRays\ and CR background events \newText{(see Sec.~3.4.3 of \cite{REF:2012.P7Perf}, in particular items 3 and 4 in the bulleted list)}.

The updated calibrations also corrected for a small ($\sim 1\%$~per year), expected degradation in the light yield of the calorimeter crystals that had 
been measured in the flight data \newText{(see Fig. 73 of \cite{REF:2012.P7Perf} and the accompanying text)}.  Consequently, the absolute energy scale has shifted up by a few percent in an energy- and time-dependent way.  \newText{This has caused the feature reported at 130~GeV with \pseven\ data to shift to 133~GeV with \psevenrep\ data.}

%% file: ROI_Optimization.tex
\section{REGION OF INTEREST OPTIMIZATION}\label{app:method_roi}  

Following~\citet{REF:Weniger:2012tx,Bringmann:2012vr}, we have adopted a method for
defining optimized ROIs by comparing the signal \gammaRays\ expected
from WIMP annihilation or decay, assuming a specific density profile, to the background \gammaRays\ expected from astrophysical processes.  Unlike~\citet{REF:Weniger:2012tx,Bringmann:2012vr}, who estimated the expected \gammaRayHyph\
background from the LAT \gammaRayHyph\ data, we used the
\newText{\stool{gtobssim} {\emph{ScienceTool}}} to
simulate 2-years of LAT observation of the expected backgrounds based on the standard LAT models of diffuse emission from the Galaxy and isotropic
emission. The differential \gammaRayHyph\ flux \newText{from the
  annihilation of self-conjugate WIMPs is:}

\begin{equation}\label{eq:flux_ann}
\left(\frac{d\Phi_\gamma}{dE d\Omega}\right)_{\rm ann} = \frac{1}{8\pi} \, \frac{\langle \sigma v\rangle}{m_\chi^2}\,
\left(\frac{dN_\gamma}{dE}\right)_{\rm ann} \frac{dJ_{\rm ann}}{d\Omega},
\end{equation}

\noindent with

\begin{equation}\label{eq:J_ann}
\frac{dJ_{\rm ann}}{d\Omega} = \int ds~ \rho(r)^2,
\end{equation}

\noindent where the integration is performed over the square of the DM mass density
($\rho$) along the line-of-sight, $\langle\sigma v\rangle$ is the velocity-averaged
annihilation cross section, $m_\chi$ is the WIMP mass, and
$dN_\gamma/dE$ is the differential \gammaRayHyph\ yield per
annihilation.  
The differential \gammaRayHyph\ flux from WIMP decays is 

\begin{equation}\label{eq:flux_decay}
\left(\frac{d\Phi_\gamma}{dE d\Omega}\right)_{\rm decay} = \frac{1}{4\pi} \,\frac{1}{\tau}\,\frac{1}{m_\chi}\,
\left(\frac{dN_\gamma}{dE}\right)_{\rm decay} \frac{dJ_{\rm decay}}{d\Omega},
\end{equation}

\noindent with 

\begin{equation}\label{eq:J_decay}
\frac{dJ_{decay}}{d\Omega} = \int ds~ \rho(r),
\end{equation}

\noindent where $\tau$ is the DM particle lifetime.  The total signal from DM is given by an integration over the ROI,

\begin{equation}\label{eq:annSigTot}
  \left(\frac{d\Phi_\gamma}{dE}\right)_{\rm ann} = \frac{1}{8\pi} \, \frac{\langle \sigma v\rangle}{m_\chi^2}\,
\left(\frac{dN_\gamma}{dE}\right)_{\rm ann} \intROI \frac{dJ_{\rm ann}}{d\Omega}~d\Omega
\end{equation}

\noindent for annihilations and

\begin{equation}\label{eq:decSigTot}
\left(\frac{d\Phi_\gamma}{dE}\right)_{\rm decay} = \frac{1}{4\pi} \,\frac{1}{\tau}\,\frac{1}{m_\chi}\,
\left(\frac{dN_\gamma}{dE}\right)_{\rm decay} \intROI \frac{dJ_{\rm decay}}{d\Omega}~d\Omega
\end{equation}

\noindent for decays.  The integral, $\intROI \frac{dJ}{d\Omega}~d\Omega$, is commonly referred to as the ``J-factor'',
and represents the astrophysical component of the DM flux calculation.

We define our ROI as a circular region of radius \Rgc\ centered on the GC.  We
additionally mask a rectangular region along the Galactic plane with
$|b| < \Delta b$ and $|l| > \Delta l$. This ROI definition excludes emission from the off-center Galactic plane where the
astrophysical background is largest and the expected DM contribution
is relatively small.  

To remove the Galactic plane, we set $\Delta b = 5^{\circ}$.  We then optimized the remaining ROI parameters (\Rgc\ and $\Delta l$) for each of our four models of the MW DM halo. Note that because our fits are background-dominated at all but the highest energies, the optimization is insensitive to any prefactor in the signal model. Specifically, we maximized the signal-to-noise ratio ($S/N$) defined
for a given ROI as

\begin{equation}
  S/N_{\rm ROI} =
  \frac{\intROI \intFOV \; S(\pTrue) \; \mathcal{E}(\pTrue) d\OmegaVhat d\Omega} 
  {\sqrt{\intROI \intFOV \; B(\pTrue) \;
      \mathcal{E}(\pTrue)  d\OmegaVhat d\Omega}},
\end{equation}

\noindent where $\mathcal{E}$ is the exposure, and $S$ and $B$ are the intensities of \gammaRays\ in the direction \pTrue\ from DM and diffuse backgrounds, respectively. We modeled the spatial distribution of the diffuse background by integrating the \gammaRays\ between 10 and 100 GeV from a simulation of 2-years of LAT observations using \textit{gtobssim} and the recommended templates for the isotropic and Galactic diffuse emission.  The optimization results did not change
significantly when only background events with energies from 10 to 30~GeV or 30 to
100~GeV were used.  Therefore, this method appears not to have a
strong dependence on the energy range (above 10~GeV). 

The value of $\Delta l = 6^{\circ}$ is close to optimal for all but the contracted NFW profile,
which is not affected by the Galactic plane mask.  In the case of
annihilating DM, there is a strong dependence of \Rgc\ on the
shape of the inner profile. The smallest ROIs are preferred for the
profiles with the largest central densities. \Figureref{roi} shows
$S/N_{\rm ROI}$ normalized to its maximum value as a function of the ROI parameters $\Delta l$ and \Rgc\ evaluated from $0.5^{\circ}-30^{\circ}$ and $0.5^{\circ}-180^{\circ}$ respectively. Note that in \Figref{roi}, \Rgc\ (a) and $\Delta l$ (b) have been fixed at their optimal values.

\twopanel{ht!}{\colSw{Figure_22a}}{\colSw{Figure_22b}}
{\caption{\label{fig:roi}Evaluation of the optimal values of $\Delta l$ and \Rgc. The plots show $S/N$ as a function of (a) $\Delta l$ and (b) \Rgc\ normalized to the maximum for  different DM density profiles (see legend) for DM annihilations (solid  lines) and decays (dashed lines).  For (a) $\Delta b$ has been fixed to 5$^\circ$ and \Rgc\ has been fixed to the optimal value for the associated DM density profile (see text).  The vertical
  dashed line shows the fixed value of $\Delta l$ that was chosen for all optimized
  ROIs.  (We do not show the case for DM annihilation with a
  contracted NFW profile, for which the optimal ROI is smaller than
  our Galactic plane mask.)  
  For (b) $\Delta b$ and $\Delta l$ have been fixed to 5$^\circ$ and 6$^\circ$ respectively. }}

For the contracted NFW profile, the optimal \Rgc\ is found at the
smallest radius considered ($0.5^{\circ}$), which is at the characteristic scale
of the LAT PSF at 1 GeV.  Optimization of the ROI with $\Rgc \lesssim
0.5^{\circ}$ would require convolving the DM signal profile
with the LAT PSF and require a different, more complicated analysis than the one presented in this paper. In the case of decaying DM, the optimal ROI parameters are nearly independent of the shape of the
DM distribution, preferring a large optimal \Rgc\ for all
four profiles studied. 

We define a set of 5 ROIs with a fixed Galactic plane mask ($\Delta l =
6^{\circ}$ and $\Delta b = 5^{\circ}$) and the following values of
\Rgc: $3^{\circ}$ (R3), $16^{\circ}$ (R16), $41^{\circ}$ (R41),
$90^{\circ}$ (R90), and $180^{\circ}$ (R180).  We use the smallest ROI
(R3) to search for a signal compatible with the contracted NFW
profile.  In this instance, the ROI size of 3$^{\circ}$ was intentionally
chosen to be larger than the region with best $S/N$.  For the
contracted NFW profile, the S/N of R3 is reduced by $40\%$ when
compared to the smallest circular region in our optimization scan
($\Rgc = 0.5^{\circ}$).  On the other hand, by using a larger search 
region, the analysis is less dependent on the LAT PSF.  Additionally, we 
limit the search in R3 to spectral lines above 30~GeV.  At these high 
energies, emission from known \gammaRayHyph\ sources is at least 
an order of magnitude dimmer than the Galactic diffuse emission 
integrated over R3.  Thus, we also avoid complications from point 
sources and no longer need to apply a source mask (see
\Tabref{event_samples}).  
This allows us to use all of the events in this already small,
event-limited, ROI.

The ROIs R16, R41, and R90 were chosen to optimize the sensitivity to annihilating DM assuming
the Einasto, NFW, and Isothermal halo models respectively.  Finally, we
chose a large ROI (R180), which is close to optimal for decaying DM
models.  R180 is also similar to the ROI used in the previous LAT Collaboration 
line search\footnote{The previous LAT Collaboration line search used a Galactic plane mask with $\Delta l =
10^{\circ}$ and $\Delta b = 10^{\circ}$.}~\cite{REF:2012.LineSearch}.

%% file: Formalism.tex
\section{LIKELIHOOD FORMALISM}\label{app:Likelihood_Formalism}

Many \fermi-LAT analyses, and many of the \stools, follow the
likelihood formalism of~\citet{REF:1996.Mattox} to 
test hypotheses about the spatial and spectral distribution
of observed \gammaRays.  In this paper, we made several assumptions
and approximations to derive the parametrized likelihood function that we used for our analysis.  In this appendix, we discuss those
assumptions and approximations.

\subsection{General formalism}\label{app:Formalism_General}

We test the hypothesis that the distribution of measured 
energies, \Ereco, directions, \pReco, and
arrival times, $t$, and other observable parameters, \sVec,
comes from a model of the total source fluxes, $S(\Etrue,\pTrue)$, where 
\Etrue\ and \pTrue\ are the true energies and directions of the
incident \gammaRays.  This testing requires a representation of the response of the LAT: 
$R(\Ereco,\pReco;\Etrue,\pTrue,\sVec,t)$. Note that \sVec\ represents event
parameters (e.g., \probE, or the tracker layer at which the \gammaRayHyph\ converted),
as well as the event selection (e.g., \irf{P7REP\_CLEAN}).

In practice, we parametrize the instrument response as a function of
the \gammaRayHyph\ direction in the LAT reference frame, \vTrue.
Furthermore, since the LAT performance has changed
little over the course of the mission, we use a single
representation for the entire mission.   Therefore, we express the instrument
response as
$R(\Ereco,\pReco;\Etrue,\vTrue(t;\pTrue),\sVec)$, where we
have explicitly indicate that the reference frame of the LAT rotates with respect to the
celestial frame by writing the time dependence of \vTrue.   From here on
we suppress the dependencies of \vTrue\ on $t$ and \pTrue.

The parametrization provided with the \stools\ factors the instrument response 
into three parts:
\begin{fermienumerate}
\item{The effective collecting area for \gammaRays\ for a given
    \Etrue, \vTrue\ and \sVec: $\aeff(\Etrue,\vTrue,\sVec)$.} 
\item{The \newText{PSF}, i.e., the probability density to reconstruct an incident direction \pReco,
for a given \Etrue, \pTrue, \vTrue\ and \sVec:  $P(\pReco;\pTrue,\Etrue,\vTrue,\sVec)$.}
\item{The energy dispersion, i.e., the probability density to measure \Ereco, for a given 
\Etrue, \vTrue\ and \sVec: $D(\Ereco;\Etrue,\vTrue,\sVec)$.}
\end{fermienumerate}
Note that by factoring the instrument response in this way we
implicitly assumed that the spatial dispersion and energy dispersion are uncorrelated for given \Etrue, \vTrue\ and
\sVec.  

We can obtain the predicted \gammaRayHyph\ distribution, $M(\Ereco,\pReco;\sVec)$, by convolving the source flux models with the IRFs and integrating over the spatial, energy, and time ranges of interest:

\begin{align}\label{eq:Predicted_Counts}
   M(\Ereco,\pReco;\sVec) = & \intTime \intEnergy \intPTrue
   S(\Etrue,\pTrue) \aeff(\Etrue,\vTrue,\sVec)  \nonumber \\
   & \times P(\pReco;\pTrue,\Etrue,\vTrue,\sVec)
   D(\Ereco;\Etrue,\vTrue,\sVec) d\Omega dE dt.
\end{align}

We make a few approximations to simplify this expression, and to improve the
computational efficiency of the analysis:
\begin{fermienumerate}
\item{The performance variation as a 
    function of the angle with respect the boresight ($\theta$) is much larger 
    than the performance variation as a function of the azimuthal
    angle ($\phi$).  In fact, for long-term observations, averaging
    the  LAT response over azimuth is a very good approximation.  
    Therefore, although the standard IRFs used are
    parametrized in terms of $\theta$ and $\phi$: we ignore the
    $\phi$-dependence, i.e.,
    $\aeff(\Etrue,\theta,\sVec)$ and $D(\Ereco;\Etrue,\theta,\sVec)$.}
\item{We calculate the ``observing profile'', $\tobs(\theta;\pTrue)$,
    i.e., distribution of observing time with incident angle,
    by integrating the time that a particular direction in the sky is
    at a particular direction in the LAT reference
    frame\footnote{The \stool{gtltcube} tool calculates 
      observing profile for each direction in the sky, which is often called
      the ``livetime cube''.}.  We can also precompute the exposure 
    as as a function of $\theta$ for each direction in the sky, 
    $\mathcal{E}(\Etrue,\theta,\sVec) = \aeff(\Etrue,\theta,\sVec) \tobs(\theta;\pTrue)$.}
\item{We assume that we can neglect the effect of the PSF.  
    This is equivalent to assuming the PSF is small
    compared to changes in the product of the exposure and the source intensity.
    Since we are masking bright point sources, this is a reasonable
    approximation.}

\end{fermienumerate}

We can then express the predicted counts spectrum in terms of the livetime cube:

\begin{align}\label{eq:Predicted_From_Livetime_Cube}
M(\Ereco,\pTrue,\theta;\sVec) = & \intEnergy 
S(\Etrue,\pTrue)\aeff(\Etrue,\theta,\sVec)\tobs(\theta;\pTrue)
\nonumber \\
& \times D(\Ereco; \Etrue, \theta, \sVec) dE.
\end{align}

Furthermore, we do not have particularly strong \textit{a priori}
knowledge about the morphological details of
the DM-line signal and the astrophysical backgrounds have
substantial uncertainties.  Therefore, for each ROI we analyze, we choose to integrate over the ROI and perform the fit only in the energy domain. We note that while some authors have chosen to retain the spatial information in their fitting procedures~\citep[e.g.,][]{Su:2012ft}, while others have not~\citep[e.g.,][]{REF:Weniger:2012tx}.  While including spatial information in the fit increases sensitivity, one must chose a specific DM hypothesis to test.  By integrating over the ROI, we are able to test for the existence of a monochromatic source generally.

After integrating the model over the ROI \newText{and the FOV}, we obtain a predicted counts spectra that 
we can compare with observations:

\begin{equation}\label{eq:Predicted_Energy_Spectrum}
   C(\Ereco;\sVec) = \intROI \intFOV M(\Ereco,\pTrue,\theta;\sVec) d\OmegaVhat d\Omega.
\end{equation}

\noindent By \newText{further integrating over the fit energy band}, we can
obtain the total number counts predicted by the model:

\begin{equation}\label{eq:npred_model}
  \nPred =  \intEnergy C(\Ereco;\sVec) d\Ereco.
\end{equation}

\noindent This also allows us to split the predicted counts spectrum
into a normalized probability density function
$F(\Ereco;\sVec)$ times the number of counts ($n$, which we will
treat as a free parameter of the fit);

\begin{equation}\label{eq:relation_C_F}
   C(\Ereco;\sVec) = n F(\Ereco;\sVec), 
\end{equation}

\noindent where $F(\Ereco;\sVec)$ is defined by:

\begin{equation}\label{eq:def_norm_PDF}
   F(\Ereco;\sVec) = \frac{C(\Ereco;\sVec)}{\nPred}.
\end{equation}

\noindent  Note that while $n$ is just a scalar quantity that is
varied by the likelihood minimizer, \nPred\ is a normalization 
integral that must be calculated from \Eqref{eq:npred_model}.

In the particular case of a line search, we separate the source model
into the contributions from a \gammaRayHyph\ line, \SSig, 
and those from all other astrophysical sources, \SBkg, such that:

\begin{equation}\label{eq:Source_Model_Total}
S(\Etrue,\pTrue) = \SSig(\Etrue,\hat{p}) + \SBkg(\Etrue,\pTrue).
\end{equation}

\noindent And likewise the predicted counts distributions:

\begin{equation}\label{eq:Predicted_Model_Total}
C(\Ereco;\sVec) = \CSig(\Ereco;\sVec) + \CBkg(\Ereco;\sVec).
\end{equation}

For a binned likelihood analysis we compute
the log-likelihood as the sum of the logarithm of the Poisson probability to 
observe $n^{obs}$ events in a particular bin in \Ereco\ and \probE\
given that the model predicts $n^{\rm pred}$:

\begin{equation}\label{eq:Binnned_LH}
  \ln \mathcal{L}_{b} = \sum_{i}^{\rm bins} \nObs_{i} 
  \ln \nPred_{i} - \sum_{i}^{\rm bins} \nPred_{i}
\end{equation}

\noindent For an unbinned likelihood analysis we instead compute the sum of the
log-likelihood of the individual events based on the predicted distribution:

\begin{equation}\label{eq:Unbinnned_LH}
  \ln\mathcal{L}_{u} = \sum_{i}^{\rm events} \ln
  C(\Ereco_i;\sVec|\vec{\alpha}) - C_{\rm tot},
\end{equation}

\noindent where $C_{\rm tot}$ is the total number of \gammaRays\ predicted by the
model, and $\vec{\alpha}$ represents the model parameters, such as \Egamma\ and $\GamBkg$.

\subsection{Line search signal model}\label{app:Formalism_Signal}

We can factor the signal model into photon spectrum and spatial
intensity $\ISig(\pTrue)$, 
and explicitly write the photon spectrum as a delta function at the line
energy \Egamma:

\begin{equation}\label{eq:Signal_Flux_Model}
\SSig(\Etrue,\pTrue|\Egamma) = I_{\rm sig}(\pTrue)\delta(\Etrue - \Egamma).
\end{equation}

\noindent We then express the model in terms of the total
number of signal counts, \nSig, which will become a free parameter in
our fit, and the total predicted number of counts, \nPredSig:

\begin{equation}\label{eq:TwoD_Model_Signal}
\CSig(\Ereco;\sVec|\Egamma) = \nSig \intROI \intFOV \frac{D(\Ereco;
  \Egamma, \theta, \sVec) I_{\rm sig}(\pTrue)
  \mathcal{E}(\Egamma,\pTrue,\theta;\sVec)}{\nPredSig} d\OmegaVhat d\Omega,
\end{equation}

\noindent where the normalization term \nPredSig\ must be calculated using:

\begin{equation}\label{eq:npred_sig_integral}
\nPredSig = \intEnergy \intROI \intFOV
D(\Ereco; \Egamma, \theta, \sVec) I_{\rm sig}(\pTrue) \mathcal{E}(\Egamma,\pTrue,\theta;\sVec)
d\OmegaVhat d\Omega d\Ereco.
\end{equation}

\subsection{Line search background model}\label{app:Formalism_Background}

Empirically, at GeV energies, the spectrum of diffuse emission for
relatively large regions of the sky is quite smooth.  Thus, in our
ROIs, it can be well-modeled as a power law for the relatively 
narrow ($\sim 1/2$ decade) energy intervals we are fitting.
Furthermore, by design the energy dispersion is much smaller than the
fit energy ranges (recall, we fit in $\pm 6\sigma_E$ ranges).  
Thus, for the background model, we approximate $\SBkg(\Etrue,\pTrue)$ to 
have a single power-law dependence and write the spatial dependence as $\IBkg$.  
Also, the energy resolution varies fairly slowly with energy and changes
only slightly across any given fit range, therefore we treat the
energy dispersion as $\delta(\Ereco - \Etrue)$.

With these approximations, we can express the background model in
terms of the total number of counts, \nBkg, which will become a free parameter in our fit:

\begin{equation}\label{eq:TwoD_Model_Background}
\CBkg(\Ereco;\sVec|\GamBkg) = \nBkg
\frac{1}{\nPredBkg}\left(\frac{\Ereco}{\Epivot}\right)^{-\GamBkg}
\intROI \intFOV \IBkg(\pTrue)\mathcal{E}(\Ereco,\pTrue,\theta;\sVec)
d\OmegaVhat d\Omega.
\end{equation}

\noindent note that normalization \nPredBkg\ depends on \GamBkg\ and
must be calculated using:

\begin{equation}\label{eq:Normalization_Bkg}
\nPredBkg = \intEnergy \intROI \intFOV
\left(\frac{\Ereco}{\Epivot}\right)^{-\GamBkg}
\IBkg(\pTrue)\mathcal{E}(\Ereco,\pTrue,\theta;\sVec)
d\OmegaVhat d\Omega d\Ereco.
\end{equation}

\subsection{Energy dispersion parametrization}\label{app:Formalism_EDisp}

In this paper we use two parametrizations of the energy dispersion, depending on the study being performed.
\begin{fermienumerate}
\item{We use an energy quality estimator, \probE,
  in our predicted counts model and integrating over $\theta$; in this case
  our predicted ``2D'' counts model depends on \Ereco\ and \probE:
  $C(\Ereco,\probE;\sVec)$.   We use this parametrization for all of the
  fits \newText{except those listed below}.}
\item{We obtain a simpler ``1D'' model of the energy dispersion at the
  price of a 15\% loss of senstivity by averaging the energy dispersion over $\theta$.  
  We use this approach when \newText{finely} scanning the Galactic plane for spectral
  features (\Secref{sec:GalaxySmoothness}) and \newText{for the pseudo-experiments we used to estimate the
  effective trials factor (\Secref{subsec:LEE})  and the effects of the 
  dips in efficiency we observed in the Limb data (\Secref{subsec:133GeV_Feature_Earth_Limb})
  as it is computationally much faster than the 2D parametrization}.}  
\end{fermienumerate}
We derive the energy dispersion models for both formulations below.

\subsubsection{Energy dispersion parametrized by energy only}\label{app:Formalism_PDF_E_Only}

We can obtain the simpler ``1D'' form of the energy dispersion model \newText{by averaging the 
instrument response across the FOV as well as the ROI.}

For the signal model, the spatial integrals give the factor needed to re-weight
the contributions to the energy dispersion model. \newText{However, since the intensity of the signal is  
brightest toward the Galactic center, and in fact markedly so for some
of the DM models considered, we simply pick the Galactic center out of
the integral (i.e., $I_{\rm sig}(\pTrue) = I_{0} \delta(\pTrue =
\hat{p}_{\rm GC})$).   Therefore, we can define an effective energy dispersion model:}

\begin{equation}\label{eq:Effective_Dispersion_1D}
\Deff(\Ereco; \Egamma, \sVec) = \intFOV 
D(\Ereco;\Egamma,\theta, \sVec) \frac{I_{0}
  \mathcal{E}(\Egamma,\theta;\hat{p}_{\rm GC},\sVec)}{\nPredSig} d\OmegaVhat.
\end{equation}

\noindent \newText{In practice we perform the integration by re-weighting an
isotropically generated MC sample of \gammaRays\ to match the
$\theta$-distribution and fitting for the energy dispersion parameters
(see \Appref{app:Formalism_Correction_Terms}).}

\newText{Given the effective energy dispersion we} can write the signal model as:
\begin{equation}\label{eq:Predicted_Signal_Model_1D_Final}
\CSig(\Ereco;\sVec|\Egamma) = \nSig \Deff(\Ereco; \Egamma, \sVec).
\end{equation}

For the background model the spatial integrals give us the 
energy-dependent exposure correction:

\begin{equation}\label{eq:Expsosure_Correction_1D}
\eta(\Ereco;\sVec) = \intFOV \intROI \frac{\IBkg(\pTrue)\mathcal{E}(\Ereco,\pTrue,\theta;\sVec)}{\nPredBkg} d\Omega d\OmegaVhat.
\end{equation}

\noindent So that we can write the background model as:

\begin{equation}\label{eq:Predicted_Bkg_Model_1D_Final}
\CBkg(\Ereco;\sVec|\GamBkg) = \nBkg \left(\frac{\Ereco}{\Epivot}\right)^{-\GamBkg} \eta(\Ereco;\sVec).
\end{equation}

\noindent Putting together the signal and background models, we have:

\begin{equation}\label{eq:Predicted_Model_1D_Final}
C(\Ereco;\sVec|\vec{\alpha}) = \nSig \Deff(\Ereco; \Egamma, \sVec)
+ \nBkg\left(\frac{\Ereco}{\Epivot}\right)^{-\GamBkg} \eta(\Ereco,\sVec),
\end{equation}

\noindent where the model parameters are
\Egamma(held fixed), \GamBkg, \nSig\ and \nBkg.  While \Egamma\ and \GamBkg\ are physical quantities,
we must use the exposure and intensity maps to extract the source fluxes from 
\nSig\ and \nBkg.

\subsubsection{Energy dispersion parametrized by energy and
  \texorpdfstring{\probE}{P\_E} }\label{app:Formalism_PE_PDF}

If we are considering an energy dispersion model that includes 
the energy quality estimator \probE, then we must factor out \probE\ from the
instrument response. In particular, we consider the distribution 
of \probE, $w(\probE;\Etrue,\theta,\sVec)$, such that:

\begin{equation}\label{eq:Split_P_E_Distribution}
\aeff(\Etrue,\theta,\probE,\sVec) = \aeff(\Etrue,\theta,\sVec) w(\probE;\Etrue,\theta,\sVec),
\end{equation}

\noindent with the normalization constraint:

\begin{equation}\label{eq:Normalize_P_E_Distribution}
\intPE w(\probE;\Etrue,\theta,\sVec) d\probE = 1,
\end{equation}

\noindent for all \Etrue,$\theta$ and \sVec.

We can now include \probE\ in the expression for our expected counts spectrum, 
and integrate over the FOV, the ROI, and \Etrue.   In this way, we can define an effective energy dispersion model for the ROI:

\begin{align}\label{eq:Effective_Dispersion_2D}
\Deff(\Ereco;\Egamma,\probE,\sVec) = & \intFOV \intROI
D(\Ereco;\Egamma,\theta,\probE,\sVec) \nonumber \\
 & \times \frac{I_{\rm sig}(\pTrue)
  \mathcal{E}(\Egamma,\pTrue,\theta;\sVec)}{\nPredSig} w(\probE;\Egamma,\theta,\sVec)
d\Omega d\OmegaVhat.
\end{align}

At this point we assume that distribution of \probE\ for \Egamma\ and all $\theta$ is
adequately modeled by the total observed distribution of \probE\ in the
ROI; i.e., we replace $w(\probE;\Egamma,\theta,\sVec)$ by $\wSig(\probE,\sVec)$, and remove it 
from the spatial integrals.   \newText{Rather than obtaining $\Deff(\Ereco;\Egamma,\probE,\sVec)$
by explicitly performing the above integrals we assume that it is
reasonably well modeled by a distribution of isotropically-generated
\gammaRays\  (i.e., we obtain the model by interpolating in energy between the parameters obtained
from fitting to the ``isotropic-monochromatic'' samples
described in \Secref{sec:MC}).}

With these approximations we can write the predicted counts distribution as:

\begin{equation}\label{eq:Predicted_Signal_Model_Approx_1}
\CSig(\Ereco,\probE;\sVec|\Egamma) =  \nSig \Deff(\Ereco;\Egamma,\probE,\sVec) \wSig(\probE;\sVec).
\end{equation}

On the other hand, for the background we are neglecting the energy dispersion, and modeling the spectrum as a power law:

\begin{align}\label{eq:Predicted_Bkg_Model_2D}
\CBkg(\Ereco,\probE;\sVec|\GamBkg) =  & \nBkg \left(\frac{\Ereco}{\Epivot}\right)^{-\GamBkg} \nonumber \\
& \times \intFOV \intROI \frac{\IBkg(\pTrue)\mathcal{E}(\Ereco,\pTrue,\theta;\sVec)}{\nPredBkg}
w(\probE;\Ereco,\theta,\sVec) d\Omega d\OmegaVhat.
\end{align}

\noindent As for the signal case, we assume distribution of \probE\ for all \Ereco\ and $\theta$ is
adequately modeled by the total observed distribution of \probE\ in the
ROI, and remove it from the spatial integrals, which we then replace with the energy-dependent exposure correction
from \Eqref{eq:Expsosure_Correction_1D}.  This gives use the following background model:

\begin{equation}\label{eq:Predicted_Bkg_Model_2D_Final}
\CBkg(\Ereco,\probE;\sVec|\Egamma) = \nBkg \left(\frac{\Ereco}{\Epivot}\right)^{-\GamBkg} \eta(\Ereco;\sVec)\wBkg(\probE;\sVec).
\end{equation}

Combining the signal and background models, we obtain:

\begin{align}\label{eq:Predicted_Model_2D_Combined}
C(\Ereco,\probE;\sVec|\vec{\alpha}) = & 
\nSig \Deff(\Ereco;\Egamma,\probE,\sVec) \wSig(\probE;\sVec) \nonumber \\
 & + \nBkg
 \left(\frac{\Ereco}{\Epivot}\right)^{-\GamBkg} \eta(\Ereco;\sVec) \wBkg(\probE,\sVec).
\end{align}

\noindent where the fit parameters $\vec{\alpha}$ are the same as for the previous case.

In practice, we take the model for the distribution of \probE\ for both 
the signal and background from the flight data in the ROI, i.e.,

\begin{equation}\label{eq:PE_Distribution_Data}
\wBkg(\probE,\sVec) = \wSig(\probE,\sVec) = \wROI(\probE,\sVec).
\end{equation}

\noindent A subtlety exists in this last approximation:
the $\theta$ distribution of \gammaRayHyph\ directions differ
for the signal and background \gammaRays, because of differences
in the spatial morphology, or because of CR contamination in the
background \gammaRayHyph\ sample.  This means that this last approximation
might be wrong in slightly different ways when applied to signal or
background.  This is the so called ``Punzi effect''\cite{REF:Punzi:2004wh}.   We consider 
this further in \Secref{sec:systematic_errors_edispmodel}.

\subsection{Calculating the effective energy dispersion and exposure corrections}{\label{app:Formalism_Correction_Terms}

We absorbed many details about the morphology of the flux models and spatial variations 
of the exposure into the calculations of effective energy dispersion and the energy-dependent exposure corrections.

Practically speaking, we can create an effective energy dispersion model with MC simulations by \newText{re-weighting
events from an isotropically generated sample to match} a particular observing profile $\tobs(\theta;\pTrue)$, applying the event selection criteria
and fitting the parameters of $\Deff(\Ereco; \Egamma,\probE,\sVec)$ to
the resulting energy dispersion distribution.    The observing
profiles and corresponding effective energy dispersion models for
several different directions in the sky are shown in
\Figref{ObsProfilesAndPDFS}.   \newText{Since the observing profile
  for the Galactic center is so close to uniform, and since the
  variation in resolution is already described by the \probE\ parameter,  when generating the
  2D PDF we elected not to re-weight the events and simply interpolated the
  parameters obtained from the fits to the isotropic-monochromatic
  MC samples (\Secref{sec:MC}).}

\twopanel{ht}{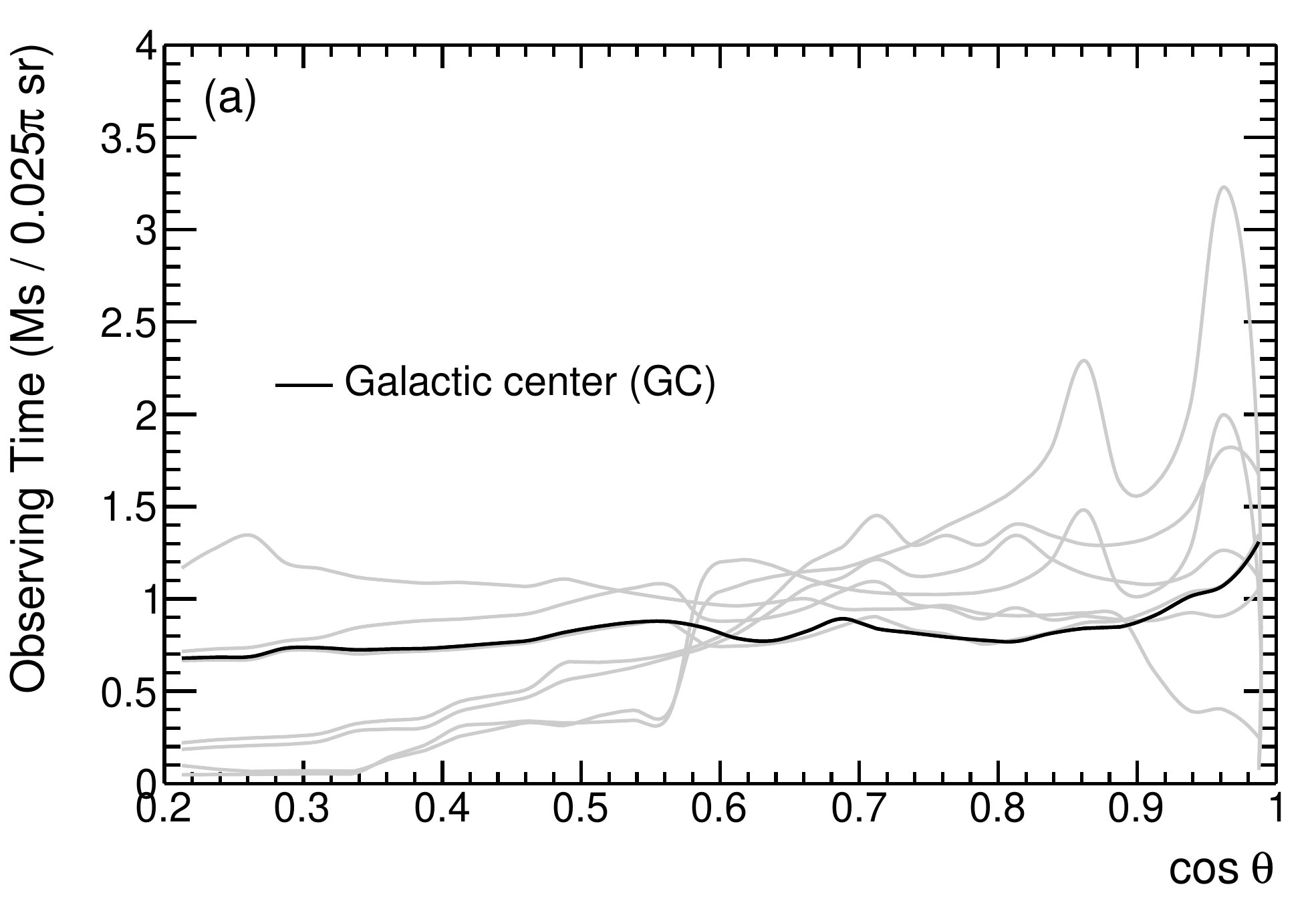}{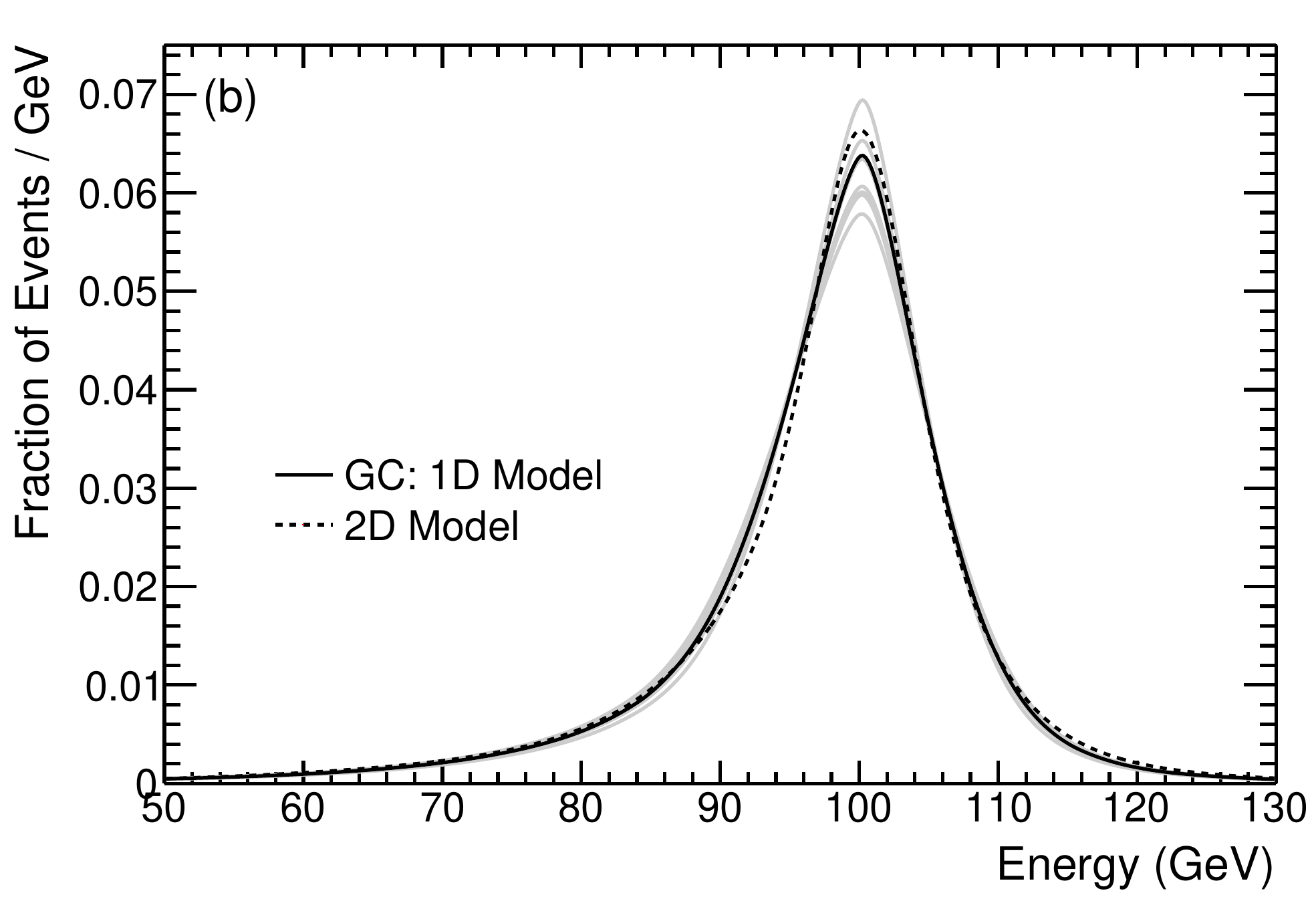}
{\caption{\label{fig:ObsProfilesAndPDFS}Effect of variations in observing profiles on the energy
  dispersion model: (a) observing profiles, $\tobs(\theta;\pTrue)$ for
  several directions with the same right ascension as the Galactic center, but
  different declinations ($\pm30^{\circ}, \pm60^{\circ}, and \pm90^{\circ}$);  (b) the corresponding energy dispersion models for $\Etrue = 100$~GeV.}}

%% file: App_Systematic_Errors.tex
\section{STUDIES OF SYSTEMATIC UNCERTAINTIES}\label{app:systematics_details}

In this appendix we provide details about sources of systematic uncertainty and describe of studies we performed to quantify the particular effects.

\subsection{Uncertainties and approximations of the exposure}\label{subsec:syst_exposure}

We convert our counts limits to flux limits by using the average exposure in the ROI (see \Eqref{equ:FluxToCnt}).  The LAT observes the sky with approximately uniform exposure, and the RMS variation of the exposure in each ROI ranges from $|\delta \Exposure / \Exposure|  < 0.01$
in R3, up to $|\delta \Exposure / \Exposure| = 0.10\ (0.13)$ in R180 at 5~GeV (300~GeV).   (The \newText{off-axis effective area} decreases at higher energies, causing the exposure to be slightly less uniform.)

The \fermi-LAT Collaboration has estimated that the overall uncertainty of the effective area is $10\%$ for energies $> 10$~GeV, and decreases to $8\%$ at $\sim 5$~GeV
\cite{REF:2012.P7Perf}.  For simplicity, in this analysis we have chosen to assign a $10\%$ uncertainty from the overall effective area at all energies.  This uncertainty causes a corresponding $|\delta\Exposure / \Exposure| = 0.10$ uncertainty in the exposure.  Adding this effect in quadature with the variation in exposure between the ROIs yields $0.10 < |\delta \Exposure / \Exposure| < 0.16$ as the overall range of relative uncertainty.

\subsection{Uncertainties in the energy resolution}\label{sec:systematic_errors_unmodeled_features}
The error in the measurement of the energy resolution was measured in beam tests and found to be better than 10\% for energies up to 280 GeV~\cite{REF:2012.P7Perf}.  To test 
how a different energy resolution would affect our limits, we scaled the standard deviations of all the Gaussians in the 2D energy dispersion model 
(see \Secref{sec:method_2DPDF}) by a common scale factor of 1.1 or 0.9, while also scaling the means to preserve the shape, and then fit
to MC simulations containing a line.   We found that the relative error on the best-fit number of signal events ($\delta \nSig / \nSig$) was proportional to the width scale factor used in the fit.  If the fit model was too narrow, it underfit the number of signal counts.  However, if the fit model was too wide, then it overfit the number of signal counts.  The constant of proportionality between $\delta \nSig / \nSig$ and the model scale factor was 0.7.  Therefore, fitting with a model that was 10\% too narrow would, on average, underfit the number of signal counts by 7\%.   We found similar variation in the expected limits in background-only MC simulations.

\subsection{Intrinsic width of the \texorpdfstring{\gammaRayHyph}{gamma-ray} emission}\label{sec:systematic_errors_decay_width}
 In the context of this analysis, any intrinsic width of the \gammaRayHyph\ emission, e.g., from the $Z$ width in $Z \gamma$ final states, 
would manifest very similarly as an unmodeled increase in the energy resolution.  We note that even if the intrinsic width of the emission 
were $50 \%$ of the energy resolution, when convolved with the energy resolution, it would only increase the width of the observed spectral feature
by $11\%$.  As stated in \Secref{sec:systematic_errors_unmodeled_features}, this would cause us to underestimate the signal  by $7\%$.
Furthermore, for $m_{\chi} = m_{\rm Z} = 91$~GeV the \gammaRayHyph\ threshold energy is $\Egamma = 68$~GeV. 
Given that the $Z$ width is 2.5~GeV, and the energy resolution is $\Delta E \sim 5$~GeV at $68$~GeV we estimate that at worst
$\delta \nSig / \nSig = 0.07$ for $Z\gamma$ final states at $E_{\gamma} \simeq 68$~GeV.

\subsection{Approximations in the energy dispersion modeling}\label{sec:systematic_errors_edispmodel}
The \probE\ distribution of the events in a specific ROI and energy interval ($\wROI(\probE)$) influences the energy dispersion model, $\Deff(\Ereco;\Egamma,\probE)$ used in each fit.  The same $\wROI(\probE)$ was used for both the signal and background pieces of the total counts model, see \Eqref{eq:SigPlusBkgCnts}.  However, if the true \probE\ distribution of the signal events is slightly different, e.g., because of differences in the observing profile, or because of CR contamination in the background sample, then the approximation $\wSig(\probE) = \wROI(\probE)$ would be incorrect and wrongly neglect the ``Punzi effect''\cite{REF:Punzi:2004wh}.  We created 1000 MC simulations with a signal where the ``true'' $\wSig(\probE)$ were taken from the 50~GeV ``isotropic monochromatic'' MC dataset, but the fit assumed the $\wROI(\probE)$ from the \irf{P7CLEAN} data with $\Ereco\approx 50$~GeV. The difference is very similar to the discrepancy shown in \Figref{ProbDist_DataVsMC} between the \irf{P7CLEAN} data and the all-sky MC.   \newText{We elected to use the \irf{P7CLEAN} dataset for this study as the discrepency is larger in that dataset than in \irf{P7REP\_CLEAN} and thus provides a more conservative estimate of the magnitude of this effect.}  Also, the difference between the ``true'' $\wSig(\probE)$ and the fit $\wROI(\probE)$ was somewhat larger than the \probE\ distribution variation we see in the data.  On average, using the incorrect $\wROI(\probE)$ in the fit resulted in an error on the total number of signal counts of $\delta \nSig / \nSig \leq 0.01$.  Therefore, the approximation $\wBkg(\probE) = \wSig(\probE) = \wROI(\probE)$ in the fit does not result in a large systematic effect.

Though the event incidence angle ($\theta$) and \probE\ are correlated, the expected 2D energy dispersion, $D(\Ereco;\Etrue,\probE)$, varies only moderately with $\theta$.  In a given \probE\ bin, the energy resolution for events with large $\theta$ tends to be better than for on-axis events.  Since the $\theta$ distribution in the monochromatic MC we used to derive the 2D energy dispersion model is very similar to the $\theta$ distribution in the flight data, we do not expect differences in the $\theta$ distribution compared to the MC to introduce a large systematic effect.  We reweighted the monochromatic MC in each \probE\ bin to match the $\theta$ distribution in the flight data and 
re-derived the energy dispersion model.  On average, the scale factor to convert the average widths of the nominal model to the widths of the reweighted model is 0.97.  Using the scaling relation derived in the previous section, the resulting relative uncertainty on \nSig\ is $2\%$ on average.

\subsection{Cosmic-ray background contamination}\label{sec:systematic_errors_backgrounds}

Our energy reconstruction algorithm is based on the assumption that the incoming particle is a \gammaRay, and therefore that the energy deposited in the calorimeter is
well described as an electromagnetic shower.  For hadronic CRs the energy reconstruction is therefore incorrect, and can create spectral artifacts.   Furthermore, although care was taken to ensure that the selection criteria vary smoothly with energy for
\gammaRays, this was not the case for hadronic CRs.    This second point implies that any spectral features caused by CRs are likely to be different for different event classes.

The \irf{P7REP\_CLEAN} event selection rejects CR background at the level of $10^5$ or higher.   This makes it difficult to use MC to study the
spectra of background contamination for two reasons: 1) the need to generate prohibitively large samples to have reasonable statistics for the CR
backgrounds leaking through the \gammaRayHyph\ event selection; 2) by definition, the background events that survive \gammaRayHyph\ event selection are very unusual events, so that small problems with the fidelity of the MC simulation can easily contribute at a large enough level to invalidate predictions.    

To investigate the possible effects of background contamination in our sample, we considered the set of events in the \irf{P7REP\_SOURCE} class
that did not enter the \irf{P7REP\_CLEAN} class.  (We note that the \irf{P7REP\_CLEAN} event sample is a strict subset of the \irf{P7REP\_SOURCE} sample.)   This allows us to estimate the CR contamination in the \irf{P7REP\_SOURCE} class.   

\Figureref{syst_CR_Contamination} shows the fraction of events in \irf{P7REP\_SOURCE} that 
survive in the \irf{P7REP\_CLEAN} sample for the various ROIs, as well as the counts spectrum of the events which do not survive into the \irf{P7REP\_CLEAN} sample for the R180 ROI.  
When we fit the counts spectrum of these events for a line-like signal using the \irf{P7REP\_CLEAN} to estimate the size of potential induced signals, we find that several of the fits show $> 2\sigma$ induced signals, with the induced fractional signal for this CR-rich sample reaching $f_{\rm CR}=0.05$. 

\twopanel{ht!}{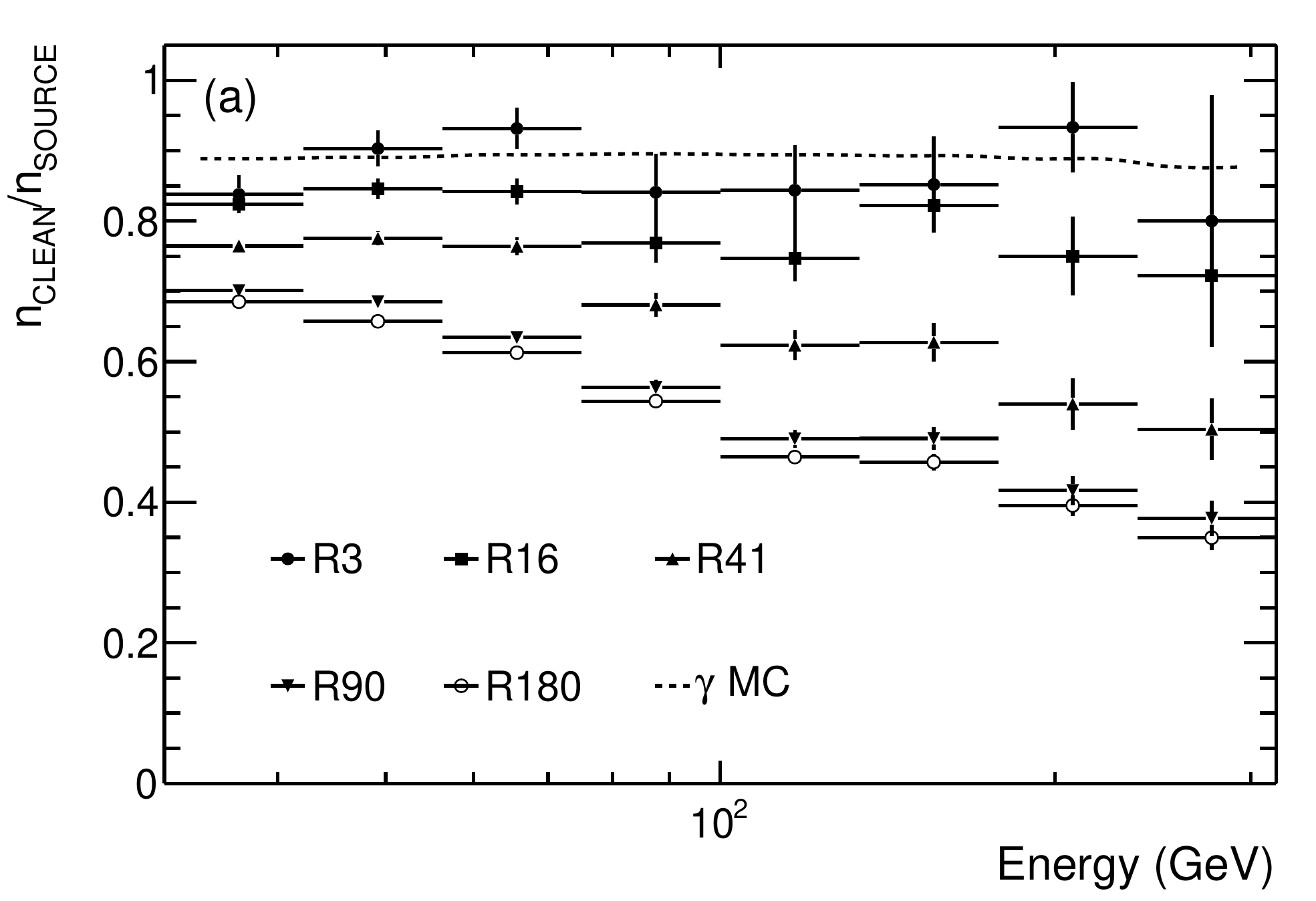}{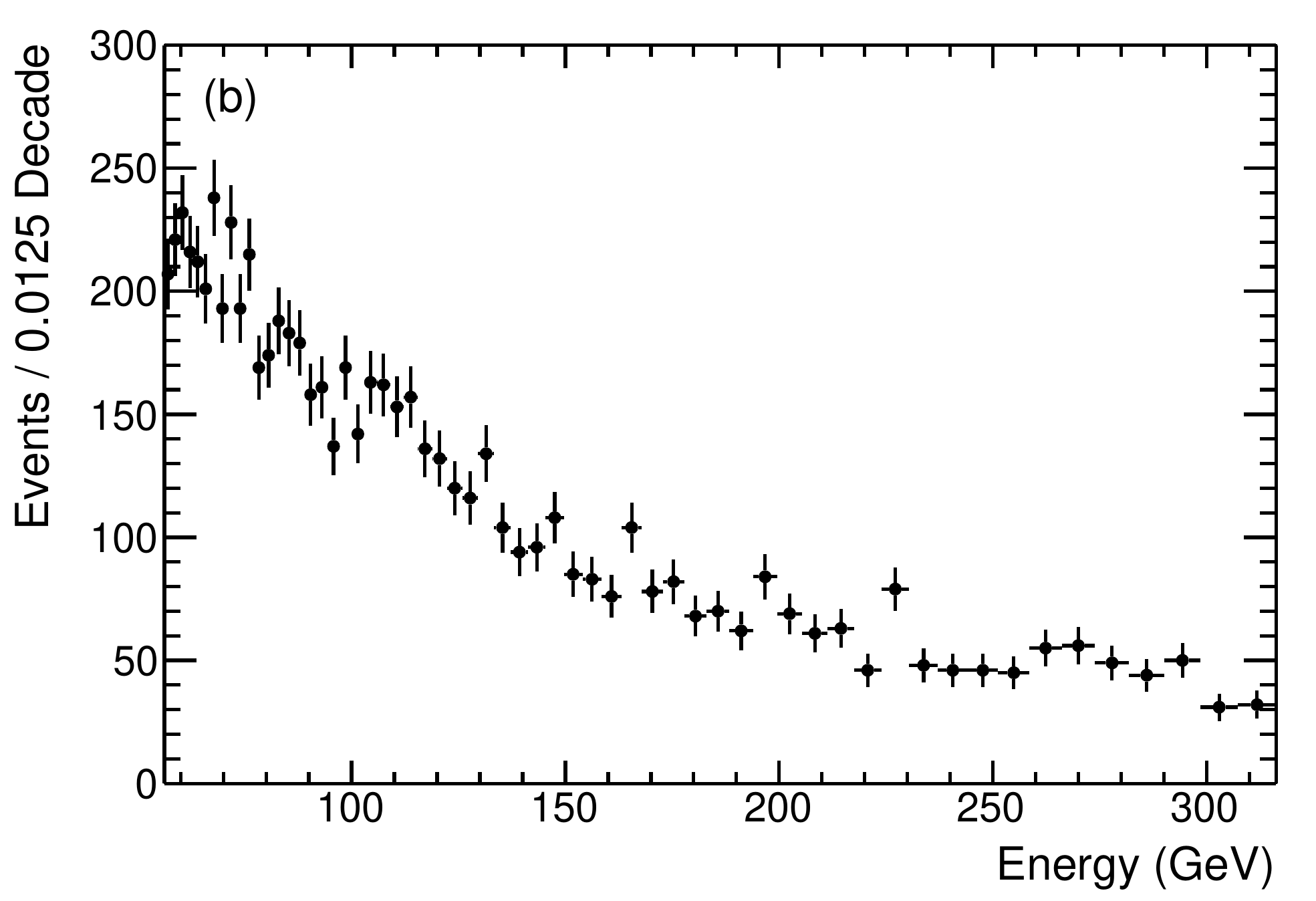}
{\caption{\label{fig:syst_CR_Contamination} CR background contamination estimation: (a) the fraction of events in the \irf{P7REP\_SOURCE} sample also present in the \irf{P7REP\_CLEAN} sample as a function of energy for all the ROIs;  (b) the counts spectrum of events in the \irf{P7REP\_SOURCE} sample but not the \irf{P7REP\_CLEAN} sample for the R180 ROI.   The dashed line in (a) is the ratio $Acc_{\rm SOURCE}(\Ereco)/Acc_{\rm CLEAN}(\Ereco)$.}}

To estimate the effect this CR contamination might have on the analysis performed with \irf{P7REP\_CLEAN} class events, we estimated the amount of background 
contamination in the \irf{P7REP\_SOURCE} sample by comparing the fraction of events in the \irf{P7REP\_SOURCE} sample that survive in to the \irf{P7REP\_CLEAN} 
sample relative to the ratio of the acceptances, $Acc(\Ereco)$, i.e., the effective area integrated over the field of view:

\begin{equation}
  \frac{n_{\rm CR}(\Ereco)}{n_{\gamma}(\Ereco)} \simeq \frac{Acc_{\rm CLEAN}(\Ereco)}{Acc_{\rm SOURCE}(\Ereco)}\frac{n_{\rm SOURCE}(\Ereco)}{n_{\rm CLEAN}(\Ereco)} - 1.
\end{equation}

\noindent The induced fractional signal from CR contamination in the total \irf{P7REP\_SOURCE} sample is smaller than in the CR-rich sub-sample that does
not survive in the \irf{P7REP\_CLEAN} event class:

\begin{equation}
  f_{\rm SOURCE}(\Ereco) = \frac{n_{\rm CR}(\Ereco)}{n_{\rm CR}(\Ereco) + n_{\gamma}(\Ereco)} f_{\rm CR}.
\end{equation}

\noindent Combining the last two equations gives us an estimate of the induced fractional signal in \irf{P7REP\_SOURCE} class:

\begin{equation}
  f_{\rm SOURCE}(\Ereco) =  f_{\rm CR} - \frac{Acc_{\rm SOURCE}(\Ereco)}{Acc_{\rm CLEAN}(\Ereco)}\frac{n_{\rm CLEAN}(\Ereco)}{n_{\rm SOURCE}(\Ereco)}f_{\rm CR}.
\end{equation}

\noindent Based on MC simulations, we estimated that the CR contamination fraction for the \irf{P7REP\_CLEAN} class is less than 10\% of the \irf{P7REP\_SOURCE} 
class fraction above 10~GeV.   This suggests that for the \irf{P7REP\_CLEAN} class CR-contamination is 
a negligible ($\delta f<0.01$) effect.  However, residual CRs surviving from the \irf{P7REP\_SOURCE} to the \irf{P7REP\_CLEAN} dataset over a
narrow ranges of energies could induce or mask a line signal.  We have not seen any evidence of such contamination, and have chosen to assign $50\%$ of the 
estimated induced fractional signal for the \irf{P7REP\_SOURCE} event class, $f_{\rm SOURCE}$, as the uncertainty for the \irf{P7REP\_CLEAN} event class.  With this assignment, we
found that the CR-background contamination is negligible at all energies for the R3 and R16 ROIs, but rises to $f = 0.014$ for the R180 ROI at 
high energies.

\subsection{Point-source contamination}\label{sec:systematic_errors_point_sources}

We estimated the effect of point-source contamination using a similar method to the one described above for CR-background contamination.   First
we fit the composite spectrum of the events removed with source masking.   We found fractional signals of $\delta f = 0.010$ on average.
Independently, we estimated that the residual contamination of the Celestial dataset from point sources in our energy range is $\lesssim 10\%$ (see \Secref{subsec:method_event}).  Taken together, these imply that potential induced fractional signals from point-source contamination is negligible ($\delta f < 0.001$).

\subsection{Spectral smoothness of control samples}\label{sec:systematic_errors_spectral_smoothness}

\subsubsection{Spectral smoothness of the Earth Limb}\label{sec:LimbSmoothness}

We used the counts spectrum of \gammaRays\ from the Limb to estimate the size of induced fractional signals from variations in the effective area. As stated in \Secref{subsec:method_event}, the Limb dataset is obtained by selecting times when $|\theta_{\rm r}| > 52^\circ$.  

Given that the Limb photon spectrum is expected to be a
featureless power law, it is an excellent control region for a spectral line search where one looks for narrow deviations from power-law behavior.  
We expect any line-like features observed in the Limb to be due to statistical fluctuations or variations in the effective area of the LAT over narrow ranges of energy. To estimate the
size of the latter, we fit for spectral lines with our standard fit energy spacing and compared the measured fractional
signals with the expected statistical variation given the number of events in the Limb dataset.

Most of the narrow features measured in the Limb are consistent with statistical fluctuations from the power-law-only hypothesis.  
However, more than 5\% of the features have a fractional size larger than the statistical 95\% containment band, suggesting that 
variations in the effective area are contributing as well.  
We approximated the size of the effective area contribution by calculating the required variation in the effective area ($f_{\rm Aeff}$) that 
allows all observed features to lie within the 95\% containment band and assigning half of that variation as an estimate of the $1\sigma$
systematic uncertainty.   We found that $f_{\rm Aeff} = 0.005$ for low energies ($<10$ GeV) and increases to $f_{\rm Aeff} = 0.015$ at 100~GeV.  Above
100~GeV the statistics from the Limb are marginal;  we assign $f_{\rm Aeff} = 0.020$ and $f_{\rm Aeff} = 0.025$ as the magnitude of the potential
fractional signals at 150 and 300~GeV respectively.   We note in passing that the fit at 133~GeV gave an anomalously large fractional signal, $f = 0.14$, see
\Secref{subsec:133GeV_Feature_Earth_Limb} for more details and discussion.

\subsubsection{Spectral smoothness along the Galactic plane}\label{sec:GalaxySmoothness}

Representing the complex \gammaRayHyph\ emission from the Galaxy as a power law is an oversimplification, and any deviations from a power law will
induce signals at some level in the likelihood fit.  However, it is generally assumed that any spectral features in the Galactic emission are 
much wider than the LAT energy resolution, and therefore that the magnitude of the induced signal is negligible.  

To test this hypothesis with data, we systematically scanned across the Galactic plane and inner Galaxy, $|b| < 8^\circ$, $|l| < 90^\circ$.   We used ROIs of $2^\circ\times 2^\circ$, $4^\circ\times 4^\circ$, and $8^\circ\times 8^\circ$, and fit for a line in each energy interval.  We compared these results to a second scan performed
with the measured energies randomly redistributed amongst the events to remove any correlation between energy and direction.   

\newText{In this study only}, we allowed for both positive and negative deviations from a power law \newText{(i.e., we allowed \nSig\ to be negative) to estimate the extent to which a true signal might be masked by non power-law behavior of the background}.  Accordingly, we define the signed significance as $\slocal = \pm\sqrt{\rm{TS}}$, where the sign matches the sign of the deviation.   For this study, we adopted an upper limit of 56~GeV (100~GeV) for the energy range in the $2^\circ\times 2^\circ$ ($8^\circ\times 8^\circ$) ROIs to avoid having the minimizer step into a parameter ranges where the likelihood function is negative.  Given the large number of fits performed, we used the simpler ``1D'' energy dispersion model, see \Secref{sec:method_2DPDF}), which does not include \probE, 
and performed binned likelihood fits.   Finally, we scanned in $b$ and $l$ using step sizes of $\frac{1}{2}$ the ROI width; thus each ROI overlaps by $50\%$ with the four nearest neighbors.   However, all the results shown here were made using only a set of non-overlapping ROIs obtained by removing every other step from the scan.

\Figureref{ScanSignifDistribution} shows the distribution of signed significances for the scan along the Galactic plane using $2^\circ \times 2^\circ$ ROIs.  
For comparison, we have overlaid the distribution for the energy-shuffled data.   We see that the flight data match the shuffled data very well;  this was also true of the scan using $4^\circ \times 4^\circ$ and $8^\circ \times 8^\circ$ ROIs.  Furthermore, in each case the distributions were consistent with Gaussians with unit width and zero mean, suggesting that describing the background as a power law is a good approximation.

\twopanel{ht}{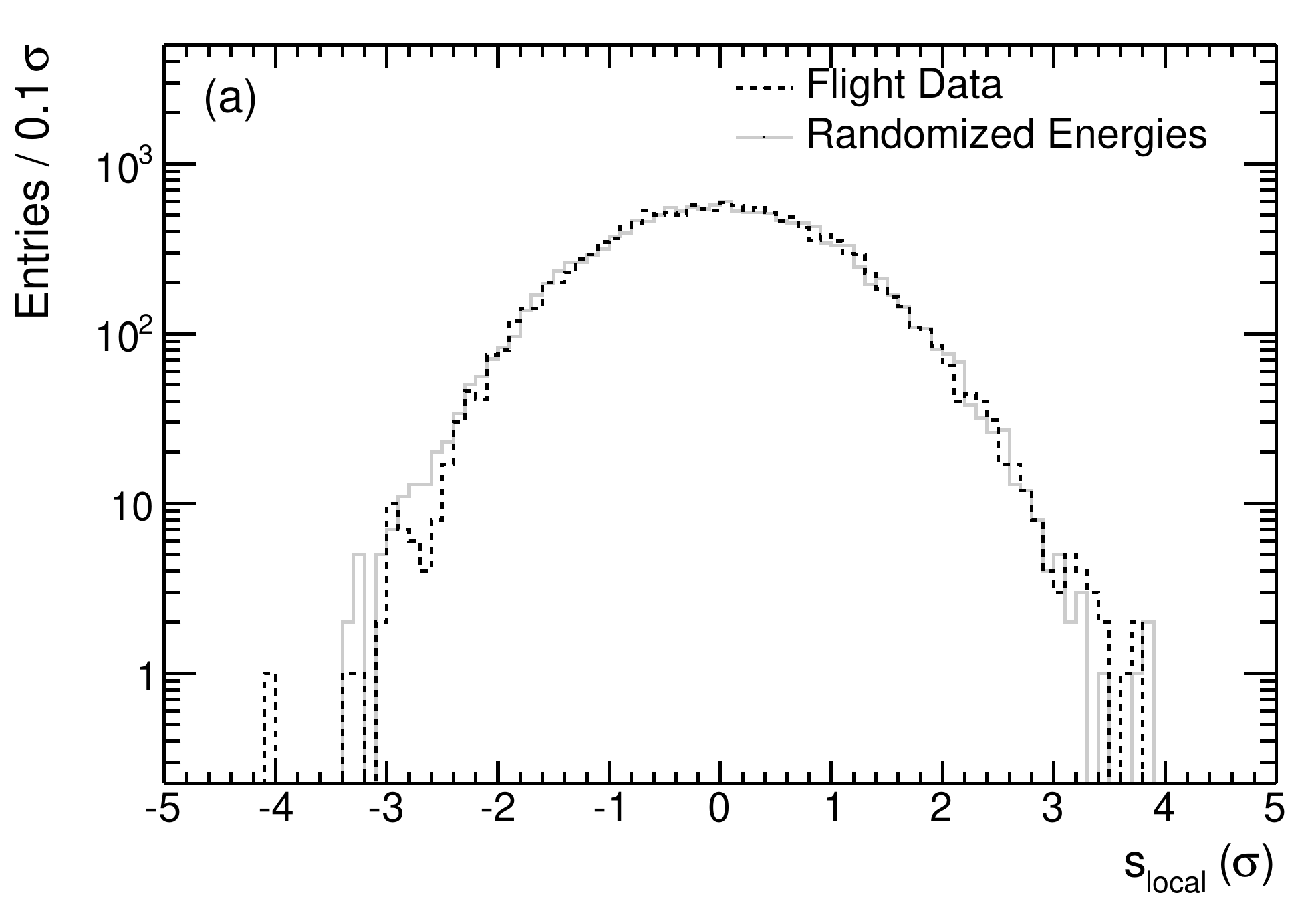}{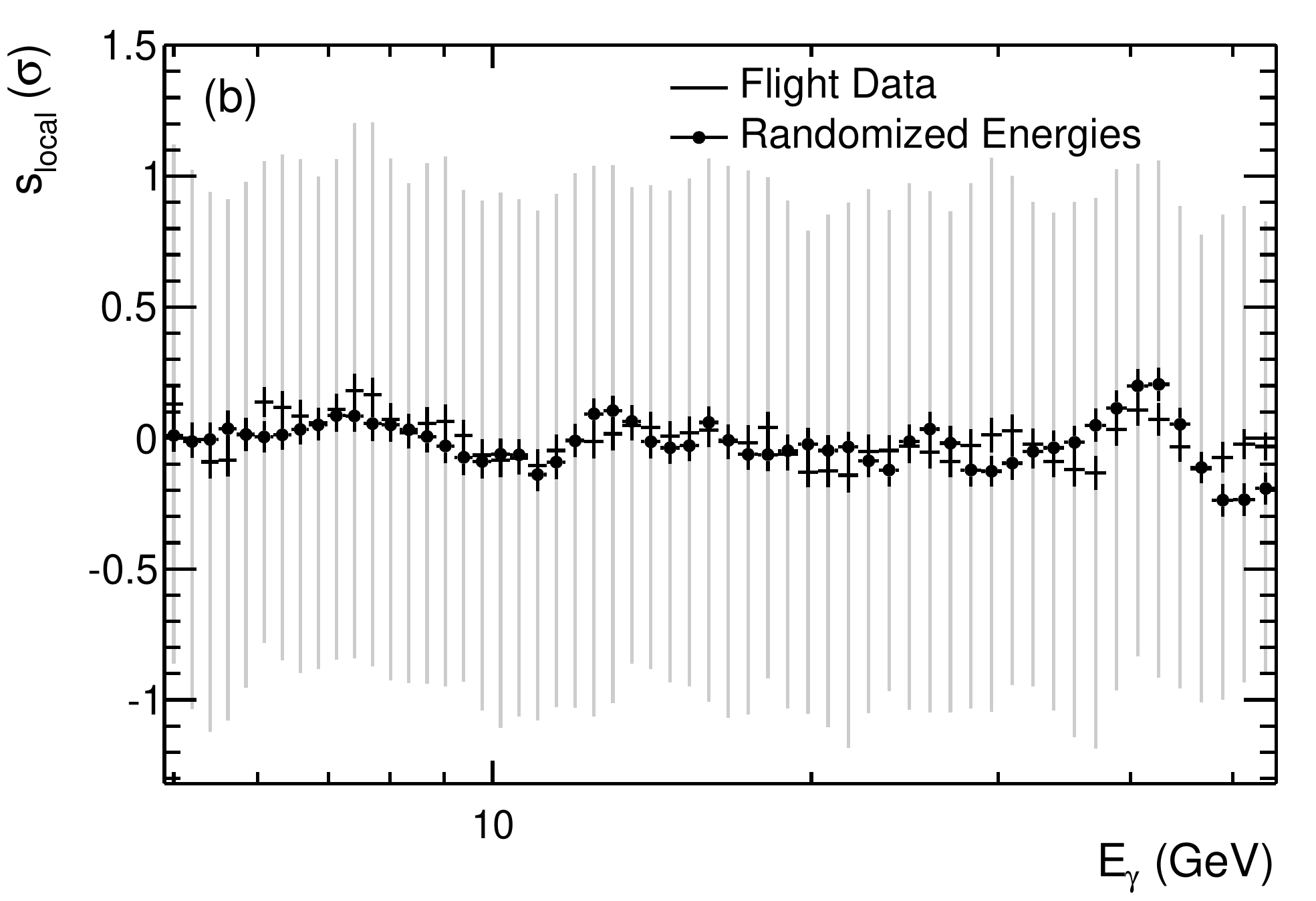}{
\caption{\label{fig:ScanSignifDistribution}Distribution of signed significances for ROIs along the Galactic plane for all $\Egamma < 56$~GeV: (a) for $2^\circ \times 2^\circ$ ROIs, (b) mean \slocal\ versus \Egamma; the large gray error bars show the RMS of the distribution at a given energy, and the small black error bars show the error on the mean.   Note that these results made using only a set of non-overlapping ROIs. }}

On the other hand, the distribution of \slocal\ at any given energy for 
ROIs of a particular size tended to be slightly narrower than for the full distribution, and the means were inconsistent with zero at many energies.  

For $2^\circ \times 2^\circ$ ROIs the means of the \slocal\ distributions were $|\overline{\slocal}| < 0.2\sigma$~for $\Egamma < 56$~GeV, while 
for the $8^\circ \times 8^\circ$ ROIs they were $|\overline{\slocal}| < 0.8\sigma$ for~$\Egamma < 100$~GeV.  The corresponding means of the 
fractional signal distribution were small, but inconsistent with zero at many energies: 
rising from $|\overline{f}| = 0.008$ below 10~GeV, to $|\overline{f}| = 0.018$ at 30~GeV, and to $|\overline{f}| <  0.042$ at $\Egamma = 56$~GeV.   
Interestingly, this effect is present both the flight-data and in the sample of events with shuffled energies, suggesting that it maybe be related to 
overall distribution of counts with energy, rather than to correlations between energies and directions.   We also note that the overlapping energy ranges introduce correlations in the means of the \slocal\ distributions at different fit energies.   However, we have chosen to assign 25\% of the magnitude of the deviations of $|\overline{f}|$ from zero as a potential systematic 
uncertainty for the larger ROIs (R41, R90, R180), rising to $\delta f = 0.02$ at 300~GeV.  This is an empirically motivated choice. We believe that this study
gives a reasonable measure of the non-power-law-like behavior for relatively large regions of the sky.

\subsubsection{Induced signals from limitations in the background modeling}\label{subsec:syst_background_model}
Here we quantify the fractional signal expected if the true spectrum is a broken power law, with the spectral index
changing from $\Gamma_1$ and $\Gamma_2$ at the line-fit energy.   We created 1000 MC simulations with broken power law spectra.  The fractional 
signal size for various break sizes is given in \Tabref{FracSig_BPL}.   We see that a true broken power law spectrum could mimic a line-like feature, though 
a relatively large break would be needed to induce a fractionally large signal.   

\begin{table}[ht]
\caption{\label{tab:FracSig_BPL} Fractional signal $f$ from fits to 1000 broken power law MC simulations with various break sizes.}
\begin{tabular}{ccc}
\hline\hline
\ $\Gamma_1$\ & \ $\Gamma_2$\  & \ $f$\  \\
\hline
2.50 & 2.55 & $0.019$ \\
2.50 & 2.60 & $0.021$ \\
2.50 & 2.70 & $0.062$ \\
2.00 & 3.00 & $0.085$ \\
2.00 & 4.00 & $0.147$ \\
2.00 & 6.00 & $0.233$ \\
\hline\hline
\end{tabular}
\end{table}

Although we could in principle distinguish between a broken power law and line-like signal on a \newText{power-law} background given 
large statistics, in practice this is only possible at the lowest energies and for the largest ROI because of the relatively narrow energy ranges used in our fits.
Since the smaller ROIs are more likely to be dominated by local variations in the 
diffuse \gammaRayHyph\ emission, and thus more likely to depart from the generally power-law-like behavior described in \Secref{sec:GalaxySmoothness} we have chosen to assign the induced fraction signal of $\delta f=0.019$ for a broken power law with a small change in index ($\Gamma_1 = 2.5$ to $\Gamma_2 = 2.55$)  as the potential systematic uncertainty for our smaller ROIs (R3 and R16).

%% file: Results_Tables.tex
\section{95\% CONFIDENCE LEVEL LIMITS}\label{app:results}
We present the 95\% CL flux upper limits derivied for each of our
5 ROIs in Tabs.~\ref{tab:allLimits_lowerEn}--\ref{tab:allLimits_higherEn}.
We also give the annihilation cross section
upper limits for the DM profiles in each ROI where sensitivity to that DM model has been optimized (see \Secref{sec:method_roi}): R3 (contracted
NFW profile),  R16 (Einasto profile), R41 (NFW profile) and R90
(Isothermal profile), and the decay lifetime lower limit for R180.

\begin{table}[ht]
\caption{\label{tab:allLimits_lowerEn}95\% confidence level limits from all ROIs for fit energies from 5--11.48 GeV.  The first column for each ROI is the ($\Phi_{\gamma\gamma}$) upper limit in $10^{-9}$ cm$^{-2}$s$^{-1}$.  The second column for each is the upper limit on $\langle\sigma v\rangle_{\gamma\gamma}$ in $10^{-28}$ cm$^{3}$s$^{-1}$ for the DM profile for which that ROI is optimal.  For R180, we give lower limit on $\tau_{\gamma\nu}$(NFW) in $10^{29}$ s.  Note that for $\tau_{\gamma\nu}$, the energy is $m_{\chi}/2$.}
\begin{ruledtabular}
\begin{tabular}{ccccccccc}
& \multicolumn{2}{c}{R16} & \multicolumn{2}{c}{R41} & \multicolumn{2}{c}{R90} & \multicolumn{2}{c}{R180} \\
Energy &  $\Phi_{\gamma\gamma}$ & $\langle\sigma v \rangle_{\gamma\gamma}$ & $\Phi_{\gamma\gamma}$ & $\langle\sigma v \rangle_{\gamma\gamma}$ & $\Phi_{\gamma\gamma}$ & $\langle\sigma v \rangle_{\gamma\gamma}$ & $\Phi_{\gamma\gamma}$ & $\tau_{\gamma\nu}$ \\
(GeV) & & Ein &  & NFW &  & Iso &  & NFW \\
 \hline
5.00 & 3.97 & 0.15 & 4.59 & 0.17 & 4.32 & 0.20 & 3.54 & 5.52 \\
5.20 & 2.96 & 0.12 & 2.53 & 0.10 & 3.09 & 0.15 & 3.36 & 5.59 \\
5.41 & 2.25 & 0.10 & 3.44 & 0.15 & 3.60 & 0.19 & 4.28 & 4.22 \\
5.62 & 1.83 & 0.09 & 4.58 & 0.21 & 4.24 & 0.24 & 5.16 & 3.37 \\
5.85 & 1.90 & 0.10 & 5.82 & 0.29 & 5.55 & 0.34 & 7.91 & 2.11 \\
6.08 & 2.03 & 0.11 & 6.21 & 0.34 & 7.19 & 0.48 & 11.69 & 1.37 \\
6.33 & 2.22 & 0.13 & 5.47 & 0.32 & 6.75 & 0.49 & 11.83 & 1.30 \\
6.58 & 2.21 & 0.14 & 4.11 & 0.26 & 6.28 & 0.49 & 9.21 & 1.61 \\
6.84 & 2.06 & 0.14 & 2.90 & 0.20 & 5.60 & 0.47 & 7.99 & 1.79 \\
7.12 & 2.05 & 0.15 & 2.17 & 0.16 & 5.48 & 0.50 & 7.37 & 1.86 \\
7.40 & 1.16 & 0.09 & 1.46 & 0.12 & 3.90 & 0.39 & 5.41 & 2.44 \\
7.70 & 0.56 & 0.05 & 1.49 & 0.13 & 2.16 & 0.23 & 2.67 & 4.74 \\
8.01 & 0.86 & 0.08 & 2.07 & 0.20 & 2.02 & 0.23 & 2.25 & 5.42 \\
8.33 & 1.51 & 0.16 & 1.39 & 0.14 & 1.38 & 0.17 & 2.12 & 5.54 \\
8.67 & 1.36 & 0.15 & 1.05 & 0.12 & 1.15 & 0.16 & 1.52 & 7.41 \\
9.02 & 1.08 & 0.13 & 0.91 & 0.11 & 1.41 & 0.21 & 1.18 & 9.15 \\
9.39 & 0.89 & 0.12 & 1.27 & 0.17 & 1.64 & 0.26 & 1.20 & 8.68 \\
9.77 & 0.73 & 0.10 & 1.45 & 0.20 & 1.36 & 0.24 & 1.14 & 8.78 \\
10.17 & 0.51 & 0.08 & 1.18 & 0.18 & 1.94 & 0.36 & 1.50 & 6.40 \\
10.59 & 0.87 & 0.15 & 1.51 & 0.25 & 1.60 & 0.33 & 1.54 & 5.98 \\
11.02 & 1.82 & 0.33 & 1.84 & 0.33 & 1.31 & 0.29 & 1.54 & 5.74 \\
11.48 & 1.48 & 0.29 & 1.85 & 0.36 & 1.86 & 0.44 & 1.93 & 4.42 \\

\end{tabular}
\end{ruledtabular}
\end{table}

\begin{table}[ht]
\caption{\label{tab:allLimits_lowEn}95\% confidence level limits from all ROIs for fit energies from 11.96--29.57 GeV.  The first column for each ROI is the ($\Phi_{\gamma\gamma}$) upper limit in $10^{-9}$ cm$^{3}$s$^{-1}$.  The second column for each is the upper limit on $\langle\sigma v\rangle_{\gamma\gamma}$ in $10^{-28}$ cm$^{3}$s$^{-1}$ for the DM profile for which that ROI is optimal.  For R180, we give lower limit on $\tau_{\gamma\nu}$(NFW) in $10^{29}$ s.  Note that for $\tau_{\gamma\nu}$, the energy is $m_{\chi}/2$.}
\begin{ruledtabular}
\begin{tabular}{ccccccccc}
& \multicolumn{2}{c}{R16} & \multicolumn{2}{c}{R41} & \multicolumn{2}{c}{R90} & \multicolumn{2}{c}{R180} \\
Energy &  $\Phi_{\gamma\gamma}$ & $\langle\sigma v \rangle_{\gamma\gamma}$ & $\Phi_{\gamma\gamma}$ & $\langle\sigma v \rangle_{\gamma\gamma}$ & $\Phi_{\gamma\gamma}$ & $\langle\sigma v \rangle_{\gamma\gamma}$ & $\Phi_{\gamma\gamma}$ & $\tau_{\gamma X}$ \\
(GeV) & & Ein &  & NFW &  & Iso &  & NFW \\
 \hline
11.96 & 0.66 & 0.14 & 1.04 & 0.22 & 1.67 & 0.43 & 2.52 & 3.24 \\
12.46 & 0.50 & 0.12 & 0.92 & 0.21 & 2.26 & 0.63 & 4.80 & 1.63 \\
12.98 & 1.26 & 0.31 & 1.14 & 0.28 & 2.47 & 0.75 & 5.19 & 1.45 \\
13.53 & 1.55 & 0.42 & 1.32 & 0.36 & 1.54 & 0.51 & 3.46 & 2.09 \\
14.10 & 0.97 & 0.29 & 0.87 & 0.25 & 0.73 & 0.26 & 1.74 & 3.99 \\
14.70 & 0.43 & 0.14 & 0.84 & 0.27 & 0.74 & 0.29 & 1.32 & 5.03 \\
15.33 & 0.40 & 0.14 & 0.95 & 0.33 & 1.31 & 0.56 & 1.61 & 3.95 \\
15.99 & 0.37 & 0.14 & 1.13 & 0.42 & 1.85 & 0.86 & 2.70 & 2.27 \\
16.69 & 0.39 & 0.16 & 1.56 & 0.64 & 1.72 & 0.87 & 2.47 & 2.37 \\
17.42 & 0.61 & 0.27 & 1.93 & 0.86 & 2.46 & 1.35 & 2.93 & 1.92 \\
18.18 & 0.70 & 0.34 & 1.78 & 0.87 & 2.15 & 1.28 & 2.45 & 2.20 \\
18.99 & 0.55 & 0.30 & 1.55 & 0.83 & 1.78 & 1.16 & 1.91 & 2.70 \\
19.84 & 0.46 & 0.27 & 1.50 & 0.87 & 1.90 & 1.35 & 2.04 & 2.41 \\
20.73 & 0.45 & 0.29 & 0.71 & 0.45 & 0.80 & 0.62 & 0.88 & 5.35 \\
21.66 & 0.42 & 0.29 & 0.53 & 0.36 & 0.55 & 0.47 & 0.87 & 5.18 \\
22.64 & 0.50 & 0.38 & 0.36 & 0.28 & 0.40 & 0.37 & 0.75 & 5.76 \\
23.66 & 0.90 & 0.74 & 0.71 & 0.59 & 0.69 & 0.70 & 1.00 & 4.12 \\
24.74 & 0.77 & 0.70 & 0.82 & 0.74 & 1.61 & 1.78 & 1.80 & 2.19 \\
25.86 & 0.81 & 0.80 & 1.50 & 1.48 & 1.96 & 2.37 & 2.25 & 1.68 \\
27.04 & 0.72 & 0.78 & 1.21 & 1.30 & 1.37 & 1.81 & 1.14 & 3.18 \\
28.28 & 0.37 & 0.43 & 0.92 & 1.09 & 1.68 & 2.43 & 1.34 & 2.58 \\
29.57 & 0.24 & 0.32 & 0.62 & 0.79 & 1.21 & 1.91 & 1.43 & 2.31 \\
\end{tabular}
\end{ruledtabular}
\end{table}

\begin{table}[ht]
\caption{\label{tab:allLimits_hiEn}95\% confidence level limits from all ROIs for fit energies from 30.93--83.43 GeV.  The first column for each ROI is the ($\Phi_{\gamma\gamma}$) upper limit in $10^{-9}$ cm$^{3}$s$^{-1}$.  The second column for each is the upper limit on $\langle\sigma v\rangle_{\gamma\gamma}$ in $10^{-28}$ cm$^{3}$s$^{-1}$ for the DM profile for which that ROI is optimal.  For R180, we give lower limit on $\tau_{\gamma\nu}$(NFW) in $10^{29}$ s.  Note that for $\tau_{\gamma\nu}$, the energy is $m_{\chi}/2$.}
\begin{ruledtabular}
\begin{tabular}{ccccccccccc}
& \multicolumn{2}{c}{R3} & \multicolumn{2}{c}{R16} & \multicolumn{2}{c}{R41} & \multicolumn{2}{c}{R90} & \multicolumn{2}{c}{R180} \\
Energy &  $\Phi_{\gamma\gamma}$ & $\langle\sigma v \rangle_{\gamma\gamma}$ & $\Phi_{\gamma\gamma}$ & $\langle\sigma v \rangle_{\gamma\gamma}$ & $\Phi_{\gamma\gamma}$ & $\langle\sigma v \rangle_{\gamma\gamma}$ & $\Phi_{\gamma\gamma}$ & $\langle\sigma v \rangle_{\gamma\gamma}$ & $\Phi_{\gamma\gamma}$ & $\tau_{\gamma X}$ \\
(GeV) &  & NFWc & & Ein &  & NFW &  & Iso &  & NFW \\
 \hline
30.93 & 0.08 & 0.07 & 0.18 & 0.26 & 0.32 & 0.45 & 0.36 & 0.62 & 0.68 & 4.67 \\
32.36 & 0.07 & 0.07 & 0.19 & 0.29 & 0.24 & 0.36 & 0.29 & 0.56 & 0.59 & 5.13 \\
33.85 & 0.07 & 0.08 & 0.22 & 0.37 & 0.21 & 0.35 & 0.30 & 0.62 & 0.36 & 7.97 \\
35.42 & 0.12 & 0.13 & 0.58 & 1.07 & 0.72 & 1.33 & 1.10 & 2.51 & 0.95 & 2.90 \\
37.07 & 0.29 & 0.36 & 0.66 & 1.34 & 1.04 & 2.10 & 1.38 & 3.44 & 0.97 & 2.73 \\
38.80 & 0.23 & 0.31 & 0.45 & 1.00 & 1.07 & 2.38 & 0.80 & 2.17 & 0.89 & 2.82 \\
40.62 & 0.25 & 0.38 & 0.36 & 0.89 & 1.07 & 2.60 & 0.77 & 2.29 & 1.27 & 1.89 \\
42.54 & 0.46 & 0.75 & 0.65 & 1.74 & 1.40 & 3.73 & 1.35 & 4.43 & 1.98 & 1.16 \\
44.55 & 0.43 & 0.78 & 0.46 & 1.35 & 0.87 & 2.53 & 0.70 & 2.52 & 0.66 & 3.32 \\
46.66 & 0.34 & 0.67 & 0.50 & 1.62 & 0.66 & 2.13 & 0.57 & 2.27 & 0.56 & 3.73 \\
48.88 & 0.27 & 0.58 & 0.35 & 1.24 & 0.29 & 1.00 & 0.43 & 1.86 & 0.59 & 3.37 \\
51.22 & 0.14 & 0.33 & 0.16 & 0.63 & 0.24 & 0.94 & 0.38 & 1.82 & 0.73 & 2.62 \\
53.69 & 0.15 & 0.39 & 0.21 & 0.90 & 0.24 & 1.02 & 0.37 & 1.91 & 0.73 & 2.49 \\
56.30 & 0.14 & 0.41 & 0.50 & 2.35 & 0.69 & 3.23 & 0.99 & 5.68 & 1.32 & 1.32 \\
59.05 & 0.11 & 0.34 & 0.32 & 1.63 & 0.46 & 2.39 & 0.60 & 3.78 & 0.50 & 3.32 \\
61.96 & 0.09 & 0.33 & 0.35 & 1.98 & 0.34 & 1.94 & 0.66 & 4.57 & 0.62 & 2.56 \\
65.04 & 0.14 & 0.55 & 0.22 & 1.38 & 0.36 & 2.26 & 0.38 & 2.89 & 0.66 & 2.28 \\
68.29 & 0.22 & 0.91 & 0.37 & 2.54 & 0.47 & 3.23 & 0.41 & 3.50 & 0.87 & 1.64 \\
71.75 & 0.18 & 0.82 & 0.44 & 3.32 & 0.52 & 3.95 & 0.73 & 6.78 & 1.12 & 1.21 \\
75.41 & 0.13 & 0.64 & 0.29 & 2.47 & 0.25 & 2.08 & 0.57 & 5.88 & 0.70 & 1.85 \\
79.30 & 0.07 & 0.39 & 0.10 & 0.93 & 0.20 & 1.85 & 0.28 & 3.21 & 0.32 & 3.91 \\
83.43 & 0.06 & 0.39 & 0.09 & 0.97 & 0.22 & 2.30 & 0.29 & 3.67 & 0.42 & 2.77 \\

\end{tabular}
\end{ruledtabular}
\end{table}

\begin{table}[ht]
\caption{\label{tab:allLimits_higherEn}95\% confidence level limits from all ROIs for fit energies from 87.82--300 GeV.  The first column for each ROI is the ($\Phi_{\gamma\gamma}$) upper limit in $10^{-9}$ cm$^{3}$s$^{-1}$.  The second column for each is the upper limit on $\langle\sigma v\rangle_{\gamma\gamma}$ in $10^{-28}$ cm$^{3}$s$^{-1}$ for the DM profile for which that ROI is optimal.  For R180, we give lower limit on $\tau_{\gamma\nu}$(NFW) in $10^{29}$ s.  Note that for $\tau_{\gamma\nu}$, the energy is $m_{\chi}/2$.}
\begin{ruledtabular}
\begin{tabular}{ccccccccccc}
& \multicolumn{2}{c}{R3} & \multicolumn{2}{c}{R16} & \multicolumn{2}{c}{R41} & \multicolumn{2}{c}{R90} & \multicolumn{2}{c}{R180} \\
Energy &  $\Phi_{\gamma\gamma}$ & $\langle\sigma v \rangle_{\gamma\gamma}$ & $\Phi_{\gamma\gamma}$ & $\langle\sigma v \rangle_{\gamma\gamma}$ & $\Phi_{\gamma\gamma}$ & $\langle\sigma v \rangle_{\gamma\gamma}$ & $\Phi_{\gamma\gamma}$ & $\langle\sigma v \rangle_{\gamma\gamma}$ & $\Phi_{\gamma\gamma}$ & $\tau_{\gamma X}$ \\
(GeV) &  & NFWc & & Ein &  & NFW &  & Iso &  & NFW \\
\hline
87.82 & 0.08 & 0.56 & 0.18 & 2.02 & 0.27 & 3.07 & 0.25 & 3.53 & 0.71 & 1.57 \\
92.51 & 0.04 & 0.34 & 0.10 & 1.29 & 0.34 & 4.28 & 0.40 & 6.25 & 0.67 & 1.59 \\
97.50 & 0.06 & 0.48 & 0.08 & 1.18 & 0.33 & 4.66 & 0.55 & 9.44 & 0.54 & 1.86 \\
102.82 & 0.11 & 1.05 & 0.14 & 2.22 & 0.48 & 7.47 & 0.57 & 11.00 & 0.46 & 2.09 \\
108.49 & 0.06 & 0.67 & 0.22 & 3.81 & 0.45 & 7.85 & 0.32 & 6.73 & 0.45 & 2.01 \\
114.51 & 0.10 & 1.13 & 0.33 & 6.49 & 0.37 & 7.19 & 0.19 & 4.52 & 0.28 & 3.03 \\
120.89 & 0.15 & 2.01 & 0.42 & 9.01 & 0.30 & 6.50 & 0.23 & 6.00 & 0.46 & 1.77 \\
127.66 & 0.28 & 4.09 & 0.37 & 8.94 & 0.42 & 10.15 & 0.51 & 15.08 & 0.50 & 1.52 \\
134.86 & 0.31 & 5.05 & 0.38 & 10.32 & 0.51 & 13.59 & 0.63 & 20.86 & 0.63 & 1.15 \\
142.51 & 0.25 & 4.52 & 0.28 & 8.28 & 0.35 & 10.46 & 0.52 & 19.23 & 0.47 & 1.46 \\
150.66 & 0.11 & 2.16 & 0.14 & 4.78 & 0.24 & 7.91 & 0.39 & 16.01 & 0.47 & 1.37 \\
159.32 & 0.06 & 1.42 & 0.18 & 6.80 & 0.16 & 5.97 & 0.28 & 12.87 & 0.28 & 2.19 \\
168.56 & 0.06 & 1.59 & 0.20 & 8.63 & 0.15 & 6.13 & 0.36 & 18.31 & 0.37 & 1.57 \\
178.41 & 0.12 & 3.46 & 0.20 & 9.62 & 0.21 & 10.01 & 0.37 & 21.20 & 0.34 & 1.60 \\
188.92 & 0.11 & 3.50 & 0.14 & 7.50 & 0.10 & 5.17 & 0.15 & 9.65 & 0.21 & 2.41 \\
200.15 & 0.08 & 2.85 & 0.12 & 7.33 & 0.10 & 5.75 & 0.09 & 6.70 & 0.17 & 2.83 \\
212.16 & 0.05 & 2.18 & 0.14 & 9.25 & 0.13 & 8.52 & 0.08 & 6.71 & 0.13 & 3.48 \\
225.08 & 0.07 & 3.41 & 0.06 & 4.15 & 0.10 & 7.55 & 0.06 & 5.95 & 0.11 & 3.98 \\
239.01 & 0.04 & 2.02 & 0.05 & 4.41 & 0.11 & 9.18 & 0.07 & 7.74 & 0.12 & 3.48 \\
254.05 & 0.05 & 2.99 & 0.08 & 7.69 & 0.16 & 15.40 & 0.16 & 18.13 & 0.14 & 2.82 \\
270.33 & 0.04 & 2.58 & 0.09 & 10.26 & 0.12 & 12.94 & 0.13 & 16.57 & 0.14 & 2.68 \\
300.00 & 0.04 & 3.29 & 0.13 & 17.62 & 0.23 & 31.02 & 0.30 & 48.83 & 0.35 & 0.93 \\
\end{tabular}
\end{ruledtabular}
\end{table}